\documentclass[dvipsnames]{lmcs} 
\pdfoutput=1
\usepackage[utf8]{inputenc}

\usepackage{lastpage}
\lmcsdoi{22}{4}{5}
\lmcsheading{}{\pageref{LastPage}}{}{}%
{Oct.~27,~2025}{Oct.~15,~2026}{}

\keywords{simple grammars, bisimulation, session types}

\usepackage{xcolor}
\usepackage{thm-restate}
\usepackage{xspace}
\usepackage{multirow}
\usepackage{array}
\newcolumntype{M}[1]{>{\centering\arraybackslash}m{#1}}
\usepackage{mathpartir}
\usepackage{tikz}
\tikzset{MyNode/.style 2 args={draw,thick,
minimum height=0.6cm,rounded corners = 0.3cm,
label={[anchor=210]0:{\scriptsize #1}},
label={[anchor=160]0:{\scriptsize #2}}}}
\tikzset{MyFinishedNode/.style 2 args={draw,thick,double,
minimum height=0.6cm,rounded corners = 0.3cm,
label={[anchor=210]0:{\scriptsize #1}},
label={[anchor=160]0:{\scriptsize #2}}}}
\usepackage{caption}
\usepackage{subcaption}
\usepackage{booktabs}
\usepackage{siunitx}

\usepackage{doi}
\usepackage{hyperref}
\usepackage[capitalize]{cleveref}

\newcommand{\freest}{\textsc{FreeST}\xspace}

\newcommand{\typec}[1]{{\color{RoyalBlue}{#1}}} 

\newcommand{\blk}[1]{{\color{black}{#1}}}

\newcommand{\grmeq}{\; ::= \;}
\newcommand{\grmor}{\; \mid \;}

\newcommand{\keyword}[1]{\ensuremath{\mathsf{#1}}\xspace}
\newcommand{\tkeyword}[1]{\keyword{\typec{#1}}}

\newcommand{\Label}[1]{\ensuremath{\mathrm{#1}}\xspace}

\newcommand{\typename}[1]{\typec{T_{\mathrm{#1}}}}

\newcommand{\bpai}{\operatorname{BPA1}}
\newcommand{\bpaii}{\operatorname{BPA2}}
\newcommand{\aloop}{\operatorname{loop}}
\newcommand{\refl}{\operatorname{refl}}
\newcommand{\pfail}{\operatorname{pFail}}
\newcommand{\tfail}{\operatorname{tFail}}
\newcommand{\word}{\operatorname{word}}
\newcommand{\free}{\operatorname{free}}

\newcommand{\Empty}{\varepsilon} 

\newcommand{\teq}{\simeq} 

\newcommand{\Oplus}{\hspace{-.1ex}\oplus\hspace{-.1ex}} 

\newcommand{\nbb}{\mathbb{N}}
\newcommand{\B}{\mathcal{B}}
\newcommand{\G}{\mathcal{G}}

\newcommand{\ocal}{\mathcal{O}}
\newcommand{\pcal}{\mathcal{P}}
\newcommand{\R}{\mathcal{R}}
\newcommand{\scal}{\mathcal{S}}
\newcommand{\T}{\mathcal{T}}
\newcommand{\V}{\mathcal{V}}

\newcommand{\declrel}[2]{\text{#1}\hfill\fbox{{#2}}}

\newcommand{\Skip}{\tkeyword{skip}}
\newcommand{\IN}[1]{\typec{?{#1}}}
\newcommand{\OUT}[1]{\typec{!{#1}}}

\newcommand{\reck}{\mu}
\newcommand{\REC}[2]{\typec{\reck{#1}.{#2}}}
\newcommand{\recordf}[3]{\{{#1}\colon {#2}\}_{{#1}\in{#3}}} 
\newcommand{\intchoice}{\typec\Oplus} 
\newcommand{\extchoice}{\typec\&} 
\newcommand{\intchoices}[1]{\intchoice\{{#1}\}} 
\newcommand{\extchoices}[1]{\extchoice\{{#1}\}} 

\newcommand{\semit}[2]{\typec{{#1};{#2}}}
\newcommand{\Bool}{\tkeyword{bool}}
\newcommand{\Int}{\tkeyword{int}}

\newcommand{\MT}{\typec{M}}
\newcommand{\TT}{\typec{T}}
\newcommand{\UT}{\typec{U}}
\newcommand{\VT}{\typec{V}}

\newcommand{\aT}{\typec{a}}
\newcommand{\bT}{\typec{b}}
\newcommand{\botT}{\typec{\bot}}

\newcommand{\subs}[3]{\typec{#1}\blk[{\typec{#3}}\blk/{\typec{#2}}\blk]} 

\newcommand{\judgementlabel}[1]{\mathrm{#1}} 

\newcommand{\judgement}[2]{{#1} \: \judgementlabel{#2}}

\newcommand{\judgementrel}[3]{{#1} \; {#2} \; {#3}}

\newcommand{\istypecta}[2]{{#1} \vdash \typec{#2}} 

\newcommand{\isdone}[1]{\judgement{\typec{#1}}{\checkmark}}
\newcommand{\isnotdone}[1]{\judgement{\typec{#1}}{\not\!\checkmark}}

\newcommand{\isequiv}[2]{\judgementrel{\typec{#1}}{\teq}{\typec{#2}}}
\newcommand{\isnotequiv}[2]{\judgementrel{\typec{#1}}{\not\teq}{\typec{#2}}}
\newcommand{\isbisim}[2]{\judgementrel{{#1}}{\sim}{{#2}}}
\newcommand{\isbisimG}[3]{\judgementrel{{#1}}{\sim_{#3}}{{#2}}}
\newcommand{\isbisimn}[3]{\judgementrel{{#1}}{\sim_{#3}}{{#2}}}
\newcommand{\isnotbisim}[2]{\judgementrel{{#1}}{\not\sim}{{#2}}}
\newcommand{\isnotbisimn}[3]{\judgementrel{{#1}}{\not\sim_{#3}}{{#2}}}

\newcommand{\congBtwo}[2]{\equiv^{\mathsf{#1}}_{#2}}

\newcommand{\iscongcoiB}[3]{\judgementrel{{#2}}{\congBtwo{c}{#1}}{{#3}}}
\newcommand{\iscongcoi}[2]{\iscongcoiB{\B}{#1}{#2}}

\newcommand{\iscontrt}[1]{\judgement{\typec{#1}}{contr}} 

\newcommand{\ltsred}[3]{#1\overset{#2}{\rightarrow}#3}
\newcommand{\notltsred}[3]{#1\overset{#2}{\not\rightarrow}#3}
\newcommand{\normof}[1]{\|#1\|}
\newcommand{\seminormof}[1]{\keyword{s}(#1)}
\newcommand{\normdynamic}[2]{[#1]_{#2}}

\ifx\vv\undefined
\newcommand{\vv}[1]{\marginpar{\textcolor{blue}{#1}}}
\else
\renewcommand{\vv}[1]{\marginpar{\textcolor{blue}{#1}}}
\fi

\newcommand{\ie}{i.e.,} 
\newcommand{\eg}{e.g.,}  
\newcommand{\etal}{et al.\ } 
\newcommand{\vs}{vs.\ }  

\theoremstyle{plain}
\newtheorem{case}{Case}
\crefname{case}{Case}{Cases}
\crefname{thm}{Theorem}{Theorems}
\crefname{prop}{Proposition}{Propositions}
\crefname{lem}{Lemma}{Lemmas}
\crefname{defi}{Definition}{Definitions}

\begin{document}

\title[Simple grammar bisimilarity and session type equivalence]{Simple grammar bisimilarity, with an application to session type equivalence}
\thanks{This work was supported by FCT - Fundação para a Ciência e a Tecnologia, I.P., through the LASIGE Research Unit, ref.\ UID/408/2025, and Instituto de Telecomunicações, ref.\ UID/50008/2025.}

\author[D.~Poças]{Diogo Poças\lmcsorcid{0000-0002-5474-3614}}[a,b]
\author[G.~Silva]{Gil Silva\lmcsorcid{0009-0007-7051-6116}}[c]
\author[V.~Vasconcelos]{Vasco T. Vasconcelos\lmcsorcid{0000-0002-9539-8861}}[c]

\address{Departamento de Matemática, Instituto Superior Técnico, Universidade de Lisboa, Portugal}	
\email{diogo.pocas@tecnico.ulisboa.pt}  

\address{Instituto de Telecomunicações, Lisboa, Portugal}	

\address{LASIGE, Faculdade de Ciências, Universidade de Lisboa, Portugal}
\email{gasilva@ciencias.ulisboa.pt, vmvasconcelos@ciencias.ulisboa.pt}  






\begin{abstract}
We provide an algorithm for deciding simple grammar bisimilarity whose complexity is polynomial in the valuation of the grammar (maximum seminorm among production rules). 
Since the valuation is at most exponential in the size of the grammar, this gives rise to a (single) exponential running time. 
Previously only a double-exponential algorithm was known. 
As an application, we provide a conversion from context-free session types to simple grammars whose valuation is linear in the size of the type. 
In this way, we provide the first polynomial-time algorithm for deciding context-free session type equivalence.
\end{abstract}


\maketitle


\section{Introduction}

A fundamental decision problem in theoretical computer science is the equivalence problem: given two machines, are they equivalent? 
Naturally, the answer depends on what we mean by `machine' and by `equivalent'. 
For computation models, such as finite automata, pushdown automata or Turing machines, `equivalence' typically means language equivalence, \ie~whether two machines accept the same language. 
In process theory, however, instead of language equivalence one is typically interested in finer notions of equivalence based on semantics, observations, and/or behaviors. 
van Glabbeek proposed a spectrum of equivalence notions, capturing aspects such as nondeterminism, parallelism and nontermination~\cite{Glabbeek1990}. 
The finest (most discriminating) of these notions is called \emph{bisimilarity}, or bisimulation equivalence~\cite{DBLP:conf/ijcai/Milner71,DBLP:books/sp/Milner80,DBLP:conf/tcs/Park81}.
Popularized by Milner, bisimilarity has become the standard notion of equivalence in settings such as concurrent and distributed systems, \eg~the Calculus of Communicating Systems~\cite{Milner1989} and the $\pi$-calculus~\cite{DBLP:journals/iandc/MilnerPW92a,DBLP:journals/iandc/MilnerPW92b}.

This work looks at the bisimilarity problem for \emph{simple grammars}. 
Simple grammars are deterministic context-free grammars (in Greibach normal form~\cite{greibach:1965:normalform}), where determinism means that there cannot be multiple transitions from any nonterminal symbol $X$ with the same terminal symbol $a$. 
More generally, context-free grammars constitute one of the simplest extensions of finite-state automata (\ie~regular expressions) with increased expressive power. 
In process theory, context-free processes are also called Basic Process Algebra (BPA) processes. 
Ever since the classical result showing that language equivalence of context-free grammars is undecidable~\cite{barhilleletal:1961:formalpropertiesgrammars}, several authors have looked at the corresponding bisimilarity problem for BPA processes, eventually showing it to be decidable~\cite{DBLP:conf/parle/BaetenBK87,DBLP:conf/mfcs/BurkartCS95,DBLP:journals/iandc/ChristensenHS95}. 
The exact complexity of the problem is still open; it is known to lie somewhere between EXPTIME~\cite{DBLP:journals/ipl/Kiefer13} and 2-EXPTIME~\cite{DBLP:journals/corr/abs-1207-2479}. 
The fact that language equivalence is undecidable while bisimilarity is decidable (for context-free grammars) provides further incentive for considering bisimilarity in practical applications. 

Besides the numerous applications of bisimilarity in process theory, we would like to emphasize its role in efficient compiler design in type theory, specifically in the setting of \emph{session types}. 
Session types provide a framework for describing structured communication~\cite{DBLP:conf/concur/Honda93,DBLP:conf/esop/HondaVK98,DBLP:conf/parle/TakeuchiHK94}, using primitives for sending/receiving messages and offering/selecting choices. 
Although traditional session types can be described by regular languages, extensions have been proposed that move beyond the limitations of tail recursion, increasing the expressive power of the underlying theory. 
Context-free session types are one such extension~\cite{DBLP:conf/tacas/AlmeidaMV20,DBLP:journals/tcs/CostaMPV24,DBLP:conf/esop/PocasCMV23,DBLP:conf/icfp/ThiemannV16}, currently implemented in the \freest programming language~\cite{DBLP:journals/corr/abs-1904-01284}. 
In the process of building a compiler for programming with session types, a relevant subroutine is type equivalence: given two types, the compiler must determine whether they are equivalent. 
It turns out that the context-free session type equivalence problem directly translates to the simple grammar bisimilarity problem. 
This has been our main motivation to study bisimilarity, both theoretically (understanding its computational complexity) and practically (designing and implementing efficient algorithms). 
Currently an online tool exists for testing simple grammar bisimilarity, based
on the ideas behind the 2-EXPTIME algorithm for context-free
grammars~\cite{DBLP:conf/tacas/AlmeidaMV20,SGBisim}.

The main result of this paper is an efficient procedure, which we call \emph{basis-updating algorithm}, for deciding simple grammar bisimilarity.

\begin{thm}
The basis-updating algorithm decides whether two words $\gamma$, $\delta$ in a simple grammar $\G$ are bisimilar. Its time complexity is polynomial on the size of $\G$, the valuation of $\G$, and the seminorms of $\gamma,\delta$.
\end{thm} 

\noindent
The seminorm of a word is the norm of its maximal normed prefix, and the valuation of a grammar is the maximum seminorm of (the right-hand sides of)
all productions. These notions are used to measure the ``complexity'' of the grammar. Since
the valuation is at most exponential in the size of the
grammar~\cite{DBLP:journals/corr/abs-1207-2479}, we obtain an exponential
running time, improving on the double-exponential upper bound for general
context-free grammars~\cite{DBLP:conf/tacas/AlmeidaMV20}.

In addition, we present a conversion from context-free session types to simple
grammars whose valuation is linear in the size of the type. In this way, we
obtain the first polynomial-time algorithm for deciding context-free session
type equivalence.

\begin{thm}
There is a polynomial-time algorithm for determining whether two context-free
session types $\TT$ and $\UT$ are equivalent.
\end{thm} 

\subsection{Related work}

The literature on equivalence checking  for infinite-state systems is vast; we refer the reader to surveys by Jančar and Moller~\cite{DBLP:conf/concur/JancarM99}, Srba~\cite{DBLP:journals/eatcs/Srba02}, and Moller~\etal~\cite{DBLP:journals/iandc/MollerSS04}. 
Below we focus on the results pertaining language equivalence and bisimulation
equivalence for three settings of interest: finite-state automata, context-free
grammars, and pushdown automata. 
\Cref{fig:table-of-results} (left) summarizes the state of the art for language equivalence.

Finite-state automata are described by transitions of the form
$\ltsred{X}{a}{Y}$, where $X$, $Y$ are states and $a$ is a transition label.
Hopcroft and Karp~\cite{Hopcroft1971} presented a polynomial-time algorithm for
deciding language equivalence for deterministic finite-state automata. However,
for nondeterministic automata, the language equivalence problem was shown to be
PSPACE-complete by Hunt \etal~\cite{DBLP:journals/jcss/HuntRS76}. Finite-state
automata correspond, in the setting of processes, to finite labelled transition
systems. In contrast to the PSPACE-completeness result, the bisimilarity problem
of finite labelled transition systems admits a polynomial-time algorithm, even
in the nondeterministic
case~\cite{Kanellakis1990,DBLP:journals/siamcomp/PaigeT87}.

Context-free grammars (in Greibach normal form) are described by productions of
the form $\ltsred{X}{}{a\gamma}$, where $X$ is a nonterminal symbol, $a$ is
a terminal symbol (understood as a transition label) and  $\gamma$ is a word of nonterminal symbols. 
The classical work by Bar-Hillel~\etal~\cite{barhilleletal:1961:formalpropertiesgrammars} shows that the language equivalence problem for context-free grammars is undecidable. 
Several authors studied the corresponding bisimilarity problem in process theory, where context-free processes are also known as Basic Process Algebra (BPA) processes. 
Here the problem may be simplified along two different axes: one may consider
deterministic processes, or one may consider normed processes (\ie
processes that may reach the empty process). 
Baeten~\etal~\cite{DBLP:conf/parle/BaetenBK87} showed that bisimilarity for
normed processes is decidable, and later
Hirschfeld~\etal~\cite{DBLP:journals/tcs/HirshfeldJM96} found a polynomial-time
algorithm (the worst-case running time has since been
improved~\cite{Czerwinski2010}). The existence of a polynomial-time algorithm
for normed context-free processes directly implies a polynomial-time algorithm
for language equivalence of deterministic context-free grammars, also called
simple grammars (the main object of study in this paper). For the general
(unnormed) setting, the first proof that bisimilarity is decidable was given by
Christensen~\etal~\cite{DBLP:journals/iandc/ChristensenHS95}. Subsequent works
have shown that the problem is
EXPTIME-hard~\cite{DBLP:journals/ipl/Kiefer13,DBLP:conf/icalp/Srba02} (lower
bound) and lies in 
2-EXPTIME~\cite{DBLP:conf/mfcs/BurkartCS95,DBLP:journals/corr/abs-1207-2479} (upper bound). 
We remark that the lower bound results by Srba or Kiefer cannot be applied to our setting of interest (deterministic context-free grammars), since the known reductions introduce non-deterministic transitions. 
It is also worth remarking that, since simple grammars are deterministic, bisimilarity coincides with \emph{trace equivalence} (which in general is a coarser notion). 
\Cref{fig:table-of-results} (right) summarizes the state of the
art for context-free grammar bisimilarity.

We should also briefly mention pushdown automata, whose memory consists of a state and a stack. 
In this setting, transitions are of the form $\ltsred{(p,X)}{a}{(q,\gamma)}$, where $p,q$ are states, $X$ is the symbol at the top of the stack and $\gamma$ is a (possibly empty) word of stack symbols (which replaces $X$). 
Since pushdown automata generalize context-free grammars, the language equivalence problem is obviously undecidable as well. 
The long-standing equivalence problem for deterministic pushdown automata was famously resolved by Sénizergues~\cite{DBLP:journals/tcs/Senizergues01}, who received the Gödel Prize for his work. 
Further effort has been put in finding simpler proofs and improved complexity upper bounds~\cite{Jancar2010,DBLP:conf/fossacs/Jancar14,DBLP:journals/jcss/Jancar21,DBLP:conf/icalp/Stirling02}.
Regarding the bisimilarity problem for pushdown automata, it is known to be ACKERMANN-complete in the unnormed setting, and TOWER-hard in the normed setting~\cite{DBLP:conf/lics/BenediktGKM13,DBLP:conf/lics/JancarS19,DBLP:conf/focs/Senizergues98,DBLP:journals/tcs/Stirling98,DBLP:conf/icalp/0004YL020}.

\begin{figure}
\begin{minipage}{0.49\textwidth}\centering
\begin{tabular}{|M{4.6em}|M{5.8em} M{5.9em}|}
\hline
\small{Language equivalence} & \small{deterministic} & \small{non-deterministic} \\ 
\hline
  & & \\[-1em] 
\tiny{Finite-state automata} & \small{P~\cite{Hopcroft1971}} & \tiny{PSPACE-complete \cite{DBLP:journals/jcss/HuntRS76}} \\
\tiny{Context-free grammars} & \small{P~\cite{DBLP:journals/tcs/HirshfeldJM96}} & \tiny{undecidable \cite{barhilleletal:1961:formalpropertiesgrammars}} \\
\tiny{Pushdown automata} & 
\tiny{TOWER \cite{DBLP:conf/fossacs/Jancar14,DBLP:journals/tcs/Senizergues01}} & 
\tiny{undecidable}\\ 
\hline
\end{tabular}
\end{minipage}\begin{minipage}{0.49\textwidth}\centering
\begin{tabular}{|M{4.6em}|M{5.8em} M{5.9em}|}
\hline
\small{CFG bisimilarity} & \small{deterministic} & \small{non-deterministic} \\
\hline
  & & \\[-.6em] 
  \small{normed} & \small{P~\cite{DBLP:journals/tcs/HirshfeldJM96}} & \small{P~\cite{DBLP:journals/tcs/HirshfeldJM96}}  \\[.4em]
\multirow{2}{*}{\small{unnormed}} & \multirow{2}{*}{\tiny{\textbf{EXPTIME}}} & 
  \tiny{EXPTIME-hard \cite{DBLP:journals/ipl/Kiefer13}}\\ 
  & & \tiny{2-EXPTIME \cite{DBLP:conf/mfcs/BurkartCS95,DBLP:journals/corr/abs-1207-2479}}\\
\hline
\end{tabular}
\end{minipage}

\caption{Left: relevant results on language equivalence problems. Right: relevant results on bisimilarity problems over context-free grammars (CFG). Prior to our work (contribution in \textbf{bold}), the best known algorithm for simple grammar bisimilarity was the 2-EXPTIME algorithm for CFG.}\label{fig:table-of-results}
\end{figure}

\subsection{Paper outline and technical contributions}

To conclude this section, we provide a brief outline of the paper, including a short description of the technical contributions and challenges we had to overcome.

In \cref{sec:preliminaries} we formally present the simple grammar bisimilarity problem as well as relevant notions underlying our algorithm. 
The norm of a word is as proposed by Baeten~\etal~\cite{DBLP:conf/parle/BaetenBK87}; 
the seminorm of a word is inspired by Christensen~\etal~\cite{DBLP:journals/iandc/ChristensenHS95}; 
the canonical norm-reducing sequence is inspired by Hirshfeld~\etal~\cite{DBLP:journals/tcs/HirshfeldJM96};
the canonical seminorm-reducing sequence is its obvious generalization to seminorms; 
the valuation of a grammar, inspired by Caucal~\cite{DBLP:conf/oa/Caucal89} and Jančar~\cite{DBLP:journals/corr/abs-1207-2479}, is a quantity of interest for analysing the complexity of our algorithms.
\Cref{lem:polytime-valuation} shows that these notions are efficiently computable. 

In \cref{sec:comparability}, we present our first main contribution, which is a unique characterization of all solutions to equations of the form $X \alpha \sim Y \beta$, where $X,Y$ are given and $\alpha,\beta$ are unknowns (\Cref{lem:charunnormedunnormed,lem:charnormedunnormed,lem:charnormnorm}). 
To achieve this, we propose notions of reducible and comparable pairs, which are inspired by Christensen~\etal~\cite{DBLP:journals/iandc/ChristensenHS95}. 
Previously, such a result was only known for the case where $X,Y$ are normed and $(X,Y)$ is reducible, which played a role in the polynomial-time algorithm of Hirshfeld~\etal~\cite{DBLP:journals/tcs/HirshfeldJM96} (for normed context-free grammars). 
The uniqueness results do not hold for general context-free grammars.

In \Cref{sec:bases-congruence}, we present our second main contribution, which is a novel notion of congruence, called \emph{coinductive congruence} (\Cref{def:coinductive-congruence}). 
Our notion is inspired by the BPA1 and BPA2 rules of Jančar and Moller~\cite{DBLP:conf/concur/JancarM99} (for BPA processes) and Almeida~\etal~\cite{DBLP:conf/tacas/AlmeidaMV20} (for context-free session types). 
As in the case of least congruence~\cite{Caucal1990}, we define the notion of self-bisimulation bases~\cite{DBLP:conf/mfcs/BurkartCS95,Caucal1990,DBLP:journals/iandc/ChristensenHS95,DBLP:journals/tcs/HirshfeldJM96}, and show that there is always a basis for which coinductive congruence coincides with bisimilarity (\Cref{thm:basis-existence}). 
The main advantage of considering coinductive congruence instead of least congruence is that it constitutes a decidable relation under mild assumptions on the basis (\Cref{thm:coinductive-congruence}).

In \Cref{sec:algorithm}, we present our third and last main contribution, which we call the \emph{basis-updating algorithm} for determining simple grammar bisimilarity.
Comparing with other works, we can think of Christensen~\etal's algorithm (for context-free grammars) as \emph{basis-guessing}: iterate through all possible bases, until a self-bisimulation is found. 
Burkart~\etal~\cite{DBLP:conf/mfcs/BurkartCS95} and Hirshfeld~\etal~\cite{DBLP:journals/tcs/HirshfeldJM96} propose \emph{basis-refining} algorithms: start with a large basis (that might not be a self-bisimulation); delete pairs of the basis in further iterations, until a self-bisimulation is found. 
We instead start with a small basis (containing only reflexive pairs $(X,X)$),
to which we either add new pairs or update existing pairs in further iterations, until a (sufficiently rich) self-bisimulation is found. 

In \Cref{sec:correctness}, we prove that our algorithm is correct. 
For a given pair $(\gamma,\delta)$, we show that a `YES'-answer by the algorithm means that $\isbisim{\gamma}{\delta}$ (\Cref{thm:yes-soundness}), using the fact that the basis produced is indeed a self-bisimulation.  
We show that a `NO'-answer means that $\isnotbisim{\gamma}{\delta}$ (\Cref{thm:no-soundness}), using the fact that the algorithm detects incomparable pairs, propagating these detections to the root node. 
Finally, we show that the algorithm always produces an answer (\Cref{thm:termination}), by arguing that the basis is updated polynomially many times, and that between two consecutive basis updates the number of iterations is at most exponential in the size of the input.

In \Cref{sec:session-types} we present context-free session types. 
We explain how session types can be converted into simple grammars, reducing type equivalence to grammar bisimilarity. 
We then show how our basis-updating algorithm can be used to obtain a polynomial-time algorithm for type equivalence (\Cref{thm:polytime-type-equivalence}). 
Finally, we provide experimental data which shows that our implementation outperforms the current
implementation in \freest~\cite{DBLP:conf/tacas/AlmeidaMV20}.

In \Cref{sec:conclusion} we briefly discuss our results and present directions for further research. 

\section{Background}\label{sec:preliminaries}

In this section, we introduce some useful assumptions, properties and notions of context-free grammars, to be used in the proofs that follow. 

\subsection{Grammars and bisimulations}

We begin by defining simple grammars and the bisimilarity problem.

\begin{defi}[Context-free grammar, simple grammar]
A \emph{context-free grammar} (in Greibach normal form) is a triple $\G=(\V,\T,\pcal)$ where: 
$\V$ is a finite set of nonterminal symbols, denoted by $X,Y,Z,\ldots$; 
$\T$ is a finite set of terminal symbols, denoted by $a,b,c,\ldots$; 
and $\pcal\subseteq\V\times\T\times\V^\ast$ is a finite set of productions. 
We use Greek letters $\alpha,\beta,\gamma,\delta,\ldots$ to denote (possibly empty) words of nonterminals in $\V^\ast$, $\Empty$ for the empty word, and we write $\ltsred X a \gamma$ to denote a production $(X,a,\gamma)\in\pcal$. 
A \emph{simple grammar} is a context-free grammar $\G=(\V,\T,\pcal)$ where, for every nonterminal $X\in\V$ and every terminal $a\in\T$, there is at most one production of the form $\ltsred X a \gamma$. 
\end{defi}

\begin{defi}[Labeled transition system]
A \emph{labeled transition system} (LTS) is given by a (possibly infinite) set of states, and transitions between these states, labeled using a finite set of labels.
A context-free grammar $\G=(\V,\T,\pcal)$ induces an LTS on the space $\V^\ast$ of words of nonterminals: 
for every production $\ltsred X a \gamma$ and every word $\delta$, the LTS includes the labeled transition $\ltsred{X\delta}a{\gamma\delta}$. 
We extend the notation $\ltsred{\gamma}{a}{\delta}$ to allow for arbitrary words of terminal symbols $u,v,\ldots$ in $\T^\ast$: 
for every $\gamma$, we define $\ltsred{\gamma}{\Empty}{\gamma}$; 
and $\ltsred{\gamma}{au}{\delta}$ if there exists $\gamma'$ with $\ltsred{\gamma}a{\gamma'}$ and $\ltsred{\gamma'}u\delta$. 
\end{defi}

\noindent
The main notion we wish to study is \emph{bisimilarity}, or bisimulation equivalence, between two words of a context-free grammar. 

\begin{defi}[Bisimulation, bisimilarity]
Given a context-free grammar $\G=(\V,\T,\pcal)$,  a \emph{bisimulation} is a relation $\R\subseteq\V^\ast\times\V^\ast$ such that, for every pair $(\gamma,\delta)\in\R$ and every terminal $a\in\T$:
\begin{enumerate}
\item if $\ltsred\gamma a{\gamma'}$ for some $\gamma'$, then there exists some $\delta'$ such that $\ltsred\delta a{\delta'}$ and $(\gamma',\delta')\in \R$;
\item if $\ltsred\delta a{\delta'}$ for some $\delta'$, then there exists some $\gamma'$ such that $\ltsred\gamma a{\gamma'}$ and $(\gamma',\delta')\in \R$.
\end{enumerate}
We say that two words of nonterminals are \emph{bisimilar}, written $\isbisim\gamma\delta$, if there exists a bisimulation $\R$ such that $(\gamma,\delta)\in\R$. 
\end{defi}

\noindent
We recall that bisimilarity is an equivalence relation~\cite{Milner1989}. 
The main problem under consideration is \emph{simple grammar bisimilarity:} 
\begin{prob}[Simple grammar bisimilarity]
Given a simple grammar $\G=(\V,\T,\pcal)$ and two words of nonterminals $\gamma,\delta\in\V^\ast$, determine whether $\isbisim\gamma\delta$.
\end{prob}

\noindent
We say that a nonterminal $X$ is \emph{dead} if it has no productions, \ie~$\notltsred X a \gamma$ for any $a,\gamma$. 
It is useful to assume that the grammar does not have dead nonterminals, and this is without loss of generality. 
Given a grammar $\G$, we may define an extended grammar $\G'$ without dead nonterminals preserving bisimilarity in $\G$; instead of $\isbisim{\gamma}{\delta}$, we shall write $\isbisimG\gamma\delta\G$ or $\isbisimG\gamma\delta{\G'}$ in order to make explicit the underlying grammar. 
The following result appears as a remark in
Jančar~\cite{DBLP:journals/corr/abs-1207-2479}; we provide a detailed proof that
illustrates the bisimulation technique.

\begin{restatable}[Dead nonterminal removal]{prop}{deadnonterminal}
Let $\G$ be a context-free grammar with at least one dead nonterminal, denoted by $\bot$. 
Let $\G'$ be a grammar obtained from $\G$ by introducing a fresh terminal symbol $d$, as well as productions $\ltsred XdX$ for every dead nonterminal $X$ in $\G$. 
For all words $\gamma,\delta$ in $\G$, we have that $\isbisimG\gamma\delta\G$ iff $\isbisimG{\gamma \bot}{\delta \bot}{\G'}$.
\end{restatable}

\begin{proof}
In the forward direction, consider the relation
$$\R = \{(\gamma \bot,\delta \bot) : \isbisimG\gamma\delta\G\};$$
we shall show that $\R$ is a bisimulation in $\G'$. This will imply that $\R\subseteq\ \sim_{\G'}$ and thus that $\isbisimG\gamma\delta\G$ implies $\isbisimG{\gamma \bot}{\delta \bot}{\G'}$. 
Letting $(\gamma \bot,\delta \bot)\in\R$, so that $\isbisimG\gamma\delta\G$, we consider a case analysis on $\gamma$:
\begin{itemize}
\item \textbf{Case} $\gamma$ has no transitions in $\G$: then either $\gamma=\Empty$ (empty word) or $\gamma=\bot'\gamma'$ for some dead nonterminal $\bot'$ (which may or may not be the same as $\bot$). 
In the first case, $\gamma\bot$ has the unique transition $\ltsred{\gamma \bot=\bot}d {\bot=\gamma\bot}$ in $\G'$. 
In the second case, $\gamma\bot$ has the unique transition $\ltsred{\gamma \bot=\bot'\gamma'\bot}d {\bot'\gamma'\bot=\gamma\bot}$ in $\G'$. 
Since $\isbisimG\gamma\delta\G$, $\delta$ also has no transitions in $\G$. 
Therefore, by the same reasoning, $\delta\bot$ has the unique transition $\ltsred{\delta \bot}d {\delta\bot}$ in $\G'$. 
Thus, we can match the transition for $\gamma\bot$ with the transition for $\delta\bot$, arriving at the same pair $(\gamma\bot,\delta\bot)\in\R$. 

\item \textbf{Case} $\gamma$ has a transition in $\G$: then, by construction, $\gamma\bot$ has exactly the transitions $\ltsred{\gamma\bot}a{\gamma'\bot}$ in $\G'$, for each transition $\ltsred\gamma a \gamma'$ in $\G$. 
Take one such transition $\ltsred{\gamma\bot}a{\gamma'\bot}$ in $\G'$. Since $\ltsred\gamma a \gamma'$ in $\G$ and $\isbisimG\gamma\delta\G$, there must be a matching transition $\ltsred\delta a\delta'$ in $\G$, and moreover $\isbisimG{\gamma'}{\delta'}\G$. 
Again by construction, $\delta\bot$ has a transition $\ltsred{\delta\bot}a{\delta'\bot}$ in $\G$. 
Matching these transitions, we arrive at pair $(\gamma'\bot,\delta'\bot)$, which is in $\R$ since $\isbisimG{\gamma'}{\delta'}\G$. 
\end{itemize}
This proves that we can match any transition for $\gamma\bot$ with some transition for $\delta\bot$. 
A similar reasoning (with a case analysis on $\delta$) shows that we can match any transition for $\delta\bot$ with some transition for $\gamma\bot$. 
Therefore, $\R$ is a bisimulation in $\G'$ as desired. 

In the converse direction, consider the relation
$$\R=\{(\gamma,\delta) : \isbisimG{\gamma\bot}{\delta\bot}{\G'}\};$$
we shall show that $\R$ is a bisimulation in $\G$. 
This will imply that $\R\subseteq\ \sim_\G$ and thus that $\isbisimG{\gamma\bot}{\delta\bot}{\G'}$ implies $\isbisimG\gamma\delta\G$. 
Letting $(\gamma,\delta)\in\R$, so that $\isbisimG{\gamma\bot}{\delta\bot}{\G'}$, we consider a case analysis on $\gamma$:
\begin{itemize}
\item \textbf{Case} $\gamma$ has no transitions in $\G$: as before, this implies that $\gamma\bot$ has the unique transition $\ltsred{\gamma\bot}d{\gamma\bot}$ in $\G'$. 
Since $\isbisimG{\gamma\bot}{\delta\bot}{\G'}$, $\delta\bot$ must have some transition by $d$ in $\G'$. 
This implies that $\delta=\Empty$ or $\delta$ starts with a dead nonterminal, so it has no transitions in $\G$. 
Thus, the condition for bisimulation holds vacuosly, since neither $\gamma$ nor $\delta$ have transitions in $\G$.
\item \textbf{Case} $\gamma$ has a transition in $\G$: take one such transition $\ltsred\gamma a \gamma'$, so that $\ltsred{\gamma\bot}{a}{\gamma'\bot}$ in $\G'$. 
Since $\isbisimG{\gamma\bot}{\delta\bot}{\G'}$, there must be a matching transition for $\delta\bot$ in $\G'$. 
By the same reasoning as before, that transition must be of the form $\ltsred{\delta\bot}{a}{\delta'\bot}$ for $\ltsred\delta a \delta'$ in $\G$. 
Matching these transitions, we arrive at pair $(\gamma',\delta')$, which is in $\R$ since $\isbisimG{\gamma'\bot}{\delta'\bot}{\G'}$.
\end{itemize}
Again, a similar reasoning (with a case analysis for $\delta$) shows that we can match any transition for $\delta$ with some transition for $\gamma$. 
We conclude that $\R$ is a bisimulation in $\G$ as desired.
\end{proof}

\noindent
From this point on we assume that our grammars do not have dead nonterminals. 
The main advantage is that bisimulation becomes a congruence with respect to sequential composition~\cite{DBLP:conf/rex/BergstraK88}, which simplifies many of the arguments in the proofs that follow. 
\Cref{prop:congruence-2} in the following proposition appears in Hirshfeld
\etal~\cite{DBLP:journals/tcs/HirshfeldJM96}; we take the opportunity to provide
a more detailed proof.

\begin{restatable}[Congruence]{prop}{congruence}
\label{prop:congruence}
Let $\G$ be a context-free grammar without dead nonterminals. Let $\alpha,\beta,\gamma,\delta$ be words of nonterminals in $\G$.
\begin{enumerate}
\item\label{prop:congruence-1} $\isbisim\gamma\Empty$ iff $\gamma=\Empty$.
\item\label{prop:congruence-2} If $\isbisim\alpha\beta$ and $\isbisim\gamma\delta$ then $\isbisim{\alpha\gamma}{\beta\delta}$.
\end{enumerate}
\end{restatable}

\begin{proof}
To prove \Cref{prop:congruence-1}, first observe that, by reflexivity of $\sim$, $\gamma=\Empty$ implies $\isbisim\gamma\Empty$. 
Conversely, if $\gamma\neq\Empty$ then $\gamma=X\gamma'$ for some nonterminal $X$. 
Since $X$ is not dead, $X$ has some transition, so that $\gamma$ has some transition as well, and thus $\isnotbisim\gamma\Empty$. 

To prove \Cref{prop:congruence-2}, consider the relation
$$\R=\{(\alpha\gamma,\beta\delta):\isbisim\alpha\beta,\isbisim\gamma\delta\};$$
we shall show that $\R$ is a bisimulation in $\G$, yielding the desired result. 
Letting $(\alpha\gamma,\beta\delta)\in\R$, so that $\isbisim\alpha\beta$ and $\isbisim\gamma\delta$, we consider a case analysis in $\alpha$:
\begin{itemize}
\item \textbf{Case} $\alpha=\Empty$: since $\isbisim\alpha\beta$, by \Cref{prop:congruence-1} we have that $\beta=\Empty$. 
Since $\isbisim\gamma\delta$, any transition $\ltsred{\alpha\gamma=\gamma}a{\gamma'}$ has a matching transition $\ltsred{\beta\delta=\delta}a{\delta'}$.
The resulting pair $(\gamma',\delta')$ is in $\R$ by taking $\alpha'=\beta'=\Empty$ and noting that $\isbisim{\gamma'}{\delta'}$.
\item \textbf{Case} $\alpha\neq\Empty$: since $\G$ has no dead nonterminals, $\alpha$ has at least one transition. Moreover, $\alpha\gamma$ has exactly the transitions $\ltsred{\alpha\gamma}{a}{\alpha'\gamma}$ for each transition $\ltsred\alpha a{\alpha'}$. 
Taking such a transition $\ltsred{\alpha}{a}{\alpha'}$, and considering that $\isbisim\alpha\beta$, we can find a matching transition $\ltsred{\beta}{a}{\beta'}$ with $\isbisim{\alpha'}{\beta'}$. 
Therefore, we can match the transition $\ltsred{\alpha\gamma}{a}{\alpha'\gamma}$ with the transition $\ltsred{\beta\delta}{a}{\beta'\delta}$, arriving at pair $(\alpha'\gamma,\beta'\delta)$, which is in $\R$ since $\isbisim{\alpha'}{\beta'}$ and $\isbisim{\gamma}{\delta}$.
\end{itemize}
By a similar case analysis in $\beta$, we can match transitions of $\beta\delta$ with transitions of $\alpha\gamma$, yielding that $\R$ is a bisimulation.
\end{proof}

\subsection{Bisimilarity approximations}

Some results in \Cref{sec:bases-congruence}
make use of the known fact that bisimilarity can be characterized via a sequence of approximations.

\begin{defi}[Bisimilarity approximations]
\label{def:bisim-approx}
Let $\G=(\V,\T,\pcal)$ be a context-free grammar. For each $n\in\nbb$, we define the $n$-th approximant of bisimilarity, denoted $\sim_n$, inductively on $n$.
\begin{itemize}
\item $\isbisimn{\gamma}{\delta}{0}$ for all words of nonterminals $\gamma,\delta\in\V^\ast$.
\item $\isbisimn{\gamma}{\delta}{n+1}$ if, for every terminal $a\in\T$:
\begin{enumerate}
\item if $\ltsred\gamma a{\gamma'}$ for some $\gamma'$, then there exists some $\delta'$ such that $\ltsred\delta a{\delta'}$ and $\isbisimn{\gamma'}{\delta'}{n}$;
\item if $\ltsred\delta a{\delta'}$ for some $\delta'$, then there exists some $\gamma'$ such that $\ltsred\gamma a{\gamma'}$ and $\isbisimn{\gamma'}{\delta'}{n}$.
\end{enumerate}
\end{itemize}
\end{defi}

\noindent
We briefly remark that $n$-th approximants may be defined generally for any LTS. A LTS is image-finite if, for every state $\gamma$ and label $a$, there are finitely many transitions of the form $\ltsred{\gamma}{a}{\gamma'}$; for example, a context-free grammar generates an image-finite LTS. The following result, which we state for context-free grammars, also applies to any image-finite LTS.

\begin{restatable}[Bisimilarity approximations~\cite{DBLP:journals/corr/abs-1207-2479,Milner1989}]{prop}{bisimapprox}
\label{prop:bisim-approx}
For any context-free grammar $\G=(\V,\T,\pcal)$ and $n\in\nbb$, $\sim_n$ is an equivalence relation. Moreover,
$$\sim_0\ \supseteq\ \sim_1\ \supseteq\ \sim_2\ \supseteq\cdots \supseteq \bigcap_{n\in\nbb}\sim_n\ =\ \sim.$$
In other words, for any words $\gamma,\delta\in\V^\ast$, we have that $\isbisim{\gamma}{\delta}$ iff $\isbisimn{\gamma}{\delta}{n}$ for every $n\in\nbb$.
\end{restatable}

\Cref{prop:congruence-approx-2} in the following proposition appears in Jančar
\etal~\cite{DBLP:journals/corr/abs-1207-2479}, we again take the opportunity to
provide a more detailed proof.

\begin{restatable}[Congruence for bisimilarity approximants]{prop}{congruenceapprox}
\label{prop:congruence-approx}
Let $\G$ be a context-free grammar without dead nonterminals. Let $\alpha,\beta,\gamma,\delta$ be words of nonterminals in $\G$. Let $n\in\nbb$.
\begin{enumerate}
\item\label{prop:congruence-approx-1} $\isbisimn{\gamma}{\Empty}{n}$ iff $n=0$ or $\gamma=\Empty$.
\item\label{prop:congruence-approx-2} If $\isbisimn{\alpha}{\beta}{n}$ and $\isbisimn{\gamma}{\delta}{n}$ then $\isbisimn{\alpha\gamma}{\gamma\delta}{n}$.
\item\label{prop:congruence-approx-3} If $\alpha\neq\Empty$ and $\isbisimn{\gamma}{\delta}{n}$ then $\isbisimn{\alpha\gamma}{\alpha\delta}{n+1}$.
\end{enumerate}
\end{restatable}

\begin{proof}
To prove \Cref{prop:congruence-approx-1}, first observe that if $n=0$ then trivially $\isbisimn{\gamma}{\Empty}{n}$; and if $\gamma=\Empty$ then by reflexivity $\isbisimn{\gamma}{\Empty}{n}$. Conversely, if $n>0$ and $\gamma\neq\Empty$, there is a transition of $\gamma$ without a match in $\Empty$, so that $\isnotbisimn{\gamma}{\Empty}n$.

We prove \Cref{prop:congruence-approx-2} by induction on $n$. The base case ($n=0$) is trivial since $\sim_0\ =\V^\ast\times\V^\ast$. For the induction step ($n+1$), suppose $\isbisimn{\alpha}{\beta}{n+1}$ and $\isbisimn{\gamma}{\delta}{n+1}$, and consider a case analysis on $\alpha$.
\begin{itemize}
\item \textbf{Case} $\alpha=\Empty$: by \Cref{prop:congruence-approx-1} we have that $\beta=\Empty$. 
Thus $\alpha\gamma=\isbisimn{\gamma}{\delta}{n+1}=\beta\delta$ as desired.
\item \textbf{Case} $\alpha\neq\Empty$: by \Cref{prop:congruence-approx-1} we have that $\beta\neq\Empty$. 
Moreover, $\alpha\gamma$ has exactly the transitions $\ltsred{\alpha\gamma}{a}{\alpha'\gamma}$ for each transition $\ltsred\alpha a{\alpha'}$.
Taking such a transition $\ltsred\alpha a{\alpha'}$, and considering that $\isbisimn\alpha\beta{n+1}$, we can find a matching transition $\ltsred{\beta}{a}{\beta'}$ with $\isbisimn{\alpha'}{\beta'}{n}$. 
Moreover, we have the transition $\ltsred{\beta\delta}{a}{\beta'\delta}$. 
Since $\isbisimn{\gamma}{\delta}{n+1}$ implies $\isbisimn{\gamma}{\delta}{n}$ (by \Cref{prop:bisim-approx}), by induction hypothesis we have $\isbisimn{\alpha'\gamma}{\beta'\delta}{n}$. 
Via a similar reasoning, we can match transitions from $\beta\delta$ with transitions from  $\alpha\gamma$. 
We conclude that $\isbisimn{\alpha\gamma}{\gamma\delta}{n+1}$ as desired.
\end{itemize}

To prove \Cref{prop:congruence-approx-3}, suppose $\alpha\neq\Empty$ and $\isbisimn{\gamma}{\delta}{n}$. 
Since $\alpha\neq \Empty$, $\alpha\gamma$ has exactly the transitions $\ltsred{\alpha\gamma}{a}{\alpha'\gamma}$ for each transition $\ltsred\alpha a{\alpha'}$; and similarly, $\alpha\delta$ has exactly the transitions $\ltsred{\alpha\delta}{a}{\alpha'\delta}$. 
Since, by reflexivity, $\isbisimn{\alpha'}{\alpha'}{n}$, we have by \Cref{prop:congruence-approx-2} that $\isbisimn{\alpha'\gamma}{\alpha'\delta}{n}$. 
Therefore, we can match transitions from $\alpha\gamma$ with transitions from  $\alpha\delta$ and vice-versa, concluding that $\isbisimn{\alpha\gamma}{\alpha\delta}{n+1}$ as desired.
\end{proof}

\subsection{Norm, seminorm and valuation}

\begin{defi}[Norm]
Given a context-free grammar $\G=(\V,\T,\pcal)$, we say that a word of nonterminals $\gamma\in\V^\ast$ is \emph{normed} if there exists some word of terminals $u\in\T^\ast$ such that $\ltsred{\gamma}{u}{\Empty}$; otherwise we say that $\gamma$ is \emph{unnormed}. 
We define the \emph{norm} of $\gamma$, denoted by $\normof\gamma$, as the minimum length of a word $u$ for which $\ltsred{\gamma}{u}{\Empty}$; if $\gamma$ is unnormed we write $\normof\gamma=\infty$. 
\end{defi}

\noindent
Clearly, a word $\gamma=X_1\cdots X_n$ is normed iff each of $X_1,\ldots,X_n$ is normed, in which case $\normof\gamma=\normof{X_1}+\cdots+\normof{X_n}$. 

\begin{restatable}[Bisimilarity-invariance of norms]{prop}{bisimnorm}
\label{prop:bisim-eq-norm}
Let $\G$ be a context-free grammar without dead nonterminals and $\gamma,\delta$ words of nonterminals in $\G$. 
If $\isbisim{\gamma}{\delta}$ then $\normof{\gamma}=\normof{\delta}$.
\end{restatable}

\begin{proof}
Suppose that $\isbisim{\gamma}{\delta}$. 
Let us prove that $\normof{\gamma}\leq\normof{\delta}$. If $\delta$ is unnormed then the inequality holds trivially. 
Otherwise let $u$ be a word of terminals of length $|u|=\normof{\delta}$ such that $\ltsred{\delta}{u}{\Empty}$. 
Since $\isbisim{\gamma}{\delta}$, there must be a matching sequence $\ltsred{\gamma}{u}{\gamma'}$ with $\isbisim{\gamma'}{\Empty}$. 
By \Cref{prop:congruence}, $\gamma'=\Empty$. 
In particular, $\gamma$ is normed and $\normof{\gamma}\leq|u|=\normof{\delta}$, as desired. 
A similar reasoning proves that $\normof{\gamma}\geq\normof{\delta}$, so that $\normof{\gamma}=\normof{\delta}$.
\end{proof}

\noindent
If $X$ is unnormed, then the subword $\delta$ in $\alpha X \delta$ can never be `reached' by a sequence of transitions. 
Hence, for the purposes of determining bisimilarity, we can discard all nonterminals following the first unnormed nonterminal. Under the \emph{pruning convention}, we shall assume that any unnormed word is of the form $\alpha X$ with $\alpha$ normed and $X$ unnormed. 
The statement below appears in
Christensen~\cite{DBLP:journals/iandc/ChristensenHS95} without proof.

\begin{restatable}[Pruning]{lem}{pruninglemma}
\label{prop:unnormedbisim} 
Let $X$ be a nonterminal and $\alpha,\gamma$ words of nonterminals in some context-free grammar $\G$. 
If $X$ is unnormed, then $\isbisim{\alpha X\gamma}{\alpha X}$.
\end{restatable}

\begin{proof}
We shall show that $\R$ is a bisimulation, where 
$$\R=\{(\alpha X\gamma,\alpha X\delta): X\text{ unnormed}\}.$$
Letting $(\alpha X\gamma,\alpha X\delta)\in\R$ with $X$ unnormed, we consider a case analysis in $\alpha$:
\begin{itemize}
\item \textbf{Case} $\alpha=\Empty$: then $\alpha X\gamma=X\gamma$ and $\alpha X\delta=X\delta$. $X\gamma$ has exactly the transitions $\ltsred{X\gamma}a{\alpha'\gamma}$ for each production $\ltsred Xa\alpha'$. 
Similarly, $X\delta$ has exactly the transitions $\ltsred{X\delta}a{\alpha'\delta}$ for each production $\ltsred Xa\alpha'$. 
Consider one such transition $\ltsred Xa\alpha'$; clearly $\alpha'$ must be unnormed, otherwise $\normof{X}\leq 1+\normof{\alpha'}<\infty$. 
Thus $\alpha'$ must contain at least one unnormed nonterminal, \ie we can write $\alpha'=\alpha''X'\gamma'$ with $X'$ unnormed. 
The transition $\ltsred{X\gamma}a{\alpha'\gamma=\alpha''X'\gamma'\gamma}$ can be matched with transition $\ltsred{X\delta}a{\alpha'\delta=\alpha''X'\gamma'\delta}$, and vice-versa. Moreover, the resulting pair $(\alpha''X'\gamma'\gamma,\alpha''X'\gamma'\delta)$ is in $\R$ since $X'$ is unnormed.
\item \textbf{Case} $\alpha\neq\Empty$: in this case $\alpha X\gamma$ has exactly the transitions $\ltsred{\alpha X\gamma}a{\alpha' X\gamma}$ for each transition $\ltsred\alpha a{\alpha'}$. 
Similarly, $\alpha X\delta$ has exactly the transitions $\ltsred{\alpha X\delta}a{\alpha' X\delta}$ for each transition $\ltsred\alpha a{\alpha'}$. 
Consider one such transition $\ltsred\alpha a{\alpha'}$. 
The transition $\ltsred{\alpha X\gamma}a{\alpha'X\gamma}$ can be matched with transition $\ltsred{\alpha X\delta}a{\alpha'X\delta}$, and vice-versa. 
Moreover, the resulting pair $(\alpha'X\gamma,\alpha'X\delta)$ is in $\R$ since $X$ is unnormed.
\end{itemize}
Since $\R$ is a bisimulation, it follows that $\R\subseteq\ \sim$; thus for every unnormed $X$ and every $\alpha,\gamma,\delta$ we have $\isbisim{\alpha X\gamma}{\alpha X\delta}$. 
The result follows by taking $\delta=\Empty$.
\end{proof}

\noindent
It is useful to extend the notion of norm to unnormed words. 

\begin{defi}[Seminorm]
The \emph{seminorm} of a word $\gamma$, denoted by $\seminormof{\gamma}$, is defined as the norm of the maximal normed prefix $\gamma'$ of $\gamma$. 
In particular: if $\gamma$ is normed then $\seminormof\gamma=\normof\gamma$; and if $\gamma=\gamma'X$ with $\gamma'$ normed and $X$ unnormed, then $\seminormof\gamma=\normof{\gamma'}$. 
Notice also that for any unnormed $X$ we have $\seminormof X = \normof\Empty = 0$. 
\end{defi}

\noindent
We say that a transition $\ltsred\gamma a{\gamma'}$ is \emph{norm-reducing} if $\gamma$ is normed and $\normof{\gamma'} = \normof{\gamma}-1$ (and thus, $\gamma'$ is also normed). 
A \emph{norm-reducing sequence} is a sequence of norm-reducing transitions. 
Notice that if $\gamma$ is normed, then there exists a norm-reducing sequence of length $n=\normof\gamma$ starting with $\gamma$ and ending with $\Empty$,
\[\ltsred{\gamma=\gamma_0}{a_1}{\ltsred{\gamma_1}{a_2}{\ltsred{\cdots}{a_n}{\gamma_n=\Empty}}}.\]

In general, there may be multiple norm-reducing sequences. Let us fix some ordering on the terminals, $\T=\{a^{(0)} < a^{(1)} < \cdots < a^{(|\T|)}\}$. The \emph{canonical norm-reducing sequence} of a normed word $\gamma$ is the norm-reducing sequence of length $n=\normof\gamma$, as above, for which $a_1\cdots a_n$ is lexicographically smallest. 
Generally, for context-free grammars, the same terminal may have multiple norm-reducing transitions; in this case we define the canonical norm-reducing sequence by further tie-breaking according to some fixed ordering of the nonterminals. 

We can extend the notion of norm-reducing transitions and sequences to seminorm-reducing transitions and sequences in the obvious manner. 
Notice that if $\gamma=\gamma'X$ with $\gamma$ normed and $X$ unnormed, then its \emph{canonical seminorm-reducing sequence} has length $\normof{\gamma'}$ and ends with $X$. 
Finally, we use the notation $\normdynamic\gamma k$, with $k=0,1,\ldots,\seminormof\gamma$, to denote the $k$-th term in the canonical seminorm-reducing sequence of $\gamma$.
In particular, for every $\gamma$:
\begin{itemize}
\item $\normdynamic\gamma 0=\gamma$;
\item if $\gamma$ is normed with $n=\normof\gamma$, then $\normdynamic\gamma n=\Empty$; 
\item if $\gamma$ is unnormed, say $\gamma=\gamma'X$ with $\gamma'$ normed, $X$ unnormed, $n=\seminormof\gamma = \normof{\gamma'}$, then $\normdynamic\gamma n=X$.
\end{itemize}

\noindent
The following notion defines a measure of the ``complexity'' of a grammar, which will be used in the analysis of our algorithms.

\begin{defi}[Valuation]
We define the \emph{valuation} of a grammar $\G=(\T,\V,\pcal)$ as the maximum seminorm among all words $\alpha$ appearing as right-hand sides of productions $\ltsred Xa\alpha$. 
\end{defi}

\noindent
We conclude this section by summarizing some tractability results for the relevant norm-related notions of context-free grammars. 
The following results stem from in Hirshfeld \etal~\cite{DBLP:journals/tcs/HirshfeldJM96,DBLP:journals/corr/abs-1207-2479}.

\begin{restatable}[Efficient computations]{lem}{polytimevaluation}
\label{lem:polytime-valuation}
Each of the following tasks can be carried out in polynomial time, where $\G$ denotes a context-free grammar and $\gamma$ denotes a word of nonterminals in $\G$.
\begin{enumerate}
\item Given $\G$ and $\gamma$, determine whether $\gamma$ is normed.
\item Given $\G$ and $\gamma$, compute $\normof \gamma$.
\item Given $\G$ and $\gamma$, compute $\seminormof \gamma$.
\item Given $\G$, $\gamma$ and $0\leq k \leq \seminormof\gamma$, compute $\normdynamic\gamma k$.
\item Given $\G$, compute the valuation of $\G$.
\end{enumerate}
\end{restatable}

\begin{proof}
It is straightforward to obtain, in polynomial time, the value of $\normof{X}$ for each nonterminal $X$, using a small variation of Dijkstra's shortest path algorithm \cite{DBLP:journals/tcs/HirshfeldJM96}. 
Given the norms of all nonterminals, we can check whether a given word $\gamma=X_1\ldots X_n$ is normed by checking whether each of $X_1,\ldots,X_n$ are normed, in which case $\normof\gamma=\normof{X_1}+\cdots+\normof{X_n}$. 
We can also easily compute $\seminormof{\gamma}$ by finding the largest normed prefix. 
Hirshfeld~\etal~\cite{DBLP:journals/tcs/HirshfeldJM96} have shown how to compute $\normdynamic\gamma k$ efficiently. 
The idea is to precompute, for each normed $X$, its canonical norm-reducing transition, that is, the transition $\ltsred{X}{a}{\alpha_X}$ such that $\normof{X}=1+\normof{\alpha_X}$ and $(a,\alpha_X)$ is lexicographically smallest. 
Then, the following recursive definition provides for an efficient computation of $\normdynamic\gamma k$.
\begin{itemize}
\item $\normdynamic\gamma 0 = \gamma$.
\item $\normdynamic{X\gamma} k = \normdynamic{\gamma}{k-\normof{X}}$, if $\normof{X}\leq k$.
\item $\normdynamic{X\gamma} k = \normdynamic{\alpha_X}{k-1}\gamma$, if $\normof{X} > k$.
\end{itemize}
Finally, the valuation of $\G$ can be computed by simply iterating over all productions $\ltsred{X}{a}{\alpha}$ of $\G$.
\end{proof}



\section{Reducibility and comparability}
\label{sec:comparability}

In this section, we present our uniqueness results characterising the solutions to equations of the form $\isbisim{X\alpha}{Y\beta}$, where $X,Y$ are given nonterminals and $\alpha,\beta$ are unknowns. 
Although our task is to decide bisimilarity of two words (a yes-or-no question), it is useful to consider intermediate possibilities, \ie~words that are `almost bisimilar'. 

\begin{defi}[Reducibility and comparability]
Let $\G$ be a simple grammar and $\gamma,\delta$ be words of nonterminals in $\G$.
\begin{enumerate}
\item We say that $(\gamma,\delta)$ is a \emph{bisimilar pair} if $\isbisim \gamma\delta$.
\item We say that $(\gamma,\delta)$ is a \emph{reducible pair} if there exists some $\beta$ such that $\isbisim{\gamma\beta}\delta$, or such that $\isbisim \gamma{\delta\beta}$.
\item We say that $(\gamma,\delta)$ is a \emph{comparable pair} if there exist some $\alpha,\beta$ such that $\isbisim{\gamma\alpha}{\delta\beta}$.
\end{enumerate}
\end{defi}

\noindent
Some remarks are in order. 
First, it should be clear that every bisimilar pair is reducible and that every reducible pair is comparable. 
Thus, reducibility and comparability are in some way less strict notions of bisimilarity. 
Second, if $\isbisim X{Y\gamma}$ then (by \cref{prop:bisim-eq-norm}) $\normof X=\normof Y+\normof\gamma$, implying that $\normof X\geq\normof Y$. 
Alternatively, $\isbisim{X\gamma}Y$ implies $\normof X\leq\normof Y$. 
Henceforth, whenever $(X,Y)$ is a reducible pair we may assume without loss of generality that $\normof X\geq\normof Y$ and that $\isbisim X{Y\beta}$ for some $\beta$; otherwise if $\normof X < \normof Y$, then we replace $(X,Y)$ by $(Y,X)$. 
The following result appears in Christensen
\etal~\cite{DBLP:journals/iandc/ChristensenHS95}; we include the proof here.

\begin{restatable}[Unique solution of tail recursive equations]{prop}{uniquenessrecursion}
\label{prop:uniquenessrecursion}
Let $\gamma\neq\Empty$, and suppose that $\isbisim{\alpha}{\gamma\alpha}$ and $\isbisim{\beta}{\gamma\beta}$. Then $\isbisim{\alpha}{\beta}$. Moreover, $\alpha$ is unnormed (and so is $\beta$).
\end{restatable}

\begin{proof}
First consider the case that $\gamma$ is unnormed. 
By \cref{prop:unnormedbisim} we have that $\isbisim{\gamma\alpha}{\isbisim{\gamma}{\gamma\beta}}$. 
Therefore, we get $\isbisim{\alpha}{\isbisim{\gamma\alpha}{\isbisim{\gamma\beta}\beta}}$. 
Now consider the case that $\gamma$ is normed. We shall show that $\R$ is a bisimulation, where
$$\R=\{(\alpha',\beta') : \isbisim{\alpha'}{\gamma'\alpha}, \isbisim{\beta'}{\gamma'\beta}, \text{ and }\ltsred{\gamma}{u}{\gamma'}\}.$$
Then, since $\ltsred{\gamma}{u}{\Empty}$ for some $u$, we can conclude that $(\alpha,\beta)\in\R$ and thus $\isbisim{\alpha}{\beta}$. 
Letting $(\alpha',\beta')\in\R$, take $\gamma'$ such that $\isbisim{\alpha'}{\gamma'\alpha}$, $\isbisim{\beta'}{\gamma'\beta}$, and $\ltsred{\gamma}{u}{\gamma'}$. 
We consider a case analysis in $\gamma'$:
\begin{itemize}
\item \textbf{Case} $\gamma'=\Empty$: then $\isbisim{\alpha'}{\isbisim\alpha{\gamma\alpha}}$ and $\isbisim{\beta'}{\isbisim\beta{\gamma\beta}}$. 
Any transition $\ltsred{\alpha'}{a}{\alpha''}$ must be matched by a transition $\ltsred{\gamma\alpha}{a}{\alpha'''}$, with $\isbisim{\alpha''}{\alpha'''}$. 
Since $\gamma\neq\Empty$, this means that there is a transition $\ltsred{\gamma}a{\gamma''}$, and moreover $\alpha'''=\gamma''\alpha$. Thus we also have $\ltsred{\gamma\beta}a{\gamma''\beta}$. 
Since $\isbisim{\beta'}{\gamma\beta}$, there must be a matching transition $\ltsred{\beta'}{a}{\beta''}$, with $\isbisim{\beta''}{\gamma''\beta}$. 
Thus, we can match $\ltsred{\alpha'}{a}{\alpha''}$ with $\ltsred{\beta'}{a}{\beta''}$, arriving at pair $(\alpha'',\beta'')$ which is in $\R$.
\item \textbf{Case} $\gamma'\neq\Empty$: any transition $\ltsred{\alpha'}{a}{\alpha''}$ must be matched by a transition $\ltsred{\gamma'\alpha}{a}{\alpha'''}$, with $\isbisim{\alpha''}{\alpha'''}$. 
Since $\gamma'\neq\Empty$, this means that there is a transition $\ltsred{\gamma'}a{\gamma''}$, and moreover $\alpha'''=\gamma''\alpha$. 
Note also that $\ltsred{\gamma}{ua}{\gamma''}$ and $\ltsred{\gamma'\beta}a{\gamma''\beta}$. 
Since $\isbisim{\beta'}{\gamma'\beta}$, there must be a matching transition $\ltsred{\beta'}{a}{\beta''}$, with $\isbisim{\beta''}{\gamma''\beta}$. 
Thus, we can match $\ltsred{\alpha'}{a}{\alpha''}$ with $\ltsred{\beta'}{a}{\beta''}$, arriving at pair $(\alpha'',\beta'')$ which is in $\R$.
\end{itemize}
By a similar case analysis, we can match transitions of $\beta'$ with transitions of $\alpha'$, yielding that $\R$ is a bisimulation. 

Finally, $\isbisim\alpha{\gamma\alpha}$ implies (by \cref{prop:bisim-eq-norm}) $\normof\alpha=\normof{\gamma\alpha}=\normof\gamma+\normof\alpha$. If $\normof\alpha<\infty$, we would get $\normof\gamma=0$, contradicting the assumption that $\gamma\neq\varepsilon$. Therefore $\alpha$ is unnormed (and similarly for $\beta$).
\end{proof}

\noindent
We arrive at the characterization results given in \cref{lem:charunnormedunnormed,lem:charnormedunnormed,lem:charnormnorm}. They provide necessary and sufficient conditions partitioning all pairs $(X,Y)$ of nonterminals into one of three groups: reducible, irreducible but comparable, and incomparable.

\begin{restatable}[Characterization of nonterminal pairs: unnormed \vs unnormed]{lem}{charunnormedunnormed}
\label{lem:charunnormedunnormed}
Let $X,Y$ be nonterminals in a context-free grammar $\G$. Suppose that $X$, $Y$ are both unnormed. Then $(X,Y)$ is bisimilar iff it is comparable. In other words, either $(X,Y)$ is incomparable or $(X,Y)$ is bisimilar. 
Moreover, if $(X,Y)$ is bisimilar, then $(\alpha,\beta)$ is a solution to $\isbisim{X\alpha}{Y\beta}$ for any $\alpha,\beta$.
\end{restatable}

\begin{proof}
  Given that that $X,Y$ are unnormed, by \cref{prop:unnormedbisim},
  $\isbisim{X}{X\alpha}$ for every $\alpha$ and $\isbisim{Y}{Y\beta}$ for every
  $\beta$. Therefore, $(X,Y)$ is comparable iff there exist $\alpha$, $\beta$
  such that $\isbisim{X\alpha}{Y\beta}$ iff for any $\alpha$, $\beta$ it is the
  case that $\isbisim{X\alpha}{Y\beta}$ iff $\isbisim{X}{Y}$ iff $(X,Y)$ is
  bisimilar.
\end{proof}

\begin{restatable}[Characterization of nonterminal pairs: normed \vs unnormed]{lem}{charnormedunnormed}
\label{lem:charnormedunnormed}
Let $X,Y$ be nonterminals in a simple grammar $\G$. Suppose that $X$ is unnormed
and $Y$ is normed. Then $(X,Y)$ is not bisimilar; and $(X,Y)$ is reducible iff
it is comparable. In other words, either $(X,Y)$ is incomparable or $(X,Y)$ is reducible. 
Moreover, suppose that $(X,Y)$ is reducible and take $\beta_0$ such that $\isbisim X{Y\beta_0}$. Then:
\begin{enumerate}
\item $(\alpha,\beta)$ is a solution to $\isbisim{X\alpha}{Y\beta}$ iff $\isbisim\beta{\beta_0}$; in particular, $\beta_0$ is unique up to bisimilarity;
\item take $u$ such that $\ltsred Yu\Empty$. Then $\ltsred Xu\beta$ for some $\isbisim\beta{\beta_0}$.
\end{enumerate}
\end{restatable}

\begin{proof}
Given that $X$ is unnormed and $Y$ is normed, and since
$\normof{X}=\infty\neq\normof{Y}$ we must have (by \cref{prop:bisim-eq-norm})
$\isnotbisim XY$. By \cref{prop:unnormedbisim}, $\isbisim{X}{X\alpha}$ for
every $\alpha$. Therefore, $(X,Y)$ is comparable iff there exist
$\alpha,\beta$ such that $\isbisim{X\alpha}{Y\beta}$ iff there exists $\beta$
such that $\isbisim X{Y\beta}$ iff $(X,Y)$ is reducible. Now suppose
$\isbisim X{Y\beta_0}$ and $\isbisim{X\alpha}{Y\beta}$. Since $Y$ is normed,
there is $u$ such that $\ltsred Yu\Empty$, and therefore
$\ltsred{Y\beta_0}u{\beta_0}$, $\ltsred{Y\beta}u\beta$. Since
$\isbisim X{Y\beta_0}$, there must be a matching sequence of transitions
$\ltsred Xu{\beta'}$ with $\isbisim{\beta_0}{\beta'}$. Since
$\isbisim X{\isbisim{X\alpha}Y\beta}$, there must be a matching sequence of
transitions $\ltsred Xu{\beta''}$ with $\isbisim{\beta}{\beta''}$. Since the
grammar is simple, $\beta'$ is uniquely defined by $X$ and $u$,
\ie~$\beta'=\beta''$. Therefore
$\isbisim{\beta}{\beta''=\isbisim{\beta'}{\beta_0}}$ and $\beta_0$ is unique
up to bisimilarity.
\end{proof}

In the second part of the above lemma, we make use of the fact that the grammar is simple to derive uniqueness of $\beta_0$ up to bisimilarity. 
The assumption is necessary, since the result does not hold for general context-free grammars.
A counter-example is given by the context-free grammar with productions
\[\ltsred XaB\qquad\ltsred XaC\qquad\ltsred Ya\Empty\qquad\ltsred YaB\qquad\ltsred YaC\qquad\ltsred BbB\qquad\ltsred CcC\]
for which $\isbisim{X}{\isbisim{YB}{YC}}$ but $\isnotbisim BC$. 

For context-free grammars, it follows from the work of Christensen~\etal~\cite{DBLP:journals/iandc/ChristensenHS95} that there are at most $d^{\normof Y}$ different words $\beta$ such that $\isbisim X{Y\beta}$, where $d$ is the maximum number of distinct productions $\ltsred Xa\gamma$ among nonterminals $X$ and terminals $a$. 
For simple grammars, $d=1$, thus obtaining uniqueness. 
For context-free grammars the number of solutions is finite, but exponentially large. 
The uniqueness of solutions for simple grammars is a key ingredient of our efficient algorithm.

We give a small example exhibiting both cases of \cref{lem:charnormedunnormed}:
a reducible pair and an incomparable pair. 
Consider the simple grammar with nonterminals $\{X,Y,A,B,C\}$, terminals $\{a,b\}$, and production rules below.
\begin{align*}
&\ltsred{X}{a}{\varepsilon} && 
\ltsred{Y}{a}{\varepsilon} &&
\ltsred{Y}{b}{\varepsilon} &&
\ltsred{A}{a}{A} &&
\ltsred{B}{b}{B} && 
\ltsred{C}{a}{A} &&
\ltsred{C}{b}{B}
\end{align*}
Notice that $\normof{X}=\normof{Y}=1$ and $\normof{A}=\normof{B}=\normof{C}=\infty$. 
The pair $(A,X)$ is reducible, and in particular $\isbisim{A}{XA}$, corresponding to a state where the transition with label $a$ can be applied infinitely often. 
On the other hand, the pair $(C,Y)$ is incomparable. To see this, suppose that $\isbisim{C\alpha}{Y\beta}$. 
By applying the transition with label $a$ (respectively, $b$), we would get $\isbisim{A\alpha}{\beta}$ (respectively, $\isbisim{B\alpha}{\beta}$). 
By transitivity, we would infer $\isbisim{A\alpha}{B\alpha}$, which is impossible since the transitions of $A$ and $B$ do not match. 
Hence, there do not exist $\alpha,\beta$ such that $\isbisim{C\alpha}{Y\beta}$.

\begin{restatable}[Characterization of nonterminal pairs: normed \vs normed]{lem}{charnormnorm}
\label{lem:charnormnorm}
Let $X,Y$ be nonterminals in a simple grammar $\G$ with no dead nonterminals. 
Suppose that $X$, $Y$ are both normed, and, without loss of generality, that $\normof X\geq\normof Y$. Then $(X,Y)$ may be either reducible, irreducible but comparable, or incomparable. Moreover, the following claims hold.
\begin{enumerate}
\item\label{lem:charnormnorm-1} Suppose that $(X,Y)$ is reducible, and take $\beta_0$ such that $\isbisim X{Y\beta_0}$; then
\begin{enumerate}
\item\label{lem:charnormnorm-1a} $(\alpha,\beta)$ is a solution to $\isbisim{X\alpha}{Y\beta}$ iff $\isbisim{\beta_0\alpha}\beta$.
\item\label{lem:charnormnorm-1c} $\isbisim{\beta_0}{\normdynamic Xk}$ with $k=\normof Y$; in particular, $\beta_0$ is unique up to bisimilarity.
\end{enumerate}
\item\label{lem:charnormnorm-2} Suppose that $(X,Y)$ is irreducible but comparable, and take $\alpha_0$, $\beta_0$ such that $\isbisim{X\alpha_0}{Y\beta_0}$; then $\alpha_0$, $\beta_0$ are unique up to bisimilarity. Moreover, both $\alpha_0$ and $\beta_0$ are unnormed.
\end{enumerate}
\end{restatable}

\begin{proof}
Suppose that $X$, $Y$ are normed with $\normof X\geq\normof Y$ and that $\isbisim X{Y\beta_0}$. 
By \cref{prop:bisim-eq-norm}, $\normof X = \normof{Y\beta_0}=\normof Y+\normof{\beta_0}$. 
Consider the canonical norm-reducing sequence $\ltsred Xu\Empty$. 
By bisimilarity, there must be a matching sequence of transitions $\ltsred{Y\beta_0}u{\gamma'}$ with $\isbisim{\gamma'}{\Empty}$. 
By \cref{prop:congruence}, we know that $\gamma'=\Empty$. 
In other words, $\ltsred{Y\beta_0}u{\Empty}$ is also a norm-reducing sequence. 
Each pair of words at the same position along these two sequences must be bisimilar. 
By taking the $k$-th term, with $k=\normof Y$, we thus get $\isbisim{\normdynamic Xk}{\normdynamic {Y\beta_0}k=\beta_0}$. 
This proves \cref{lem:charnormnorm-1c}. 

For the forward direction of \cref{lem:charnormnorm-1a}, suppose $\isbisim{X\alpha}{Y\beta}$. Let $u'$ be the prefix of size $k$ of $u$ as above, so that $\ltsred{X}{u'}{\normdynamic Xk}$, $\ltsred{Y}{u'}\Empty$, $\ltsred{X\alpha}{u'}{\normdynamic Xk\alpha}$, $\ltsred{Y\beta}{u'}\beta$. 
Since $\G$ is simple, $\ltsred{X\alpha}{u'}{\normdynamic Xk\alpha}$ can only be matched with $\ltsred{Y\beta}{u'}\beta$, that is, $\isbisim{\normdynamic Xk\alpha}{\beta}$. 
Since $\isbisim{\normdynamic Xk}{\beta_0}$, by congruence (\cref{prop:congruence}) we get that $\isbisim{\normdynamic Xk\alpha}{\beta_0\alpha}$, which proves $\isbisim{\beta_0\alpha}\beta$. Conversely, if $\isbisim{\beta_0\alpha}\beta$ then again by congruence we get $\isbisim{X\alpha}{\isbisim{Y\beta_0\alpha}{Y\beta}}$. 

Now suppose $(X,Y)$ is irreducible but comparable. 
Take $\alpha_0$, $\beta_0$ such that $\isbisim{X\alpha_0}{Y\beta_0}$. 
Since $\isnotbisim XY$, there must be a word $u$ such that either $\ltsred{Y}u{\delta}$ has no matching sequence for $X$, or $\ltsred{X}u{\gamma}$ has no matching sequence for $Y$. 
We consider the case that $\ltsred{Y}u{\delta}$ has no matching sequence for $X$, as the other case is similar. 
Let $u'$ be the maximal prefix of $u$ for which $\ltsred{Y}{u'}{\delta'}$ does have a matching sequence $\ltsred{X}{u'}{\gamma'}$. 
By writing $u=u'au''$, this means that $\ltsred{\delta'}a{\delta''}$ is a transition without a match in $\gamma'$. 
Now notice that $\ltsred{X\alpha_0}{u'}{\gamma'\alpha_0}$ and $\ltsred{Y\beta_0}{u'}{\delta'\beta_0}$. 
Since $\isbisim{X\alpha_0}{Y\beta_0}$ and the grammar is simple, these sequences can only match with each other, implying that $\isbisim{\gamma'\alpha_0}{\delta'\beta_0}$. 
Since $\ltsred{\delta'\beta_0}a{\delta''\beta_0}$, there must be a matching transition for $\gamma'\alpha_0$. 
In other words, $\gamma'$ does not have a transition with label $a$ but $\gamma'\alpha_0$ does. 
This can only occur if $\gamma'=\Empty$. 
Therefore, we have concluded that $\isbisim{\alpha_0}{\delta'\beta_0}$. 
Thus, by congruence (\cref{prop:congruence}), also $\isbisim{X\delta'\beta_0}{Y\beta_0}$. 

Continuing with our proof, we are assuming that $(X,Y)$ is irreducible, so that in particular $\isnotbisim{X\delta'}Y$. 
There must be a word $v$ such that either $\ltsred{X\delta'}v{\gamma'}$ has no matching sequence for $Y$, or $\ltsred Yv{\delta''}$ has no matching sequence for $X\delta'$. 
We consider the case that $\ltsred{X\delta'}v{\gamma'}$ has no matching sequence for $Y$, as the other case is similar. 
Let $v'$ be the maximal prefix of $v$ for which $\ltsred{X\delta'}{v'}{\gamma''}$ does have a matching sequence $\ltsred Y{v'}{\delta''}$. 
By writing $v=v'av''$, this means that $\ltsred{\gamma''}a{\gamma'''}$ is a transition without a match in $\delta''$. 
Now notice that $\ltsred{X\delta'\beta_0}{v'}{\gamma''\beta_0}$ and $\ltsred{Y\beta_0}{v'}{\delta''\beta_0}$. 
Since $\isbisim{X\delta'\beta_0}{Y\beta_0}$ and the grammar is simple, these sequences can only match with each other, implying that $\isbisim{\gamma''\beta_0}{\delta''\beta_0}$. 
Since $\ltsred{\gamma''\beta_0}a{\gamma'''\beta_0}$, there must be a matching transition for $\delta''\beta_0$. 
In other words, $\delta''$ does not have a transition with label $a$ but $\delta''\beta_0$ does. This can only occur if $\delta''=\Empty$. 
Therefore, we have concluded that $\isbisim{\beta_0}{\gamma''\beta_0}$. 
\Cref{prop:uniquenessrecursion} then tells us that $\beta_0$ is unique up to bisimilarity and that  $\beta_0$ is unnormed. Since $\isbisim{\alpha_0}{\delta'\beta_0}$, we conclude that $\alpha_0$ is also unique up to bisimilarity and unnormed.
\end{proof}

\noindent
We give small examples exhibiting all cases of \cref{lem:charnormnorm}: a
reducible pair, an irreducible but comparable pair, and an incomparable pair. 
For the simple grammar with nonterminals $\{X,Y\}$, terminals $\{a\}$, and production rules
\begin{align*}
&\ltsred{X}{a}{Y} && 
\ltsred{Y}{a}{\varepsilon}
\end{align*}
it is immediate that $\isbisim{X}{YY}$, so that the pair $(X,Y)$ is reducible.

For the simple grammar with nonterminals $\{X,Y,A,B\}$, terminals $\{a,b\}$, and production rules
\begin{align*}
&\ltsred{X}{a}{\varepsilon} && 
\ltsred{X}{b}{B} && 
\ltsred{Y}{a}{A} && 
\ltsred{Y}{b}{\varepsilon} && 
\ltsred{A}{a}{A} && 
\ltsred{B}{b}{B}
\end{align*}
we get that the pair $(X,Y)$ is irreducible.  
To see this, suppose that $\isbisim{X}{Y\beta}$. By applying the transition with label $a$, we would get $\isbisim{\varepsilon}{A\beta}$, which is clearly impossible. 
A similar argument shows that there exists no $\beta$ such that $\isbisim{Y}{X\beta}$.
On the other hand, the pair $(X,Y)$ is comparable, since in particular $\isbisim{XA}{YB}$. 

For the simple grammar with nonterminals $\{X,Y,A,B\}$, terminals $\{a,b,c\}$, and production rules
\begin{align*}
&\ltsred{X}{a}{\varepsilon} && 
\ltsred{X}{b}{B} && 
\ltsred{X}{c}{\varepsilon} && 
\ltsred{Y}{a}{A} && 
\ltsred{Y}{b}{\varepsilon} && 
\ltsred{Y}{c}{\varepsilon} && 
\ltsred{A}{a}{A} && 
\ltsred{B}{b}{B}
\end{align*}
we get that the pair $(X,Y)$ is incomparable. To see this, suppose that $\isbisim{X\alpha}{Y\beta}$. 
By applying the transition with label $a$, $b$ and $c$ respectively, we would get $\isbisim{\alpha}{A\beta}$, $\isbisim{B\alpha}{\beta}$ and $\isbisim{\alpha}{\beta}$ respectively. 
By transitivity, we would infer $\isbisim{A\beta}{B\alpha}$, which is impossible since the transitions of $A$ and $B$ do not match. 



\section{Bases and coinductive congruence}
\label{sec:bases-congruence}

One of the key ingredients used in the known decidability results of bisimilarity for context-free grammars in general is the observation that bisimilarity can be induced by a finite set. 

\begin{defi}[Basis, basis properties]
\label{def:basis-prop}
A \emph{basis} for a given context-free grammar $\G=(\V,\T,\pcal)$ is a finite relation $\B\subseteq\V^+\times\V^+$.
\begin{enumerate}
\item A basis $\B$ is \emph{reflexive} if $(X,X)\in\B$, for every $X\in\V$.
\item A basis $\B$ is \emph{simple} if, for every $X,Y\in\V$, there is at most one pair of the form $(X\alpha,Y\beta)$ or $(Y\beta,X\alpha)$ in $\B$.
\item A basis $\B$ is \emph{functional} if for every pair $(X,Y\beta)\in\B$ such that $\normof{X}<\infty$, it is the case that $\normof X\geq\normof Y$ and $\beta=\normdynamic{X}{\normof Y}$.
\item A basis $\B$ is \emph{norm-compliant} if 
\begin{itemize}
\item for every pair $(X,Y\beta)\in\B$ such that $\normof{X}=\infty$, it is the case that $\normof{Y\beta}=\infty$;
\item for every pair $(X\alpha,Y\beta)\in\B$ such that $\alpha\neq\varepsilon$, it is the case that $\normof{X\alpha}=\normof{Y\beta}=\infty$.
\end{itemize}
\end{enumerate}
\end{defi}

\noindent
Caucal~\cite{Caucal1990} defined the notion of a least congruence induced by $\B$ and showed that, for every context-free grammar, there exists a basis $\B$ such that any two words are bisimilar iff they are congruent with respect to $\B$ (in other words, the least congruence relation coincides with bisimilarity). 
However, for a general grammar $\G$ and basis $\B$, least congruence is at best a semi-decidable relation.
In this paper, instead of using least congruence, we shall consider a novel notion, which we call \emph{coinductive congruence}, denoted $\congBtwo{c}{\B}$. 

\begin{figure}[t]
\begin{mathpar}
\inferrule*[lab=$\Empty$-Ax]
{\phantom{lala}}
{\iscongcoi{\varepsilon}{\varepsilon}}

\inferrule*[lab=BPA1]
{
\iscongcoi{\beta\alpha'}{\beta'}
\\
(X,Y\beta)\in\B
}
{\iscongcoi{X\alpha'}{Y\beta'}}

\inferrule*[lab=BPA2]
{
\iscongcoi{\alpha}{\alpha'}
\\
\iscongcoi{\beta}{\beta'}
\\
(X\alpha,Y\beta)\in\B
}
{\iscongcoi{X\alpha'}{Y\beta'}}
\end{mathpar}
\caption{Coinductive congruence. The symmetric rules of \textsc{BPA1},
  \textsc{BPA2} are implicitly assumed. Words $\alpha$ and $\beta$ are
  existentially quantified.}
\label{fig:congruence-rules}
\end{figure}


\begin{defi}[Coinductive congruence]
\label{def:coinductive-congruence}
Let $\B$ be a basis for a context-free grammar $\G=(\V,\T,\pcal)$. The \emph{coinductive congruence} relation (with respect to $\B$) is defined according to the rules in \Cref{fig:congruence-rules}, interpreted coinductively. 
That is, $\congBtwo{c}{\B}$ is the \emph{largest} relation on $\V^\ast\times\V^\ast$ such that, if $\iscongcoi\sigma\tau$, then one of the following must hold:
\begin{itemize}
\item \textsc{($\Empty$-Ax)} $\sigma=\tau=\Empty$;
\item \textsc{(BPA1)} $\sigma=X\alpha'$ and $\tau=Y\beta'$, where $\iscongcoi{\beta\alpha'}\beta'$, $(X,Y\beta)\in\B$ for some $\beta$;
\item \textsc{(BPA2)} $\sigma=X\alpha'$ and $\tau=Y\beta'$, where $\iscongcoi\alpha{\alpha'}$, $\iscongcoi\beta{\beta'}$, $(X\alpha,Y\beta)\in\B$ for some $\alpha,\beta$.
\end{itemize}
\end{defi}

The terminology \textsc{BPA1}, \textsc{BPA2} comes from Basic Process Algebra rules~\cite{DBLP:conf/concur/JancarM99}. The symmetric rules (obtained by swapping $\sigma$ and $\tau$) are assumed implicitly. 

\begin{defi}[Self-bisimulation]\label{def:selfbisim}
We say that a basis $\B$ for some given context-free grammar is a \emph{self-bisimulation} if, for every pair $(\gamma,\delta)\in\B$ and every terminal $a\in\T$:
\begin{itemize}
\item if $\ltsred\gamma a{\gamma'}$ for some $\gamma'$, then there exists some $\delta'$ such that $\ltsred\delta a{\delta'}$ and $\iscongcoi{\gamma'}{\delta'}$;
\item if $\ltsred\delta a{\delta'}$ for some $\delta'$, then there exists some $\gamma'$ such that $\ltsred\gamma a{\gamma'}$ and $\iscongcoi{\gamma'}{\delta'}$.
\end{itemize}
\end{defi}

\begin{restatable}[Bisimilarity preservation]{lem}{basisinsim}
\label{lem:basisinsim}
Let $\B$ be a basis for some given context-free grammar without dead nonterminals. 
If $\B\subseteq\ \sim$, then $\congBtwo{c}{\B}\ \subseteq\ \sim$.
\end{restatable}

\begin{proof}
Suppose that $\B\subseteq\ \sim$. We prove by induction on $n$ that $\congBtwo{c}{\B}\ \subseteq\ \sim_n$ for every $n$, where $\sim_n$ is the $n$-th approximant of bisimilarity (\Cref{def:bisim-approx}). 
The base case ($n=0$) is trivial since $\sim_0\ =\V^\ast\times\V^\ast$. 
For the induction step ($n+1$), take $(\gamma,\delta)\in\ \congBtwo{c}{\B}$. 
We consider a case analysis on the last rule in the derivation of $\iscongcoi\gamma\delta$.

(\textsc{$\Empty$-Ax}): then $\gamma=\delta=\Empty$, so that trivially $\isbisim\Empty\Empty$ and thus $\isbisimn\Empty\Empty{n+1}$.

(\textsc{BPA1}): then $\gamma=X\alpha$ and $\delta=Y\beta$, with $(X,Y\gamma')\in\B$ and $\iscongcoi{\gamma'\alpha}\beta$. 
By induction hypothesis, $\isbisimn{\gamma'\alpha}\beta{n}$. 
Since $\B\subseteq\ \sim$, $\isbisim{X}{Y\gamma'}$. 
Since the grammar has no dead nonterminals, we may use congruence (\Cref{prop:congruence}) to infer that $\isbisim{X\alpha}{Y\gamma'\alpha}$, and use \Cref{prop:congruence-approx-3} of \Cref{prop:congruence-approx} to infer that $\isbisimn{Y\gamma'\alpha}{Y\beta}{n+1}$. We conclude that $\isbisimn{X\alpha}{Y\beta}{n+1}$ as desired. 

(\textsc{BPA2}): similar as before.

We have shown that $\congBtwo{c}{\B}\ \subseteq\ \sim_n$ for every $n$. Thus, by \Cref{prop:bisim-approx}, $\congBtwo{c}{\B}\ \subseteq\ \sim$, as desired. 
\end{proof}

\begin{restatable}[Self-bisimulation]{lem}{selfbisimlemma}
\label{lem:selfbisim}
Let $\B$ be a basis for some given context-free grammar without dead nonterminals. 
If $\B$ is a self-bisimulation, then $\congBtwo{c}{\B}\ \subseteq\ \sim$.
\end{restatable}

\begin{proof}
We shall prove by induction on $n$ that $\congBtwo{c}{\B}\ \subseteq\ \sim_n$ for every $n\in\nbb$, from which the result follows due to \Cref{prop:bisim-approx}. 
The base case $n=0$ is trivial. 
For the induction step ($n+1$), take $\iscongcoi\gamma\delta$. 
We shall show that $\isbisimn\gamma\delta{n+1}$ by a case analysis on the last rule in the derivation of $\iscongcoi\gamma\delta$.

(\textsc{$\Empty$-Ax}): then $\gamma=\delta=\Empty$ and trivially $\isbisimn\Empty\Empty{n+1}$.

(\textsc{BPA1}): then $\gamma=X\alpha$ and $\delta=Y\beta$, with $(X,Y\gamma')\in\B$ and $\iscongcoi{\gamma'\alpha}{\beta}$. 
By induction hypothesis, $\isbisimn{\gamma'\alpha}{\beta}{n}$. 
Consider a transition from $X\alpha$, which must be of the form $\ltsred{X\alpha}a{\alpha'}\alpha$ for $\ltsred{X}a{\alpha'}$. 
Since $(X,Y\gamma)\in\B$, there is a matching transition $\ltsred{Y}a{\beta'}$ such that $\ltsred{Y\gamma'}a{\beta'\gamma'}$ and $\iscongcoi{\alpha'}{\beta'\gamma'}$. 
Again by induction hypothesis, $\isbisimn{\alpha'}{\beta'\gamma'}n$. 
Notice also that we have the transition $\ltsred{Y\beta}a{\beta'\beta}$. 
By congruence (\Cref{prop:congruence-approx}), we get $\isbisimn{\alpha'\alpha}{\isbisimn{\beta'\gamma\alpha}{\beta'\beta}{n}}{n}$. 
In a similar manner, we can prove that any transition $\ltsred{Y\beta}a{\beta'\beta}$ has a matching transition $\ltsred{X\alpha}a{\alpha'}\alpha$ with $\isbisimn{\alpha'\alpha}{\beta'\beta}{n}$. 
This proves that $\isbisimn{X\alpha}{Y\beta}{n+1}$, as desired.

(\textsc{BPA2}): similar as before, concluding the proof. 
\end{proof}

\noindent
\Cref{lem:selfbisim} gives us an approach for figuring out whether two given words $\gamma$, $\delta$ are bisimilar. 
First, guess a basis $\B$. 
Then, find out whether $\B$ is a self-bisimulation and whether $\iscongcoi{\gamma}{\delta}$. 
If that is the case, then by \Cref{lem:selfbisim} we can infer that $\isbisim\gamma\delta$. 
To turn the above description into a formal argument, we need two extra ingredients. 
On the one hand, we need to prove that there always exists a basis $\B$ such that $\congBtwo{c}{\B}\ =\ \sim$ (\cref{thm:basis-existence}). 
On the other hand, we need to figure out conditions under which $\congBtwo{c}{\B}$ can be efficiently decided (\cref{thm:coinductive-congruence}). 
Notice that determining whether $\B$ is a self-bisimulation can be done with polynomially many calls to $\congBtwo{c}{\B}$.

Henceforth and in the statement of the following theorem, we assume an ordering on the set of nonterminals $\V=\{X^{(0)} < X^{(1)} < \cdots < X^{(|\V|)}\}$ such that $\normof X<\normof Y$ implies $X<Y$.

\begin{restatable}[Existence of basis for bisimilarity]{thm}{basisexistence}
\label{thm:basis-existence}
For every simple grammar without dead nonterminals, there exists a basis $\B$ such that $\congBtwo{c}{\B}\ =\ \sim$. 
Moreover, such a basis can be defined as any minimal set such that, for every pair $X\geq Y$ of nonterminals:
\begin{enumerate}
\item If $X$, $Y$ are unnormed and $\isbisim XY$, then $(X,Y)\in\B$.
\item If $X$ is unnormed, $Y$ is normed and $(X,Y)$ is reducible, then $\B$ has exactly one pair $(X,Y\beta)$, with $\isbisim X{Y\beta}$.
\item If $X$, $Y$ are normed and $(X,Y)$ is reducible, then $(X,Y\normdynamic{X}{\normof Y})\in\B$.
\item If $X$, $Y$ are normed and $(X,Y)$ is irreducible but comparable, then $\B$ has exactly one pair $(X\alpha,Y\beta)$, with $\isbisim{X\alpha}{Y\beta}$.
\end{enumerate}
\end{restatable}

\begin{proof}
Let $\B$ be any minimal set as defined in the statement of the theorem. 
Note that $\B$ is finite, having at most $\ocal(n^2)$ elements, where $n$ is the number of nonterminals. 
Since, by construction, $\B\subseteq\ \sim$, we infer by \Cref{lem:basisinsim} that $\congBtwo{c}{\B}\ \subseteq\ \sim$. 

To prove that $\sim\ \subseteq\ \congBtwo{c}{\B}$, we shall show that $\sim$ is backward closed for the rules defining $\congBtwo{c}{\B}$. 
Let $\isbisim\gamma\delta$, and we consider several cases.

($\gamma=\Empty$ and $\delta=\Empty$): in this case we can apply rule \textsc{$\Empty$-Ax}, generating no descendants.

($\gamma=\Empty$ and $\delta\neq\Empty$): we can disregard this case, since by \Cref{prop:congruence} we could not have $\isbisim\gamma\delta$, contradicting our assumption. 

($\gamma\neq\Empty$ and $\delta=\Empty$): similar as before.

For the remaining cases, we can assume that both $\gamma$ and $\delta$ are non-empty. 
Let us write $\gamma=X\alpha$, $\delta=Y\beta$. 
For now, assume $X\geq Y$.

($X$, $Y$ are unnormed): by the pruning convention (\Cref{prop:unnormedbisim}), we may assume that $\alpha=\beta=\Empty$, so that $\isbisim XY$. 
By construction, $(X,Y)\in\B$. Hence we can apply rule \textsc{BPA1} with this pair, arriving at pair $(\Empty,\Empty)$. Clearly, $\isbisim\Empty\Empty$. 

($X$ is unnormed and $Y$ is normed): by the pruning convention, we may assume that $\alpha=\Empty$, so that $\isbisim X{Y\beta}$. 
Thus, $(X,Y)$ is reducible, and by construction there exists some pair $(X,Y\gamma)\in\B$. 
Applying rule \textsc{BPA1} with this pair, we arrive at pair $(\gamma,\beta)$. 
By \Cref{lem:charnormedunnormed}, we conclude that $\isbisim\gamma\beta$ as desired.

($X$, $Y$ are normed and $(X,Y)$ is reducible): by construction, $(X,Y\gamma)\in\B$, where $\gamma=\normdynamic{X}{\normof Y}$. 
Applying rule \textsc{BPA1} with this pair, we arrive at pair $(\gamma\alpha,\beta)$. 
Since $\isbisim{X\alpha}{Y\beta}$, and by \Cref{lem:charnormnorm-1} of \Cref{lem:charnormnorm}, we conclude that $\isbisim{\gamma\alpha}{\beta}$ as desired.

($X$, $Y$ are normed and $(X,Y)$ is irreducible): since $\isbisim{X\alpha}{Y\beta}$, we get that $(X,Y)$ is irreducible but comparable. 
By construction, there exists some pair $(X\alpha',Y\beta')\in\B$. 
Applying rule \textsc{BPA2} with this pair, we arrive at pairs $(\alpha',\alpha)$ and $(\beta',\beta)$. 
By \Cref{lem:charnormnorm-2} of \Cref{lem:charnormnorm}, we conclude that $\isbisim{\alpha'}{\alpha}$ and $\isbisim{\beta'}{\beta}$ as desired.

The cases with $X<Y$ can be handled in a similar manner, but using the symmetric versions of rules \textsc{BPA1} and \textsc{BPA2}.
\end{proof}

\noindent
Next, we turn our attention to the decidability of coinductive congruence. From \cref{def:coinductive-congruence}, we can naturally design an algorithm for coinductive congruence based on proof trees.

\begin{defi}[Proof tree]
Given a basis $\B$ for a context-free grammar and a pair $\gamma$, $\delta$ of words of nonterminals, a \emph{proof tree} for $\iscongcoi\gamma\delta$ is a (possibly infinite) tree where the nodes are labelled with judgements of the form $\iscongcoi{\sigma}{\tau}$. 
Moreover, the root node has label $\iscongcoi\gamma\delta$; and for each node of the tree, with label $\iscongcoi{\sigma}{\tau}$, one of the following must hold:
\begin{itemize}
\item \textsc{($\Empty$-Ax)} $\sigma=\tau=\Empty$; in which case, the node has no children (valid leaf);
\item \textsc{(BPA1)} $\sigma=X\alpha'$ and $\tau=Y\beta'$, where $(X,Y\beta)\in\B$ for some $\beta$; in which case, the node has a single child, with label $\iscongcoi{\beta\alpha'}\beta'$;
\item \textsc{(BPA2)} $\sigma=X\alpha'$ and $\tau=Y\beta'$, where $(X\alpha,Y\beta)\in\B$ for some $\alpha,\beta$; in which case, the node has exactly two children, with labels $\iscongcoi\alpha{\alpha'}$ and $\iscongcoi\beta{\beta'}$ respectively;
\item none of the above applies; in which case, the node has no children (invalid leaf).
\end{itemize} 
A proof tree is \emph{valid} if all leaves have label $\iscongcoi\Empty\Empty$, and \emph{invalid} otherwise.
A proof tree is \emph{canonical} if, whenever rule BPA2 is applied with a pair $(X\alpha,Y\beta)\in\B$, it is the case that $\alpha\neq\Empty\neq\beta$.
\end{defi}

\noindent
It follows from the definition that $\iscongcoi\gamma\delta$ iff there exists a valid proof tree for $\iscongcoi\gamma\delta$. 
We emphasize that proof trees may be infinite; for example, given the basis $\B = \{(XZ,YAB), (Z,Z), \allowbreak (Z,AWZ), (B,WAB)\}$, the following is a valid proof tree for $\iscongcoi{XZ}{YZ}$:
$$
\inferrule*[right=BPA2]
{\inferrule*[right=BPA1]
{\inferrule*[right=$\Empty$-Ax]{\phantom{lala}}{\iscongcoi{\Empty}{\Empty}}}
{\iscongcoi{Z}{Z}}
\quad
\inferrule*[right=BPA1]
{\inferrule*[right=BPA1]
{\inferrule*[right=BPA1]
{\vdots}
{\iscongcoi{Z}{AB}}}
{\iscongcoi{WZ}{B}}}
{\iscongcoi{Z}{AB}}}
{\iscongcoi{XZ}{YZ}}
$$
This is in contrast with inductive definitions, where (valid) proof trees have to be finite, and infinite branches are by definition invalid. 

We now argue that a proof tree for $\iscongcoi\gamma\delta$ is ``essentially'' unique, as long as $\B$ is a simple basis. 

\begin{lem}[Uniqueness of canonical proof trees]\label{lem:congcoi-uniqueness}
Let $\B$ be a basis for a context-free grammar. 
Suppose that $\B$ is simple. 
Then, for any pair $\gamma$, $\delta$, there exists a unique canonical proof tree for $\iscongcoi\gamma\delta$. 
Moreover, $\iscongcoi\gamma\delta$ iff the unique canonical proof tree for $\iscongcoi\gamma\delta$ is valid.
\end{lem}

\begin{proof}
Given a pair $\gamma$, $\delta$, suppose we build a (canonical) proof tree for $\iscongcoi\gamma\delta$, starting from the root and considering the rules that can be applied. 
If $\gamma=\delta=\Empty$, then the only rule that can be applied is \textsc{$\Empty$-Ax}. 
If $\gamma=\Empty$ but $\delta\neq\Empty$, then no rule can be applied (notice that rules \textsc{BPA1}, \textsc{BPA2} require both sides to be non-empty) and $\iscongcoi\gamma\delta$ does not hold. 
Similarly, if $\gamma=\Empty$ but $\delta\neq\Empty$, no rule can be applied and $\iscongcoi\gamma\delta$ does not hold. 
If $\gamma$ and $\delta$ are non-empty, write $\gamma=X\alpha'$, $\delta=Y\beta'$. 
Now, since $\B$ is simple, there is at most one rule of the form $(X\alpha,Y\beta)$ or $(Y\beta,X\alpha)$ in $\B$. 
If there is no such rule, then the proof tree cannot continue (and $\iscongcoi\gamma\delta$ does not hold). 
Otherwise, suppose without loss of generality that $(X\alpha,Y\beta)\in\B$. 
If $\alpha\neq\Empty$, then the only rule that can be applied is \textsc{BPA2}. 
Finally, if $\alpha=\Empty$, a canonical proof tree must use rule \textsc{BPA1} (but an arbitrary proof tree could instead use rule \textsc{BPA2}). 
Since at any stage of the canonical construction there is exactly one applicable rule, the canonical proof tree is unique. 

We now prove the second part of the statement ($\iscongcoi\gamma\delta$ iff the unique canonical proof tree is valid). 
The right-to-left implication is immediate: if the unique canonical proof tree for $\iscongcoi\gamma\delta$ is valid, then there exists a valid proof tree, so $\iscongcoi\gamma\delta$. 
Next, suppose that $\iscongcoi\gamma\delta$, so that by definition there is a valid proof tree for $\iscongcoi\gamma\delta$. 
By the discussion above, the only situation where a valid proof tree may differ from the canonical proof tree occurs when using rule \textsc{BPA2}
with $\alpha=\Empty$; the canonical proof tree would have to use \textsc{BPA1}. 
This yields two possible proof trees, written using the above notation:
$$
\inferrule*[right=BPA1]
{\inferrule
{\vdots}
{\iscongcoi{\beta\alpha'}{\beta'}}}
{\iscongcoi{X\alpha'}{Y\beta'}}
\quad
\text{or}
\quad\quad
\inferrule*[right=BPA2]
{\inferrule{\vdots}{\iscongcoi{\Empty}{\alpha'}}
\quad
\inferrule{\vdots}{\iscongcoi{\beta}{\beta'}}}
{\iscongcoi{X\alpha'}{Y\beta'}}
$$
Supposing that the proof tree using \textsc{BPA2} is valid, we notice that the judgement $\iscongcoi{\Empty}{\alpha'}$ holds if and only if $\alpha'=\Empty$ (in which case we can apply \textsc{$\Empty$-Ax}). 
Thus, the other judgement, $\iscongcoi{\beta}{\beta'}$, coincides with the judgement $\iscongcoi{\beta\alpha'}{\beta'}$ obtained had we used \textsc{BPA1}. 
In other words, we may exchange the application of rule \textsc{BPA2} by \textsc{BPA1}, yielding a canonical proof tree. 
Hence, if there exists a valid proof tree, then there exists a valid canonical proof tree, completing the proof.
\end{proof}

\noindent
Even though a proof tree may be infinite, we now argue that only finitely many \emph{distinct} pairs may appear in a proof tree for a given pair $\gamma$, $\delta$, as long as $\B$ is functional.

\begin{lem}[Finiteness of distinct pairs in proof trees]\label{lem:congcoi-finiteness}
Let $\B$ be a basis for a context-free grammar.
Suppose that $\B$ is functional. 
Then, for any pair $\gamma$, $\delta$, and any judgement $\iscongcoi{\gamma'}{\delta'}$ in a proof tree for $\iscongcoi{\gamma}{\delta}$, each of $\gamma'$ and $\delta'$ is either empty or a word in the seminorm-reducing sequence of $\gamma$, $\delta$, or some $\alpha$, $\beta$ such that $(\alpha,\beta)\in\B$.
\end{lem}

\begin{proof}
By induction on the depth of the node $\iscongcoi{\gamma'}{\delta'}$ in the proof tree. 
The only node at depth $0$ is the root node $\iscongcoi{\gamma}{\delta}$, which trivially satisfies the claim.
For the induction step, we suppose the claim holds for a given judgement $\iscongcoi{\gamma'}{\delta'}$ in a proof tree of $\iscongcoi{\gamma}{\delta}$, and show that the claim also holds for the children of $\iscongcoi{\gamma}{\delta}$. 

(\textsc{$\Empty$-Ax} or invalid leaf): if the node is a valid or invalid leaf, it has no children, so that the claim holds trivially.

(\textsc{BPA1}): write $\gamma'=X\alpha$, $\delta'=Y\beta$, where $(X,Y\gamma'')\in\B$. The application of the rule produces pair $\iscongcoi{\gamma''\alpha}{\beta}$. 
We split the analysis into four subcases.

($X$, $Y$ are unnormed): 
by the pruning convention (\Cref{prop:unnormedbisim}), we may assume that $\alpha=\beta=\gamma'=\Empty$. 
Hence, the pair generated is actually $\iscongcoi{\Empty}{\Empty}$, which satisfies the claim. 

($X$ is unnormed and $Y$ is normed): 
by the pruning convention, we may assume that $\alpha=\Empty$. 
Since $Y$ is normed, $\beta$ is a word in the seminorm-reducing sequence of $Y\beta$ (which satisfies the claim by induction hypothesis), and $\gamma''$ is a word in the seminorm-reducing sequence of $Y\gamma''$ (which satisfies the claim by being a word in $\B$). 
The judgement generated is $\iscongcoi{\gamma'}{\beta}$, which satisfies the claim. 

($X$ is normed and $Y$ is unnormed): 
by the pruning convention, we may assume that $\beta=\gamma''=\Empty$. 
Since $X$ is normed, $\alpha$ is a word in the seminorm-reducing sequence of $X\alpha$ (which satisfies the claim by induction hypothesis). 
The judgement generated is $\iscongcoi{\alpha}{\Empty}$, which satisfies the claim. 

($X,Y$ are normed): 
this is the only point in the proof where we use the fact that $\B$ is functional, yielding that $\gamma''=\normdynamic{X}{\normof Y}$. 
Then $\gamma''$ is in the seminorm-reducing sequence of $X$, so that $\gamma''\alpha$ is in the seminorm-reducing sequence of $X\alpha$ (which satisfies the claim by induction hypothesis). 
Moreover, $\beta$ is in the seminorm-reducing sequence of $Y\beta$ (which satisfies the claim by induction hypothesis). 
Thus, the judgement generated satisfies the claim. 
This concludes the analysis of case \textsc{BPA1}.

(\textsc{BPA2}): write $\gamma=X\alpha'$, $\delta=Y\beta'$, where $(X\alpha,Y\beta)\in\B$. The application of the rule produces pairs $\iscongcoi{\alpha}{\alpha'}$ and $\iscongcoi{\beta}{\beta'}$. 
If $X$ is unnormed, then, by the pruning convention, $\alpha=\alpha'=\Empty$, and the pair $\iscongcoi{\alpha}{\alpha'}$ is actually $\iscongcoi{\Empty}{\Empty}$, which satisfies the claim. 
If $X$ is normed, then $\alpha$ is in the seminorm-reducing sequence of $X\alpha$ (which satisfies the claim by being in $\B$) and $\alpha'$ is in the seminorm-reducing sequence of $X\alpha'$ (which satisfies the claim by induction hypothesis). 
Hence the judgement $\iscongcoi{\alpha}{\alpha'}$ satisfies the claim. 
By a similar analysis for $Y$, we get that the judgement $\iscongcoi{\beta}{\beta'}$ satisfies the claim. 
\end{proof}

\begin{restatable}[Decidability of coinductive congruence]{thm}{coinductivedecidability}
\label{thm:coinductive-congruence}
Let $\B$ be a basis for a context-free grammar. 
Suppose that $\B$ is functional and simple. 
Then $\congBtwo{c}{\B}$ is decidable. 
More precisely, given a pair $\gamma$, $\delta$, we can determine whether $\iscongcoi\gamma\delta$ in time polynomial in the size of $\B$ and the maximum seminorm among $\gamma$, $\delta$, and the words in $\B$.
\end{restatable}

\begin{proof}
Consider an algorithm that, given a pair $\gamma$, $\delta$, tries to build the canonical proof tree for $\iscongcoi{\gamma}{\delta}$ (that is, prefering \textsc{BPA1} over \textsc{BPA2}), which is unique by \cref{lem:congcoi-uniqueness}. 
Notice that such a proof tree may be infinite. 
However, since congruence is coinductively defined, the algorithm does not need to expand pairs which have already appeared in the proof tree. 
Thus we can keep track of visited pairs in order to prune off infinite branches of the tree. 
The algorithm terminates if it produces a node $(\gamma',\delta')$ such that no rule can be applied (concluding that the proof tree is invalid) or if all unexpanded pairs have already appeared in the tree (concluding that the proof tree is valid). 
The latter case also includes finite proof trees, in which there are no unexpanded pairs at the end of the algorithm.

By \cref{lem:congcoi-uniqueness}, we know that $\iscongcoi{\gamma}{\delta}$ iff the canonical proof tree is valid. Hence, the algorithm is correct; that is, if the algorithm terminates, it gives the right answer. 
To prove termination, we use \cref{lem:congcoi-finiteness}, which shows that that each word appearing in the proof tree is either empty or a word in the seminorm-reducing sequence of $\gamma$, $\delta$, or some $\alpha$, $\beta$ such that $(\alpha,\beta)\in\B$. 
There are finitely many words in $\B$, and each word $\alpha$ has (by definition) a unique seminorm-reducing sequence, of size equal to $\seminormof\alpha+1$. 
Letting $m$ denote the maximum seminorm among $\gamma$, $\delta$ and the words in $\B$, we conclude that there are at most $\ocal(|\B|m)$ distinct words and $\ocal(|\B|^2m^2)$ distinct pairs of words generated by the algorithm.
\end{proof}



\section{The basis-updating algorithm}
\label{sec:algorithm}

In this section we present our main contribution, which we call the basis-updating algorithm for deciding bisimilarity of two words in a simple grammar. 
Before we describe the algorithm, we open with a short example which illustrates the `updating' idea. 
Consider the simple grammar with nonterminals $\{X,Y,C\}$, terminals $\{a,b,c\}$, and production rules below.
\begin{align*}
&\ltsred{X}{a}{\varepsilon} && 
\ltsred{X}{b}{\varepsilon} &&
\ltsred{Y}{a}{\varepsilon} &&
\ltsred{Y}{b}{C} &&
\ltsred{C}{c}{C} 
\end{align*}
Suppose we wished to determine whether $\isbisim{XC}{YC}$, using the technique of self-bisimulation bases. 
In other words, we need to find a self-bisimulation basis $\B$ such that $\iscongcoi{XC}{YC}$.
Since $X,Y$ are the head nonterminals of each word, we need $\B$ to include some pair associated with $X,Y$. 
Since $\normof{X}=\normof{Y}=1$, we might first attempt to guess that $(X,Y)\in\B$. 
Under this guess, we could apply rule \textsc{BPA1} to judgement $\iscongcoi{XC}{YC}$, yielding $\iscongcoi{C}{C}$. 
If we further assume that $\B$ contains pair $(C,C)$, we could then apply rule \textsc{BPA1} to obtain $\iscongcoi{\varepsilon}{\varepsilon}$, which clearly holds by rule \textsc{$\Empty$-Ax}. 
We obtain the following proof tree for $\iscongcoi{XC}{YC}$, assuming that $\B=\{(X,Y),(C,C)\}$.
$$
\inferrule*[right=BPA1]
{\inferrule*[right=BPA1]
{\inferrule*[right=$\Empty$-Ax]{\phantom{lala}}{\iscongcoi{\Empty}{\Empty}}}
{\iscongcoi{C}{C}}}
{\iscongcoi{XC}{YC}}
$$
However, we still need to check if $\B$ is a self-bisimulation (\cref{def:selfbisim}), and in particular, that $(X,Y)$ was the right guess. 
The transition $\ltsred{X}{a}{\varepsilon}$ can only be matched with $\ltsred{Y}{a}{\varepsilon}$, leading to judgement $\iscongcoi{\varepsilon}{\varepsilon}$, which clearly holds by rule \textsc{$\Empty$-Ax}. 
The transition $\ltsred{X}{b}{\varepsilon}$ can only be matched with $\ltsred{Y}{b}{C}$, leading to judgement $\iscongcoi{\varepsilon}{C}$ which clearly does not hold since no rule can be applied, regardless of $\B$. 
In other words, there is no self-bisimulation $\B$ containing pair $(X,Y)$ 
(the above also proves that the pair $(X,Y)$ is irreducible, but it might still be comparable).

At this point, we backtrack to our guess that $(X,Y)\in\B$. 
Since rule \textsc{BPA1} did not work, we next attempt to guess a BPA2 rule for $X,Y$, and the obvious candidate is $(XC,YC)\in\B$. 
Under this guess, we could apply rule \textsc{BPA2} to judgement $\iscongcoi{XC}{YC}$, yielding (two copies of) $\iscongcoi{C}{C}$. 
These are again handled by assuming $(C,C)\in\B$. We obtain the following proof tree for $\iscongcoi{XC}{YC}$, assuming that $\B=\{(XC,YC),(C,C)\}$.
$$
\inferrule*[right=BPA2]
{\inferrule*[right=BPA1]
{\inferrule*[right=$\Empty$-Ax]{\phantom{lala}}{\iscongcoi{\Empty}{\Empty}}}
{\iscongcoi{C}{C}}
\\
\inferrule*[right=BPA1]
{\inferrule*[right=$\Empty$-Ax]{\phantom{lala}}{\iscongcoi{\Empty}{\Empty}}}
{\iscongcoi{C}{C}}}
{\iscongcoi{XC}{YC}}
$$
We still need to check whether $\B$ is a self-bisimulation (\cref{def:selfbisim}), and in particular, that $(XC,YC)$ was the right guess. 
The transition $\ltsred{XC}{a}{C}$ can only be matched with $\ltsred{YC}{a}{C}$, leading to judgement $\iscongcoi{C}{C}$, which holds by assuming $(C,C)\in\B$. 
The transition $\ltsred{XC}{b}{C}$ can only be matched with $\ltsred{YC}{b}{CC}$. Since $C$ is unnormed, we know that $\isbisim{CC}{C}$ (pruning lemma, \cref{prop:unnormedbisim}). Hence, we arrive at judgement $\iscongcoi{C}{C}$, which holds by assuming $(C,C)\in\B$. 
We conclude that the self-bisimulation condition holds for $(XC,YC)$. 
A similar argument shows that it also holds for $(C,C)$, implying that $\B=\{(XC,YC),(C,C)\}$ is a self-bisimulation basis. 
Hence, by \cref{lem:selfbisim}, we get $\congBtwo{c}{\B}\ \subseteq\ \sim$; and since $\iscongcoi{XC}{YC}$, we conclude that $\isbisim{XC}{YC}$.

\subsection{Rough description of the basis-updating algorithm}

Inspired by the short example above, we may now present the main ideas driving the algorithm execution.

\begin{itemize}
\item The algorithm takes as input a simple grammar $\G=(\V,\T,\pcal)$ and two words $\gamma,\delta \in\V^\ast$ and returns `YES' or `NO', indicating whether $\isbisim{\gamma}{\delta}$.
\item The algorithm works by building a derivation tree, whose nodes are pairs of words $(\gamma',\delta')$. Each node intuitively corresponds to the (sub)goal of determining whether $\isbisim{\gamma'}{\delta'}$. 
\item The algorithm keeps track of a basis $\B\subseteq\V^+\times\V^+$ and a set $\scal\subseteq\V\times\V$ of pairs of nonterminals. 
The set $\scal$ contains all pairs which are proved to be irreducible by the algorithm during its execution.
\item Initially, the tree has the single leaf node $(\gamma,\delta)$ corresponding to the pair of words given as input; $\B$ is comprised of the pairs $(X,X)$ for every nonterminal $X\in\V$; and $\scal=\emptyset$.
\item $\B$ and $\scal$ may be updated in the following ways:
\begin{enumerate}[(a)]
\item adding a pair $(X,Y\beta)$ to $\B$, where $(X,Y)$ is not in $\scal$;
\item removing a pair $(X,Y\beta)$ from $\B$, where $(X,Y)$ is not in $\scal$; furthermore, $(X,Y)$ may be added to $\scal$;
\item replacing a pair $(X,Y\beta)$ in $\B$ by a pair $(X\alpha,Y\beta')$, where $(X,Y)$ is not in $\scal$; furthermore, $(X,Y)$ is added to $\scal$;
\item removing a pair $(X\alpha,Y\beta)$ from $\B$, where $(X,Y)$ is in $\scal$.
\end{enumerate}
\item Each internal node in the derivation tree is either unmarked, marked as a BPA1 guess, or marked as a BPA2 guess.
Each leaf in the derivation tree is either finished or unfinished. 
\item If, at some iteration, all leaves are finished, the algorithm returns `YES', indicating that the initial pair is bisimilar.
\item Otherwise, the algorithm chooses the first (in a depth-first search) unfinished leaf and expands it. Suppose that the algorithm is about to examine an unfinished leaf $(\gamma,\delta)$. 
We can essentially divide the expansion cases in five categories:
\begin{enumerate}[(a)]
\item Finishing a leaf: if $\gamma=\delta$ or if $(\gamma,\delta)$ coincides with an already visited node, then the algorithm marks the leaf as finished (producing no children).
\item Applying a congruence rule: if $\gamma=X\alpha'$, $\delta=Y\beta'$ and $\B$ contains a pair of the form $(X\alpha,Y\beta)$, then the algorithm applies the corresponding congruence rule, producing either one (rule BPA1) or two (rule BPA2) children.
\item Adding a new pair: if $\gamma=X\alpha'$, $\delta=Y\beta'$, $\B$ does not contain a pair of the form $(X\alpha,Y\beta)$, and the (one-step) transitions of $X$ and $Y$ match, then the algorithm `guesses' between adding a BPA1 or a BPA2 pair to the basis. 
The algorithm prefers BPA1 guesses over BPA2 guesses. 
The guess results in a new pair $(X\alpha,Y\beta)$ added to $\B$, which is another goal that must be checked for bisimulation. 
Thus, the algorithm adds the matching transition pairs as children of the current node.
\item Partial failure: if exactly one of $\gamma$, $\delta$ is the empty word, or more generaly, if the algorithm can infer that $\isnotbisim\gamma\delta$, then the algorithm moves up the tree until the most recent BPA1 guess, and updates $\B$ by replacing the corresponding BPA1 guess with a BPA2 guess.
\item Total failure: if $\gamma=X\alpha'$, $\delta=Y\beta'$, and the (one-step) transitions of $X$ and $Y$ do not match, then the algorithm immediately returns `NO', indicating that the initial pair is not bisimilar.
\end{enumerate}
\end{itemize}

\noindent
Recall that we have assumed a fixed ordering of the nonterminals $\V=\{X^{(0)} < X^{(1)} < \cdots < X^{(|\V|)}\}$ such that $\normof X < \normof Y$ implies $X<Y$. 
We shall also adopt the pruning convention (\Cref{prop:unnormedbisim}) and assume that, whenever we add a
new pair $(\gamma,\delta)$ to the derivation tree, if $\gamma$ is unnormed then
it is of the form $\gamma=\alpha X$ with $\alpha$ normed and $X$ unnormed; and
similarly for $\delta$. In other words, we remove all nonterminals in a word after its first unnormed nonterminal.

\subsection{Full case analysis of the basis-updating algorithm}

We are now ready to describe the ten possible ways in which a leaf can be expanded. 
Suppose that the algorithm examines an unfinished leaf $(\gamma,\delta)$. 
Each of the following cases is considered in order.

\begin{case}[Loop detection]\label{subsec:loop}~\\
If $(\gamma,\delta)$ coincides with an already visited node (either an internal node, or some finished leaf), then the expansion of this leaf produces zero children and the leaf becomes finished. 
This corresponds to detecting a loop in the coinductive congruence algorithm.
\end{case}

\begin{case}[Identical words]\label{subsec:reflexive}~\\
If $\gamma=\delta$, then the expansion of this leaf produces zero children and marks the leaf as finished. 
This corresponds to successively applying rule \textsc{BPA1} with pairs of identical nonterminals, followed by rule \textsc{$\Empty$-Ax}. 
Note that this also includes the case in which $\gamma$ and $\delta$ are both empty.
\end{case}

\begin{case}[Empty \vs nonempty]\label{subsec:emptyvsnonempty}~\\
If $\gamma=\Empty$ and $\delta\neq\Empty$, or if $\gamma\neq\Empty$ and $\delta=\Empty$, then the expansion of this leaf is a partial failure (described in \Cref{subsec:partialfailure}). 
\end{case}

From now on, we can assume that $\gamma$ and $\delta$ are both non-empty and $(\gamma,\delta)$ is distinct from all internal nodes and finished leaves. 
Let us rewrite $\gamma=X\alpha'$ and $\delta=Y\beta'$. 
We assume $X\geq Y$, the symmetric cases being handled similarly. 
The next cases consider whether $\B$ already contains a pair associated with $X,Y$. 

\begin{case}[Basis includes pair, BPA1 expansion]\label{subsec:bpa1}~\\
If $\B$ contains a pair $(X,Y\beta)$, the expansion of this leaf produces a single child $(\beta\alpha',\beta')$. 
This corresponds to applying rule BPA1 to $(X\alpha',Y\beta')$. 
(In the following diagrams, we include edge labels for mere reader convenience. The algorithm does not require any edge labeling.)

\center
\begin{tikzpicture}
\node[MyNode={}{}](A){$X\alpha',Y\beta'$}
 child {node[MyNode={}{}] {$\beta\alpha',\beta'$} 
  edge from parent node[left] {$\bpai$}};
\node[right of=A,anchor=west,xshift=2cm]{
   $\B$ and $\scal$ remain unchanged};
\end{tikzpicture}
\end{case}

\begin{case}[Basis includes pair, BPA2 expansion]\label{subsec:bpa2}~\\
If $\B$ contains a pair $(X\alpha,Y\beta)$ with $\alpha\neq\Empty$, the expansion of this leaf produces two children $(\alpha,\alpha')$ and $(\beta,\beta')$. 
This corresponds to applying rule BPA2 to $(X\alpha',Y\beta')$.

\center
\begin{tikzpicture}
\node[MyNode={}{}](A){$X\alpha',Y\beta'$}
 child {node[MyNode={}{}] {$\alpha,\alpha'$} 
  edge from parent node[left] {$\bpaii$}}
 child {node[MyNode={}{}] {$\beta,\beta'$} 
  edge from parent node[right] {$\bpaii$}};
\node[right of=A,anchor=west,xshift=2cm]{
	$\B$ and $\scal$ remain unchanged};
\end{tikzpicture}
\end{case}

\noindent
For the remaining cases, we assume that $\B$ does not contain a pair associated with $X,Y$.

\begin{case}[Basis does not include pair, transitions do not match, total failure]\label{subsec:totalfailure}~\\
If there exists some $\ltsred X a {\gamma'}$ without a corresponding $\ltsred Y a {\delta'}$ or vice-versa, then the expansion of this leaf is a total failure. The algorithm returns `NO', indicating that the initial pair is not bisimilar.
\end{case}

From now on, we assume that every transition $\ltsred X a {\gamma'}$ has a matching transition $\ltsred Y a {\delta'}$ and vice-versa.

\begin{case}[Basis does not include pair, transitions match, both unnormed]\label{subsec:unnormedunnormed}~\\
Suppose that $X$ and $Y$ are both unnormed. By the pruning convention, we may assume that $\alpha'=\beta'=\varepsilon$. 
\begin{enumerate}
\item Update $\B$ by adding the pair $(X,Y)$. 
\item In the derivation tree, mark node $(X,Y)$ as a BPA1 guess.
\item For each pair of matching transitions $\ltsred X {a_i} {\gamma_i}, \ltsred Y {a_i} {\delta_i}$, add the node $(\gamma_i,\delta_i)$ as a child of $X,Y$.
\end{enumerate} 

\center
\begin{tikzpicture}
\node[MyNode={}{$\bpai$}](A){$X,Y$}
 child {node[MyNode={}{}] {$\gamma_1,\delta_1$} 
  edge from parent node[left] {$a_1$}}
 child {node {$\dots$}}
 child {node[MyNode={}{}] {$\gamma_k,\delta_k$} 
  edge from parent node[right] {$a_k$}};
\node[right of=A,anchor=west,xshift=2cm,align=left] {$(X,Y)$ enters $\B$\\$\scal$ remains unchanged};
\end{tikzpicture}
\end{case}

\begin{case}[Basis does not include pair, transitions match, unnormed \vs normed]\label{subsec:unnormednormed}~\\
Suppose that $X$ is unnormed but $Y$ is normed (the symmetric case is handled similarly).
By the pruning convention, we may assume that $\alpha'=\varepsilon$. 
We consider two subcases:
\begin{enumerate}
  \item If $Y\beta'$ is unnormed, then: 
\begin{enumerate}
\item Update $\B$ by adding the pair $(X,Y\beta')$. 
\item In the derivation tree, mark node $(X,Y\beta')$ as a BPA1 guess.
\item For each pair of matching transitions $\ltsred X {a_i} {\gamma_i}, \ltsred Y {a_i} {\delta_i}$, add the (pruning of) node $(\gamma_i,\delta_i\beta')$ as a child of $(X,Y\beta')$.
\end{enumerate}
\begin{center}\begin{tikzpicture}
\node[MyNode={}{$\bpai$}](A){$X,Y\beta'$}
 child {node[MyNode={}{}] {$\gamma_1,\delta_1\beta'$} 
  edge from parent node[left] {$a_1$}}
 child {node {$\dots$}}
 child {node[MyNode={}{}] {$\gamma_k,\delta_k\beta'$} 
  edge from parent node[right] {$a_k$}};
\node[right of=A,anchor=west,xshift=2cm,align=left] {$(X,Y\beta')$ enters $\B$\\$\scal$ remains unchanged};
\end{tikzpicture}\end{center}
 \item If $Y\beta'$ is normed, then execute the partial failure routine (\Cref{subsec:partialfailure}) on node $(X,Y\beta')$.
\end{enumerate}
\end{case}

\begin{case}[Basis does not include pair, transitions match, both normed]\label{subsec:normednormed}~\\
Suppose that $X$ and $Y$ are both normed, with $\normof X\geq\normof Y$. 
Let $\ltsred Y u \varepsilon$ be the canonical norm-reducing sequence and let $\beta=\normdynamic X {\normof Y}$ be the $\normof Y$-th word in the canonical norm-reducing sequence of $X$. We consider the following subcases:
\begin{enumerate}
\item If $(X,Y)$ is not in $\scal$ and $\ltsred X u \beta$, then:
\begin{enumerate}
\item Update $\B$ by adding the pair $(X,Y\beta)$.
\item In the derivation tree, mark node $(X\alpha',Y\beta')$ as a BPA1 guess.
\item For each pair of matching transitions $\ltsred X {a_i} {\gamma_i}, \ltsred Y {a_i} {\delta_i}$, add the (pruning of) node $(\gamma_i,\delta_i\beta)$ as a child of $(X\alpha',Y\beta')$.
\item Add the node $(\beta\alpha',\beta')$ as a child of $(X\alpha',Y\beta')$.
\end{enumerate}
\begin{center}\begin{tikzpicture}
[sibling distance=2cm]
\node[MyNode={}{$\bpai$}](A){$X\alpha',Y\beta'$}
 child {node[MyNode={}{}] {$\gamma_1,\delta_1\beta$} 
  edge from parent node[left,xshift=-.1cm] {$a_1$}}
 child {node {$\dots$}}
 child {node[MyNode={}{}] {$\gamma_k,\delta_k\beta$} 
  edge from parent node[right] {$a_k$}}
 child {node[MyNode={}{}] {$\beta\alpha',\beta'$} 
  edge from parent node[right,xshift=.2cm] {$\bpai$}};
\node[right of=A,anchor=west,xshift=3cm,align=left] {$(X,Y\beta)$ enters $\B$\\$\scal$ remains unchanged};
\end{tikzpicture}\end{center}
\item If $(X,Y)$ is in $\scal$ or $\notltsred X u \beta$, and if $X\alpha'$ and $Y\beta'$ are both unnormed, then:
\begin{enumerate}
\item Update $\B$ by adding the pair $(X\alpha',Y\beta')$.
\item Update $\scal$ by adding the pair $(X,Y)$ (if it is not yet in $\scal$).
\item In the derivation tree, mark node $(X\alpha',Y\beta')$ as a BPA2 guess.
\item For each pair of matching transitions $\ltsred X {a_i} {\gamma_i}, \ltsred Y {a_i} {\delta_i}$, add the (pruning of) node $(\gamma_i\alpha',\delta_i\beta')$ as a child of $(X\alpha',Y\beta')$.
\end{enumerate}
\begin{center}\begin{tikzpicture}
\node[MyNode={}{$\bpaii$}](A){$X\alpha',Y\beta'$}
 child {node[MyNode={}{}] {$\gamma_1\alpha',\delta_1\beta'$} 
  edge from parent node[left,xshift=-.1cm] {$a_1$}}
 child {node {$\dots$}}
 child {node[MyNode={}{}] {$\gamma_k\alpha',\delta_k\beta'$} 
  edge from parent node[right] {$a_k$}};
\node[right of=A, anchor=west, xshift=2cm, yshift=-.2cm, align=left] {
$(X\alpha',Y\beta')$ enters $\B$\\$(X,Y)$ enters $\scal$};
\end{tikzpicture}\end{center}
\item If $(X,Y)$ is in $\scal$ or $\notltsred X u \beta$, and one of $X\alpha'$, $Y\beta'$ is normed, then:
\begin{enumerate}
\item Update $\scal$ by adding the pair $(X,Y)$ (if it is not yet in $\scal$).
\item Execute the partial failure routine (\Cref{subsec:partialfailure}) on node $(X\alpha',Y\beta')$.
\end{enumerate}
\end{enumerate}
\end{case}

\begin{case}[Partial failure]\label{subsec:partialfailure}~\\
In a partial failure, the algorithm moves up the tree, removing some of the nodes and updating the basis. 
When executing a partial failure on a given node $(\gamma,\delta)$, the algorithm considers the following subcases:
\begin{enumerate}
\item If $(\gamma,\delta)$ is the root node of the tree, then the partial failure becomes a total failure. The algorithm returns `NO', indicating that the words in the initial pair are not bisimilar. Otherwise, if $(\gamma,\delta)$ is not the root node, then it has a parent, call it $(X\alpha,Y\beta)$. We assume $X\geq Y$, the symmetric cases being handled similarly.
\item If $(X\alpha,Y\beta)$ is a BPA1 guess, and $(\gamma,\delta)$ was obtained from $(X\alpha,Y\beta)$ by a pair of matching transitions, and $X\alpha,Y\beta$ are both unnormed, then:
\begin{enumerate}
\item Update $\B$ by removing the pair associated with $X,Y$ and adding the pair $(X\alpha,Y\beta)$.
\item Update $\scal$ by adding the pair $(X,Y)$.
\item In the derivation tree, mark node $(X\alpha,Y\beta)$ as a BPA2 guess.
\item Prune the tree by removing every node below $(X\alpha,Y\beta)$. 
This includes $(\gamma,\delta)$, its descendants, the sibling nodes of $(\gamma,\delta)$, and their descendants. 
If a removed node is a BPA1 or BPA2 guess, remove also the corresponding pair in $\B$ (leaving $\scal$ unchanged).
\item For each pair of matching transitions $\ltsred{X}{a_i}{\gamma_i}$, $\ltsred{Y}{a_i}{\delta_i}$, add the (pruning of) node $(\gamma_i\alpha,\delta_i\beta)$ as a child of $(X\alpha,Y\beta)$.
\end{enumerate}
\begin{center}\begin{tikzpicture}
\node[MyNode={}{$\bpaii$}](A){$X\alpha,Y\beta$}
 child {node[MyNode={}{}] {$\gamma_1\alpha,\delta_1\beta$} 
  edge from parent node[left,xshift=-.1cm] {$a_1$}}
 child {node {$\dots$}}
 child {node[MyNode={}{}] {$\gamma_k\alpha,\delta_k\beta$} 
  edge from parent node[right] {$a_k$}};
\node[right of=A, anchor=west, xshift=2cm, yshift=-.5cm, align=left] {
some $(X,Y\beta')$ leaves $\B$\\
$(X\alpha,Y\beta)$ enters $\B$\\
some other pairs might leave $\B$\\
$(X,Y)$ enters $\scal$};
\end{tikzpicture}\end{center}
\item If $(X\alpha,Y\beta)$ is a BPA1 guess and $(\gamma,\delta)$ was obtained from $(X\alpha,Y\beta)$ by a pair of matching transitions, but at least one of $X\alpha,Y\beta$ is normed, then:
\begin{enumerate}
\item Update $\B$ by removing the pair associated with $X,Y$.
\item Update $\scal$ by adding the pair $(X,Y)$.
\item Recursively execute the partial failure routine on $(X\alpha,Y\beta)$; \ie go back to subcase 1, considering node $(X\alpha,Y\beta)$ instead of $(\gamma,\delta)$.
\end{enumerate}
\item In any other cases, recursively execute the partial failure routine on $(X\alpha,Y\beta)$; \ie go back to subcase 1, considering node $(X\alpha,Y\beta)$ instead of $(\gamma,\delta)$.
\end{enumerate}
\end{case}

\subsection{Illustration of the basis-updating algorithm}
\label{sec:algorithm-example}

We illustrate the basis-updating algorithm with an example. Consider the simple grammar with nonterminals $\{X,Y,Z,W,V,C,D\}$, terminals $\{a,b,c,d\}$, and production rules below.
\begin{align*}
&\ltsred{X}{a}{\varepsilon} && 
\ltsred{X}{b}{ZC} &&
\ltsred{X}{c}{\varepsilon} &&
\ltsred{Y}{a}{\varepsilon} &&
\ltsred{Y}{b}{WC} &&
\ltsred{Y}{c}{V} \\
&\ltsred{Z}{a}{\varepsilon} &&
\ltsred{Z}{b}{XD} &&
\ltsred{W}{a}{\varepsilon} &&
\ltsred{W}{b}{YD} &&
\ltsred{V}{c}{\varepsilon} &&
\ltsred{C}{c}{C} &&
\ltsred{D}{d}{D}
\end{align*}
Let us sketch the several steps required to determine whether $\isbisim{XC}{YC}$ (the reader may first try to figure out by themselves that, in fact, these are not bisimilar words). 
The resulting derivations trees are depicted in \Cref{fig:XCYC-no}. 
As preprocessing we compute the norms $\normof X = \normof Y = \normof Z = \normof W = \normof V = 1, \normof C = \normof D = \infty$. 
The initial state is composed of a tree with a single node $(XC,YC)$, a basis
$\B = \{(X,X),(Y,Y),(Z,Z),(W,W),(V,V),(C,C),(D,D)\}$ containing all pairs of
identical non-terminals, and an empty set $\scal$ of irreducible pairs of
nonterminal symbols.

\begin{figure}\centering
\begin{tikzpicture}
[sibling distance = 2.2cm, label distance = -.1cm]
\node[MyNode={1}{$\bpai$}] (root1) at (0,0) {$XC,YC$}
 child {node[MyFinishedNode={2}{$\refl$}] {$\varepsilon,\varepsilon$} 
  edge from parent node[left,xshift=-.2cm] {$a$}}
 child {node[MyNode={3}{$\bpai$}] {$ZC,WC$}
  child {node[MyFinishedNode={4}{$\aloop$}] {$\varepsilon,\varepsilon$}
   edge from parent node[left,xshift=-.1cm] {$a$}}
  child {node[MyNode={5}{}] {$XD,YD$}
   child {node[MyFinishedNode={6}{$\refl$}] {$D,D$}
    edge from parent node[left] {$\bpai$}}
   edge from parent node[left] {$b$}}
  child {node[MyFinishedNode={7}{$\refl$}] {$C,C$}
   edge from parent node[right,xshift=.1cm] {$c$}}
  edge from parent node[left] {$b$}}
 child {node[MyFinishedNode={8}{$\pfail$}] {$\varepsilon,V$} 
  edge from parent node[right] {$c$}}
 child {node[MyNode={}{}] {$C,C$} 
  edge from parent node[right,xshift=.3cm] {$\bpai$}}
;
\tikzset{sibling distance = 2.4cm}
\node[MyNode={8}{$\bpaii$}] (root2) at (7.3,0) {$XC,YC$}
 child {node[MyFinishedNode={9}{$\refl$}] {$C,C$} 
  edge from parent node[left,xshift=-.1cm] {$a$}}
 child 
 {node[MyNode={10}{$\bpai$}] {$ZC,WC$}
  child {node[MyFinishedNode={11}{$\refl$}] {$\varepsilon,\varepsilon$}
   edge from parent node[left,xshift=-.1cm] {$a$}}
  child {node[MyNode={12}{}] {$XD,YD$}
   child {node[MyFinishedNode={13}{$\tfail$}] {$C,D$}
    edge from parent node[left,xshift=-.1cm] {$\bpaii$}}
   child {node[MyNode={}{}] {$C,D$}
    edge from parent node[right,xshift=.1cm] {$\bpaii$}}
   edge from parent node[left] {$b$}}
  child {node[MyNode={}{}] {$C,C$}
   edge from parent node[right,xshift=.2cm] {$\bpai$}}
  edge from parent node[left] {$b$}}
 child {node[MyNode={}{}] {$C,VC$} 
  edge from parent node[right,xshift=.1cm] {$c$}}
;
\draw[->,dashed] (root1.east) -- (root2.west);
\end{tikzpicture}
\caption{Derivation trees for determining that $\isnotbisim {XC}{YC}$. Double boundaries indicate finished leaves. A superscript on a node identifies the step in which it was visited. A subscript on an internal node indicates whether it corresponds to a BPA1 or BPA2 guess. A subscript on a finished leaf indicates whether it corresponds to a loop, pair of identical words, partial failure, or total failure. The right tree results from the left tree after step 8, keeping only the root node. For convenience of the reader, each edge is also labeled by either a terminal symbol (corresponding to a matching transition), BPA1 or BPA2 (corresponding to a congruence rule).\label{fig:XCYC-no}}
\end{figure}
\begin{enumerate}
\item Examining $(XC,YC)$, we see that there is no pair in $\B$ corresponding to $X,Y$ and their immediate transitions match. Since $\ltsred Ya\varepsilon, \ltsred Xa\varepsilon$ (\Cref{subsec:normednormed}.1), we `guess' that $\isbisim{X}{Y}$ and update $\B$ by adding the pair $(X,Y)$. 
We mark node $(XC,YC)$ as a BPA1 guess and add four children: one for each matching transition $(\varepsilon,\varepsilon)$, $(ZC,WC)$, $(\varepsilon,V)$, as well as $(C,C)$ obtained by the BPA1 rule.
\item Examining $(\varepsilon,\varepsilon)$, we mark this leaf as finished
  (identical words, \Cref{subsec:reflexive}) and proceed to the next leaf.
\item\label{item:zcwc} Examining $(ZC,WC)$, we again see that there is no pair in $\B$ corresponding to $Z,W$ and their immediate transitions match. Since $\ltsred Za\varepsilon, \ltsred Wa\varepsilon$ (\Cref{subsec:normednormed}.1), we `guess' that $\isbisim ZW$ and update $\B$ by adding the pair $(Z,W)$. 
We mark node $(ZC,WC)$ as a BPA1 guess and add three children: one for each matching transition $(\varepsilon,\varepsilon)$, $(XD,YD)$, as well as $(C,C)$ obtained by the BPA1 rule.
\item Examining $(\varepsilon,\varepsilon)$, we mark this leaf as finished (\Cref{subsec:loop}) and proceed to the next leaf.
\item Examining $(XD,YD)$, we see that $\B$ already contains a BPA1 pair $(X,Y)$ (\Cref{subsec:bpa1}). We apply rule \textsc{BPA1}, producing a child $(D,D)$.
\item Examining $(D,D)$, we mark this leaf as finished (\Cref{subsec:reflexive}) and proceed to the next leaf.
\item We mark the leaf $(C,C)$ as finished (\Cref{subsec:reflexive}).
\item Examining $(\varepsilon,V)$, we arrive at a partial failure (\Cref{subsec:emptyvsnonempty,subsec:partialfailure}). 
We travel up the tree to the earliest BPA1 guess, which corresponds to the parent node $(XC,YC)$.
Then, we start over using a BPA2 guess (\Cref{subsec:partialfailure}.2); that is, we update $\B$ by removing $(X,Y)$ and adding $(XC,YC)$. 
We also add $(X,Y)$ to $\scal$ (to indicate that $(X,Y)$ cannot be a BPA1 guess anymore) and mark node $(XC,YC)$ as a BPA2 guess. 
We remove every descendant of $(XC,YC)$ (which effectively removes every node except the root). 
This also means that we remove pair $(Z,W)$ from $\B$. 
Finally, we add three children to $(XC,YC)$: one for each matching transition
$(C,C)$, $(ZC,WC)$, $(C,VC)$. 
Notice in particular that the transition with label $b$ gives rise to the pair
$(ZC,WC)$ rather than $(ZCC,WCC)$. Although $\ltsred{XC}b{ZCC}$, we apply the
pruning convention, writing $ZC$ in place of $ZCC$ since $C$ is unnormed.
\item We mark the leaf $(C,C)$ as finished (\Cref{subsec:reflexive}).
\item Examining $(ZC,WC)$, we repeat the process in \Cref{item:zcwc} (\Cref{subsec:normednormed}.1). We update $\B$ by adding the pair $(Z,W)$. We mark node $(ZC,WC)$ as a BPA1 guess and add three children: one for each matching transition $(\varepsilon,\varepsilon)$, $(XD,YD)$, as well as $(C,C)$ obtained by the BPA1 rule.
\item We mark the leaf $(\varepsilon,\varepsilon)$ as finished (\Cref{subsec:reflexive}).
\item\label{item:XDYD}  Examining $(XD,YD)$, we see that $\B$ contains a pair $(XC,YC)$ (\Cref{subsec:bpa2}). We apply rule \textsc{BPA2}, producing children $(C,D)$ and $(C,D)$.
\item Examining $(C,D)$, we see that there is no corresponding pair in $\B$ and their immediate transitions do not match. Therefore we arrive at a total failure (\Cref{subsec:totalfailure}), meaning that $\isnotbisim CD$ and this non-bisimilarity propagates to the root. We terminate concluding that $(XC,YC)$ is not bisimilar.
\end{enumerate}

\noindent
For a positive answer, replace transition $\ltsred DdD$ by $\ltsred DcD$ in the example above. The algorithm follows the same steps as above until \Cref{item:XDYD}, and continues as follows (the resulting derivation tree is in \Cref{fig:XCYC-yes}).
\begin{figure}[t]\centering
\begin{tikzpicture}
[sibling distance = 4.5cm]
\node[MyNode={8}{$\bpaii$}] {$XC,YC$}
 child {node[MyFinishedNode={9}{$\refl$}] {$C,C$}
  edge from parent node[left,xshift=-.2cm] {$a$}}
 child [sibling distance = 2.5cm] {
  node[MyNode={10}{$\bpai$}] {$ZC,WC$}
  child {node[MyFinishedNode={11}{$\refl$}] {$\varepsilon,\varepsilon$}
   edge from parent node[left,xshift=-.1cm] {$a$}}
  child {node[MyNode={12}{}] {$XD,YD$}
   child {node[MyFinishedNode={13'}{$\bpai$}] {$C,D$}
    child {node[MyFinishedNode={14}{$\aloop$}] {$C,D$}
     edge from parent node[left] {$c$}}
    edge from parent node[left,xshift=-.1cm] {$\bpaii$}}
   child {node[MyFinishedNode={15}{$\aloop$}] {$C,D$}
    edge from parent node[right,xshift=.1cm] {$\bpaii$}}
   edge from parent node[left] {$b$}}
  child {node[MyFinishedNode={16}{$\aloop$}] {$C,C$}
   edge from parent node[right,xshift=.2cm] {$\bpai$}}
  edge from parent node[left] {$b$}}
 child {node[MyNode={17}{$\bpai$}] {$C,VC$}
  child {node[MyFinishedNode={18}{$\aloop$}] {$C,C$}
   edge from parent node[right] {$c$}} 
  edge from parent node[right,xshift=.2cm] {$c$}}
;
\end{tikzpicture}
\caption{Derivation tree for determining that $\isbisim {XC}{YC}$, in the case that $\isbisim CD$. Superscripts, subscripts and edge labels are as in \Cref{fig:XCYC-no}.\label{fig:XCYC-yes}}
\end{figure}
\begin{enumerate}
\setcounter{enumi}{13}
\item[(13')] Examining $(C,D)$, we see that there is no corresponding pair in $\B$ and the immediate transitions match. Since $C$, $D$ are both unnormed (\Cref{subsec:unnormedunnormed}), we `guess' that $\isbisim CD$ and update $\B$ by adding pair $(C,D)$. We mark node $(C,D)$ as a BPA1 guess and add a child $(C,D)$ corresponding to the only matching transition.
\item We mark the leaf $(C,D)$ as finished, as it is coincides with a previously visited pair of words (\Cref{subsec:loop}).
\item We mark the leaf $(C,D)$ as finished (\Cref{subsec:loop}).
\item We mark the leaf $(C,C)$ as finished (\Cref{subsec:loop}).
\item Examining $(C,VC)$, we see that there is no corresponding pair in $\B$ and the immediate transitions match. Since $C$ is unnormed and $V$ is normed (\Cref{subsec:unnormednormed}), we `guess' that $\isbisim C{VC}$ and update $\B$ by adding pair $(C,VC)$. We mark node $(C,VC)$ as a BPA1 guess and add a child $(C,C)$ corresponding to the only matching transition.
\item We mark the leaf $(C,C)$ as finished (\Cref{subsec:loop}).
\item Finally, there are no more unfinished leaves. The algorithm terminates deducing that pair $(XC,YC)$ is bisimilar. At the end of the algorithm, $\B$ contains the pairs $(XC,YC)$, $(Z,W)$, $(C,D)$, $(C,VC)$, and $\scal=\{(X,Y)\}$.
\end{enumerate}



\section{Correctness of the algorithm}
\label{sec:correctness}

In this section we prove that the basis-updating algorithm is correct. We split the proof in three parts.
\begin{itemize}
\item If the algorithm returns `YES' (by finishing all leaves), then the given pair is bisimilar (\cref{thm:yes-soundness}).
\item If the algorithm returns `NO' (by a total failure), then the given pair is not bisimilar (\cref{thm:no-soundness}).
\item The algorithm terminates for every input (\cref{thm:termination}).
\end{itemize}

The algorithm shall preserve as an invariant the property that $\B$ is reflexive, norm-compliant, functional and simple (\Cref{lem:basis-invariant}). 
This will ensure that expansion steps which leave $\B$ unchanged cannot go on forever, by the reasoning presented in \Cref{thm:coinductive-congruence}. 
As such, we shall prove termination of the algorithm by showing that the basis can be updated at most polynomially many times ($\ocal(n^4)$ where $n$ is the number of nonterminals), and that the number of expansion steps between two consecutive updating steps is bounded by a polynomial on the current size of the basis and the maximum seminorm among all words encountered thus far. 

\begin{restatable}[Basis invariant]{lem}{basisinvariant}
\label{lem:basis-invariant}
At each stage during the execution of the basis-updating algorithm, the basis $\B$ is norm-compliant, reflexive, functional and simple.
\end{restatable}

\begin{proof}
At the start of the algorithm $\B$ consists of all pairs $(X,X)$ of identical non-terminals, which is trivially a norm-compliant, reflexive, functional and simple basis.

\textbf{Norm-compliant}: 
New pairs of the form $(X,Y\beta)$ are added to $\B$ in: 
\Cref{subsec:unnormedunnormed}, where $\beta=\varepsilon$ and $\normof X=\normof Y=\infty$; 
\Cref{subsec:unnormednormed}, where it is enforced that $\normof X=\normof{Y\beta}=\infty$; 
or \Cref{subsec:normednormed}, where it is enforced that $\beta=\normdynamic{X}{\normof{Y}}$ and thus $\normof{Y\beta}=\normof Y + (\normof{X}-\normof{Y})=\normof X$.
New pairs of the form $(X\alpha,Y\beta)$ with $\alpha\neq\Empty$ are added to $\B$ in \Cref{subsec:normednormed} or \Cref{subsec:partialfailure}: either way, it is enforced that $\normof{X\alpha}=\normof{Y\beta}=\infty$.

\textbf{Reflexive}: 
Pairs $(X,X)$ are never removed from $\B$ since they are never marked as BPA1 guesses in the derivation tree.

\textbf{Functional}: 
Suppose that a pair $(X,Y\beta)$ is added to $\B$ at some stage of the algorithm, with $\normof X<\infty$. This only occurs in \Cref{subsec:normednormed}. By construction, $\beta=\normdynamic X {\normof Y}$.

\textbf{Simple}:
New pairs are only added to $\B$ in: \Cref{subsec:unnormedunnormed,subsec:unnormednormed,subsec:normednormed}, where it is ensured that $\B$ did not yet have a pair associated with $X,Y$; or \Cref{subsec:partialfailure}, where an existing pair $(X,Y\beta')$ is replaced by a pair $(X\alpha,Y\beta)$.
\end{proof}

\subsection{`YES'-soundness}

Intuitively, to prove `YES'-soundness we shall claim that the basis $\B$ produced by the algorithm is a self-bisimulation and that the root node in the tree is in the coinductive congruence of $\B$. 
The set of all pairs appearing in the derivation tree is a witness for both claims.

\begin{restatable}{lem}{yesallcongruent}
\label{lem:yes-allpairscongruent}
Suppose that the basis-updating algorithm returns `YES' for a given input $(\gamma,\delta)$. Let $\B$ be the basis at the end of the algorithm, and let $R$ be the set of all pairs in the derivation tree at the end of the algorithm. Then $R\ \subseteq\ \congBtwo{c}{\B}$.
\end{restatable}

\begin{proof}
We shall show that the set
$$R' = R \cup \{(\gamma',\gamma') : \gamma'\in\V^\ast\}$$
is backward closed for the rules defining coinductive congruence $\congBtwo{c}{\B}$, and thus $R\subseteq R' \subseteq\ \congBtwo{c}{\B}$. 

For any pair $(\gamma',\gamma')$, either $\gamma'=\varepsilon$, and we can apply rule \textsc{$\Empty$-Ax} producing no descendants; or $\gamma'=X\gamma''$ for some non-terminal $X$. 
In the latter case, since $\B$ is reflexive (\Cref{lem:basis-invariant}) we can apply rule \textsc{BPA1} with pair $(X,X)$, producing the single descendant $(\gamma',\gamma')$ which is still in $R'$. 

We may now consider a pair $(\gamma',\delta')\in R$ with $\gamma'\neq\delta'$. 
Consider the first (according to depth-first search) occurrence of this pair in the derivation tree. 
We perform a case analysis according to which way the basis-updating algorithm expanded $(\gamma',\delta')$:
\begin{itemize}
\item (\Cref{subsec:loop} - loop) 
This cannot be the case at the first occurrence of $(\gamma',\delta')$.
\item (\Cref{subsec:reflexive} - identical) 
This cannot be the case since $\gamma'\neq\delta'$.
\item (\Cref{subsec:emptyvsnonempty} - empty \vs nonempty)  
This cannot be the case since $(\gamma',\delta')$ would have been removed from the derivation tree.
\item (\Cref{subsec:bpa1} - BPA1 expansion) 
Then $\gamma'=X\alpha'$, $\delta'=Y\beta'$, there was a pair $(X,Y\beta)$ in $\B$, and the expansion produced $(\beta\alpha',\beta')$ in the derivation tree.
We argue that the pair $(X,Y\beta)$ remains in $\B$ at the end of the algorithm: it was added prior to expanding the current node $(\gamma',\delta')$, and if it was removed by a subsequent partial failure, then the current node would necessarily have the same common ancestor and would have been removed as well (this is a consequence of depth-first search traversal).
Since $(X,Y\beta)\in\B$, we can apply rule \textsc{BPA1} to $(X\alpha',Y\beta')$, arriving at $(\beta\alpha',\beta')$ which is in $R$.
\item (\Cref{subsec:bpa2} - BPA2 expansion) 
Then $\gamma'=X\alpha'$, $\delta'=Y\beta'$, there was a pair $(X\alpha,Y\beta)$ in $\B$, and the expansion produced $(\alpha,\alpha')$ and $(\beta,\beta')$ in the derivation tree. 
The pair $(X\alpha,Y\beta)$ remains in $\B$ at the end of the algorithm, since otherwise $(X\alpha',Y\beta')$ would have been removed as well (by the same reasoning as the previous case). 
Thus, we can apply rule \textsc{BPA2} to $(X\alpha',Y\beta')$, arriving at $(\alpha,\alpha')$, $(\beta,\beta')$ which are in $R$.
\item (\Cref{subsec:totalfailure} - total failure) 
This cannot be the case since the algorithm would have returned `NO'.
\item (\Cref{subsec:unnormedunnormed} - both unnormed)
Then $\gamma'=X$, $\delta'=Y$ and the algorithm added $(X,Y)$ to $\B$ at this stage. The pair $(X,Y)$ remains in $\B$ at the end of the algorithm, since the node $(X,Y)$ remains in the derivation tree. 
Thus, we can apply rule \textsc{BPA1} to $(X,Y)$, arriving at $(\varepsilon,\varepsilon)$ which is in $R'$.
\item (\Cref{subsec:unnormednormed} - unnormed \vs normed)
Similarly to the previous case, $(\gamma',\delta')=(X,Y\beta')$ remains in $\B$ at the end of the algorithm. Applying rule \textsc{BPA1} to $(X,Y\beta')$, we arrive at $(\beta',\beta')$ which is in $R'$.
\item (\Cref{subsec:normednormed} - both normed)
Then $\gamma'=X\alpha'$, $\delta'=Y\beta'$. If the algorithm added a pair $(X,Y\beta)$ at this stage, then its expansion produced $(\beta\alpha',\beta')$ in the derivation tree. The pair $(X,Y\beta)$ remains in $\B$ at the end of the algorithm, otherwise we would have backtracked to this node and expanded it according to \Cref{subsec:partialfailure} (partial failure). 
Thus, we can apply rule \textsc{BPA1} to $(\gamma',\delta')$, arriving at $(\beta\alpha',\beta')$, which is in $R'$. 
Otherwise, the algorithm added a pair $(X\alpha',Y\beta')$ at this stage, which remains at the end of the algorithm. Applying rule \textsc{BPA2} to $(X\alpha',Y\beta')$, we arrive at $(\alpha',\alpha')$ and $(\beta',\beta')$ which are in $R'$.
\item (\Cref{subsec:partialfailure} - partial failure) 
Then $\gamma'=X\alpha$, $\delta'=Y\beta$, the algorithm removed every node below $(X\alpha,Y\beta)$, and $(X\alpha,Y\beta)$ was added to $\B$. The pair $(X\alpha,Y\beta)$ remains in $\B$ at the end of the algorithm, since the node $(X\alpha,Y\beta)$ remains in the derivation tree. 
Thus, we can apply rule \textsc{BPA2} to $(X\alpha,Y\beta)$, arriving at $(\alpha,\alpha)$, $(\beta,\beta)$ which are in $R'$.\qedhere
\end{itemize}
\end{proof}

\begin{restatable}{lem}{yesselfbisim}
\label{lem:yes-selfbisimulation}
Suppose that the basis-updating algorithm returns `YES' for a given input $(\gamma,\delta)$. Let $\B$ be the resulting basis. Then $\B$ is a self-bisimulation.
\end{restatable}

\begin{proof}
Let $R$ be the set of all pairs appearing in the resulting derivation tree, so that, by \Cref{lem:yes-allpairscongruent}, we get that $R\ \subseteq\ \congBtwo{c}{\B}$. 
Since $\B$ is reflexive (\Cref{lem:basis-invariant}), trivially $\iscongcoi{\gamma'}{\gamma'}$ for any word $\gamma'$ (by successive application of rule \textsc{BPA1} with pairs of identical nonterminals, followed by rule \textsc{$\Empty$-Ax}). 
For any given pair in $\B$, we consider three cases:
\begin{itemize}
\item if the pair is of the form $(X,X)$ (identical non-terminals) then clearly any transition $\ltsred X a {\gamma'}$ matches with itself, and $\iscongcoi{\gamma'}{\gamma'}$ by the above observation.
\item if the pair is of the form $(X,Y\beta)$ (with $X\neq Y$), then there is a node in the derivation tree corresponding to the step in which this pair was added to $\B$, according to either \Cref{subsec:unnormedunnormed}, \Cref{subsec:unnormednormed} or \Cref{subsec:normednormed}. 
In either case, for each pair of matching transitions $\ltsred X {a_i} {\gamma_i}, \ltsred Y {a_i} {\delta_i}$, there is a corresponding node $(\gamma_i,\delta_i\beta)$ in $R\ \subseteq\ \congBtwo{c}{\B}$. 
Therefore $\iscongcoi{\gamma_i}{\delta_i\beta}$.
\item if the pair is of the form $(X\alpha,Y\beta)$ (with $X\neq Y$ and $\alpha\neq\varepsilon$), then there is a node in the derivation tree corresponding to the step in which this pair was added to $\B$, according to either \Cref{subsec:normednormed} or \Cref{subsec:partialfailure}. 
In either case, for each pair of matching transitions $\ltsred X {a_i} {\gamma_i}, \ltsred Y {a_i} {\delta_i}$, there is a corresponding node $(\gamma_i\alpha,\delta_i\beta)$ in $R\ \subseteq\ \congBtwo{c}{\B}$. 
Therefore $\iscongcoi{\gamma_i\alpha}{\delta_i\beta}$.\qedhere
\end{itemize}
\end{proof}

\begin{restatable}[`YES'-soundness]{thm}{yessoundness}
\label{thm:yes-soundness}
If the basis-updating algorithm returns `YES' for a given input $(\gamma,\delta)$, then $\isbisim\gamma\delta$.
\end{restatable}

\begin{proof}
Let $R$ be the set of all pairs appearing in the resulting derivation tree. 
By \Cref{lem:yes-allpairscongruent}, we get that $R\subseteq\ \congBtwo{c}{\B}$.
By \Cref{lem:yes-selfbisimulation}, $\B$ is a self-bisimulation and thus, by \Cref{lem:selfbisim}, $\congBtwo{c}{\B}\ \subseteq\ \sim$. 
Therefore $(\gamma,\delta)\in R\subseteq\ \congBtwo{c}{\B}\ \subseteq\ \sim$, that is, $\isbisim\gamma\delta$.
\end{proof}

\subsection{`NO'-soundness}

Intuitively, to prove `NO'-soundness we shall claim that, whenever a node $(\gamma',\delta')$ results in a partial failure, the corresponding words are not bisimilar, and this non-bisimilarity can be propagated through the tree until the first BPA1 guess ancestor. 
On the other hand, whenever a node $(\gamma',\delta')$ results in a total failure, the corresponding words are incomparable (there is no way to restore bisimilarity by adding a tail $\alpha$ to $\gamma'$ and $\beta$ to $\delta'$), and this incomparability can be propagated through the tree until reading the root node.

\begin{restatable}{lem}{partialfailure}
\label{lem:partial-irreducible} 
At each iteration during the basis-updating algorithm, the following claims hold.
\begin{enumerate}
\item If the partial failure routine is called at some node $(\gamma',\delta')$, then $\isnotbisim{\gamma'}{\delta'}$ and at least one of $\gamma'$, $\delta'$ is normed.
\item If a pair $(X,Y)$ is in $\scal$, then there is no word $\gamma$ such that $\isbisim X{Y\gamma}$ or $\isbisim {X\gamma} Y$.
\end{enumerate}
\end{restatable}

\begin{proof}
We prove that both properties hold after $k$ iterations of the algorithm, for each $k\in\nbb$, by induction on $k$. At the start of the algorithm ($k=0$) both properties hold, since there have ocurred no partial failures and $\scal=\emptyset$.

Suppose both properties hold after $k$ iterations. To prove that the first property holds after $k+1$ iterations, we may assume that, in iteration $k+1$, a partial failure is called at some node $(\gamma',\delta')$, and consider all cases in which this might happen. 

\begin{itemize}
\item (\Cref{subsec:emptyvsnonempty} - empty \vs nonempty) 
If $\gamma'=\varepsilon$ and $\delta'\neq\varepsilon$ or vice-versa, since the grammar has no dead symbols, it is immediate that $\isnotbisim{\gamma'}{\delta'}$ and at least one of $\gamma'$, $\delta'$ is normed.
\item (\Cref{subsec:unnormednormed} - unnormed \vs normed)
If $\gamma'=X$ with $X$ unnormed and $\delta'=Y\beta'$ with $Y\beta'$ normed, it is immediate that $\isnotbisim{X}{Y\beta'}$.
\item (\Cref{subsec:normednormed} - both normed) 
Suppose that $\gamma'=X\alpha'$ with $X$ normed, $\delta'=Y\beta'$ with $Y$ normed, $\ltsred Y u \varepsilon$, $\beta=\normdynamic X {\normof Y}$, $(X,Y)$ is in $\scal$ or $\notltsred X u \beta$, and one of $X\alpha'$, $Y\beta'$ is normed. If $(X,Y)\in\scal$, then by induction hypothesis $(X,Y)$ is irreducible. Similarly, if $\notltsred X u \beta$, then $\isnotbisim{X}{Y\beta}$, as the canonical norm-reducing sequences of both words do not coincide. In either case, by \Cref{lem:charnormnorm} (\Cref{lem:charnormnorm-2}) we get $\isnotbisim{X\alpha'}{Y\beta'}$.
\item (\Cref{subsec:partialfailure} - partial failure) 
If a partial failure routine is recursively called on $(\gamma',\delta')$, then we can write $\gamma'=X\alpha'$, $\delta'=Y\beta'$; moreover, $(\gamma',\delta')$ is the parent of a node $(\gamma'',\delta'')$ which itself was deemed a partial failure. By induction hypothesis, $\isnotbisim{\gamma''}{\delta''}$ and at least one of $\gamma''$, $\delta''$ is normed. We perform a case analysis according to which way the basis-updating algorithm expanded $(\gamma',\delta')$:
\begin{itemize}
\item \Cref{subsec:loop} (loop), \Cref{subsec:reflexive} (identical), \Cref{subsec:emptyvsnonempty} (empty \vs nonempty), \Cref{subsec:totalfailure} (total failure) cannot be the case since they produce no children.
\item (\Cref{subsec:bpa1} - BPA1 expansion) 
Then there was a pair $(X,Y\beta)$ in $\B$ and $\gamma''=\beta\alpha'$, $\delta''=\beta'$. 
If $X$, $Y$ are both unnormed then, by the pruning convention we get that $\alpha'=\beta=\beta'=\gamma''=\delta''=\varepsilon$, contradicting the induction hypothesis that $\isnotbisim{\gamma''}{\delta''}$. 
If $X$ is unnormed and $Y$ is normed, then by the pruning convention $\alpha'=\varepsilon$. 
Since $\B$ is norm-compliant (\Cref{lem:basis-invariant}), $\beta$ is unnormed. 
By the induction hypothesis, $\delta''=\beta'$ must be normed, so that $Y\beta'$ is also normed. 
It is then clear that $\isnotbisim{X}{Y\beta'}$. Finally, suppose that $X$, $Y$ are both normed. Since $\B$ is functional, $\ltsred Y u \varepsilon$ and $\ltsred X u \beta$ for some $u$. 
Then also $\ltsred{X\alpha'}u{\beta\alpha'=\gamma''}$ and $\ltsred{Y\beta'}u{\beta'=\delta''}$. It is then immediate that $\isnotbisim{X\alpha'}{Y\beta'}$ (otherwise $\isbisim{\gamma''}{\delta''}$) and at least one of $X\alpha'$, $Y\beta'$ is normed (otherwise both $\gamma''$, $\delta''$ would be unnormed).
\item (\Cref{subsec:bpa2} - BPA2 expansion) 
Then there was a pair $(X\alpha,Y\beta)$ and $(\gamma'',\delta'')$ is either $(\alpha,\alpha')$ or $(\beta,\beta')$. 
Without loss of generality suppose it is the former. 
Since $\B$ is norm-compliant (\Cref{lem:basis-invariant}), $\alpha$ is unnormed, so that by induction hypothesis, $\alpha'$ is normed (as well as $X\alpha'$). 
Also by induction hypothesis (second property), since $(X,Y)$ is in $\scal$, it is irreducible. 
By \Cref{lem:charnormnorm} (\Cref{lem:charnormnorm-2}), $\isnotbisim{X\alpha'}{Y\beta'}$ as desired.
\item (\Cref{subsec:unnormedunnormed} - both unnormed) 
Then $\alpha'=\beta'=\varepsilon$ and there is some matching transition $\ltsred X a{\gamma''}$, $\ltsred Y a {\delta''}$. 
This cannot be the case since both $\gamma''$ and $\delta''$ would be unnormed, contradicting the induction hypothesis.
\item (\Cref{subsec:unnormednormed} - unnormed \vs normed) 
Then $\alpha'=\varepsilon$ and there is some matching transition $\ltsred X a{\gamma''}$, $\ltsred {Y\beta'} a {\delta''}$. Since $(X,Y\beta')$ was added to $\B$, which is norm-compliant (\Cref{lem:basis-invariant}), $\beta'$ is unnormed. 
This leads to a contradiction since both $\gamma''$ and $\delta''$ would be unnormed.
\item (\Cref{subsec:normednormed} - both normed) 
First consider the subcase in which $(X,Y)$ is not in $\scal$ and $\ltsred Y u \varepsilon$, $\ltsred X u \beta$. 
If $(\gamma'',\delta'')$ is obtained from $(X\alpha',Y\beta')$ by a pair of matching transitions, then at least one of $X\alpha'$, $Y\beta'$ is normed (otherwise the partial failure routine would not have been called at $(X\alpha',Y\beta')$). 
Moreover, we have $\ltsred X a {\gamma''}$, $\ltsred {Y\beta} a {\delta''}$, implying that $\isnotbisim{X}{Y\beta}$. 
By \Cref{lem:charnormnorm} (\Cref{lem:charnormnorm-1c}), we conclude that $(X,Y)$ is irreducible, and by \Cref{lem:charnormnorm} (\Cref{lem:charnormnorm-2}) we further conclude that $\isnotbisim{X\alpha'}{Y\beta'}$. 
Otherwise if $(\gamma'',\delta'')$ is not obtained from $(X\alpha',Y\beta')$ by a pair of matching transitions, then $\gamma''=\beta\alpha'$ and $\delta''=\beta'$. 
Moreover, we have $\ltsred{X\alpha'}u{\beta\alpha'=\gamma''}$ and $\ltsred{Y\beta'}u{\beta'=\delta''}$. 
It is then immediate that $\isnotbisim{X\alpha'}{Y\beta'}$ (otherwise $\isbisim{\gamma''}{\delta''}$) and at least one of $X\alpha'$, $Y\beta'$ is normed (otherwise both $\gamma''$, $\delta''$ would be unnormed). 

Next, consider the subcase in which $(X,Y)$ is in $\scal$ or $\ltsred Y u \varepsilon$ but $\notltsred X u \beta$. 
Either way, $X\alpha'$ and $Y\beta'$ are both unnormed (otherwise we would not have produced children $(\gamma'',\delta'')$). 
Moreover, there is a pair of matching transitions $\ltsred{X\alpha'}a{\gamma''}$, $\ltsred{Y\beta'}a{\delta''}$, contradicting the hypothesis that one of $\gamma''$, $\delta''$ is normed.
\item (\Cref{subsec:partialfailure} - partial failure) 
Here the only way for children $(\gamma'',\delta'')$ to be produced is if $(X\alpha',Y\beta')$ was a BPA1 guess turned into a BPA2 guess and $X\alpha'$, $Y\beta'$ are both unnormed. 
Moreover, there is a pair of matching transitions $\ltsred{X\alpha'}a{\gamma''}$, $\ltsred{Y\beta'}a{\delta''}$, contradicting the hypothesis that one of $\gamma''$, $\delta''$ is normed.
\end{itemize}
\end{itemize}

\noindent
We now prove that the second property holds after $k+1$ iterations. We may assume that, in iteration $k+1$, a pair $(X,Y)$ enters $\scal$, and consider all cases in which this might happen.
\begin{itemize}
\item (\Cref{subsec:normednormed} - both normed)
In this case, $X$ and $Y$ are both normed, $\ltsred Yu\varepsilon$, $\beta=\normdynamic X{\normof Y}$, and either $(X,Y)$ already was in $\scal$ or $\notltsred X u \beta$. 
If $X,Y$ already was in $\scal$, by induction hypothesis it is an irreducible pair. 
If $\notltsred X u \beta$, then $\isnotbisim X{Y\beta}$ (since the canonical norm-reducing sequences do not coincide) and thus, by \Cref{lem:charnormnorm} (\Cref{lem:charnormnorm-1c}), $X,Y$ is irreducible.
\item (\Cref{subsec:partialfailure} - partial failure) 
In this case, node $(X\alpha,Y\beta)$ is a BPA1 guess having a children $(\gamma',\delta')$ obtained by a pair of matching transitions; moreover, $(\gamma',\delta')$ was the target of a partial failure routine. 
Let $(X,Y\beta')$ be the corresponding pair in $\B$ so that $\ltsred X a {\gamma'}$ and $\ltsred {Y\beta'}{a}{\delta'}$. 
By induction hypothesis, $\isnotbisim{\gamma'}{\delta'}$ and one of $\gamma'$, $\delta'$ is normed. 
Therefore, $\isnotbisim X{Y\beta'}$ and one of $X$, $Y\beta'$ is normed. 
Since $\B$ is norm-compliant (\Cref{lem:basis-invariant}), both $X$ and $Y\beta$ are normed. 
Since $\B$ is functional (\Cref{lem:basis-invariant}), $\beta'=\normdynamic X{\normof Y}$. 
By \Cref{lem:charnormnorm} (\Cref{lem:charnormnorm-1c}), $X,Y$ is irreducible.
\qedhere
\end{itemize}
\end{proof}

\begin{restatable}[`NO'-soundness]{thm}{nosoundness}
\label{thm:no-soundness}
If the basis-updating algorithm returns `NO' for a given input $(\gamma,\delta)$, then $\isnotbisim\gamma\delta$.
\end{restatable}

\begin{proof}
The basis-updating algorithm returns `NO' only when it encounters a total failure. 
One way to encounter a total failure is whenever a partial failure (\Cref{subsec:partialfailure}) is called at the root node $(\gamma,\delta)$. 
By \Cref{lem:partial-irreducible} it is immediate that $\isnotbisim\gamma\delta$.

The only other possibility for a total failure occurs whenever the algorithm encounters a pair $(X\alpha',Y\beta')$ where the immediate transitions of $X$ and $Y$ do not match. We shall focus on this case for the remainder of the proof. 
Clearly, node $(X\alpha',Y\beta')$ is incomparable. 
We argue that if a non-root node $(\gamma',\delta')$ is incomparable, then there exists an earlier node in the derivation tree which is also incomparable. 
This immediately implies (by an induction argument) that the root node is incomparable, completing the proof.

With the above idea in mind, suppose $(\gamma',\delta')$ is an incomparable non-root node in the derivation tree.
Let $(X\alpha',Y\beta')$ be its parent, and consider all the possible cases by which $(X\alpha',Y\beta')$ could have produced $(\gamma',\delta')$ as a children.

\begin{itemize}
\item \Cref{subsec:loop} (loop), \Cref{subsec:reflexive} (identical), \Cref{subsec:totalfailure} (total failure) cannot be the case since they produce no children.
\item (\Cref{subsec:bpa1} - BPA1 expansion) 
Here there is a pair $(X,Y\beta)$ in $\B$ and $\gamma'=\beta\alpha'$, $\delta'=\beta'$. 
If $X,Y$ are both unnormed then, by the pruning convention we get $\alpha'=\beta=\beta'=\gamma'=\delta'=\varepsilon$, contradicting the hypothesis that $(\gamma',\delta')$ is incomparable. 

If $X$ is unnormed and $Y$ is normed, then by the pruning convention $\alpha'=\varepsilon$. 
Since $\B$ is norm-compliant, $\beta$ is unnormed. 
Moreover, there must be an earlier node $(X,Y\beta)$ in the derivation tree corresponding to the step at which this pair was added to $\B$. 
Now suppose that both $(X,Y\beta)$ and $(X,Y\beta')$ were comparable. 
Since $X$ and $\beta$ are both unnormed, it follows that $\isbisim{X}{Y\beta}$ and $\isbisim{X}{Y\beta'\beta''}$ for some $\beta''$. 
By \Cref{lem:charnormedunnormed} we get $\isbisim{\beta}{\beta'\beta''}$, contradicting the hypothesis that $(\beta,\beta')$ is incomparable. 
Thus we deduce that one of $(X,Y\beta)$, $(X,Y\beta')$ is incomparable. 

Finally, suppose $X$ and $Y$ are both normed. 
Since $\B$ is functional, $\ltsred Y u \varepsilon$ and $\ltsred X u {\beta}$. 
Then also $\ltsred {X\alpha'}u{\beta\alpha'=\gamma'}$ and $\ltsred {Y\beta'} u {\beta'=\delta'}$, so that $(X\alpha',Y\beta')$ is incomparable. 
\item (\Cref{subsec:bpa2} - BPA2 expansion) 
Here there is a pair $(X\alpha,Y\beta)$ in $\B$, $\alpha\neq\Empty$, and $(\gamma',\delta')$ is either $(\alpha,\alpha')$ or $(\beta,\beta')$. Without loss of generality suppose it is the former. Since $\B$ is norm-compliant, $\alpha$ and $\beta$ are unnormed. 
There must be an earlier node $(X\alpha,Y\beta)$ in the derivation tree corresponding to the step at which this pair was added to $\B$. 
When doing so, we must have also added $(X,Y)$ to $\scal$. 
Thus, by \Cref{lem:partial-irreducible}, $(X,Y)$ is irreducible. 
Suppose that both $(X\alpha,Y\beta)$ and $(X\alpha',Y\beta')$ are comparable. 
This means that $\isbisim{X\alpha}{Y\beta}$ (since $\alpha$, $\beta$ are unnormed) and that $\isbisim{X\alpha'\alpha''}{Y\beta'\beta''}$ for some $\alpha'',\beta''$. 
By \Cref{lem:charnormnorm} (\Cref{lem:charnormnorm-2}), we get that $\isbisim{\alpha}{\alpha'\alpha''}$, contradicting the hypothesis that $(\alpha,\alpha')$ is incomparable. 
Thus, at least one of $(X\alpha,Y\beta)$, $(X\alpha',Y\beta')$ is incomparable.
\item (\Cref{subsec:unnormedunnormed} - both unnormed) 
Here $\alpha'=\beta'=\varepsilon$ and there is some matching transition $\ltsred X a {\gamma'}$, $\ltsred Y a {\delta'}$. 
It is immediate that $(X,Y)$ is incomparable.
\item (\Cref{subsec:unnormednormed} - unnormed \vs normed) 
Here $\alpha'=\varepsilon$ and there is some matching transition $\ltsred X a {\gamma'}$, $\ltsred {Y\beta'} a {\delta'}$. 
It is immediate that $(X,Y\beta')$ is incomparable. 
\item (\Cref{subsec:normednormed} - both normed) First, consider the subcase in which $(X,Y)$ is not in $\scal$ and $\ltsred Y u \varepsilon$, $\ltsred X u \beta$. If $(\gamma',\delta')$ is obtained by a pair of matching transitions $\ltsred{X}a{\gamma'}$, $\ltsred{Y\beta}{a}{\delta'}$, then $(X,Y\beta)$ is incomparable. 
Suppose that $(X\alpha',Y\beta')$ is comparable, that is, $\isbisim{X\alpha'\alpha''}{Y\beta'\beta''}$ for some $\alpha'',\beta''$. Since $\ltsred{X\alpha'\alpha''}u{\beta\alpha'\alpha''}$ and $\ltsred{Y\beta'\beta''}u{\beta'\beta''}$, we get that $\isbisim{\beta\alpha'\alpha''}{\beta'\beta''}$. By congruence (\Cref{prop:congruence}), we get $\isbisim{Y\beta'\beta''}{Y\beta\alpha'\alpha''}$, implying $\isbisim{X\alpha'\alpha''}{Y\beta\alpha'\alpha''}$, contradicting the hypothesis that $(X,Y\beta)$ is incomparable. 
Otherwise, if $(\gamma',\delta')$ is not obtained by a pair of matching transitions, then $\gamma'=\beta\alpha'$ and $\delta'=\beta'$. Moreover, we have $\ltsred{X\alpha'}u{\gamma'}$ and $\ltsred{Y\beta'}u{\delta'}$, from which it is immediate that $(X\alpha',Y\beta')$ is incomparable. 

Next, consider the subcase in which $(X,Y)$ is in $\scal$ or $\ltsred Y u \varepsilon$ but $\notltsred X u \beta$. In either case, there is a pair of matching transitions $\ltsred{X\alpha'}a{\gamma'}$, $\ltsred{Y\beta'}a{\delta'}$, so that $(X\alpha',Y\beta')$ is incomparable. 
\item (\Cref{subsec:partialfailure} - partial failure) 
Here $(X\alpha',Y\beta')$ was a BPA1 guess turned into a BPA2 guess, and there is a pair of matching transitions $\ltsred {X\alpha'}a{\gamma'}$, $\ltsred{Y\beta'}a{\delta'}$. Since $(\gamma',\delta')$ is incomparable, so is $(X\alpha',Y\beta')$.
\qedhere
\end{itemize}
\end{proof}

\subsection{Termination}

Intuitively, to prove termination we argue that the basis $\B$ changes polynomially many times. Moreover, between any two changes of $\B$, the basis-updating algorithm is essentially verifying coinductive congruence, and we can bound the number of pairs generated in the same way as in \Cref{thm:coinductive-congruence}.

\begin{restatable}{lem}{basischanges}
\label{lem:basis-changes}
During the basis-updating algorithm, the basis $\B$ changes (by adding or removing pairs) at most $\ocal(n^4)$ times, where $n$ is the number of non-terminals in the grammar.
\end{restatable}

\begin{proof}
We begin by noticing that:
\begin{itemize}
\item initially, $\scal=\emptyset$;
\item at every partial failure, there is at least one pair $(X,Y)$ not in $\scal$ which enters $\scal$;
\item there is no way for a pair in $\scal$ to leave $\scal$.
\end{itemize}

\noindent
Let $n$ be the number of non-terminals in the grammar. 
Since there are $\ocal(n^2)$ many pairs of non-terminals, we get that there can only be at most $\ocal(n^2)$ partial failures during the execution of the algorithm. 
Moreover, since the basis is simple (\cref{lem:basis-invariant}), at each stage it must have at most $\ocal(n^2)$ pairs. 
During a partial failure, the basis is changed by removing pairs in $\B$ (corresponding to removed BPA1 or BPA2 guesses) and by adding a single BPA2 pair $(X\alpha,Y\beta)$ (corresponding to one removed BPA1 pair associated with $X,Y$). 
Thus, in total, during a single partial failure there are at most $\ocal(n^2)$ removals to $\B$ and exactly one addition to $\B$.

Next, let us bound the number of times $\B$ is changed between one partial failure and the next. 
In such situations, the only way to change $\B$ is by adding new pairs (\Cref{subsec:unnormedunnormed,subsec:unnormednormed,subsec:normednormed}). 
Since $\B$ is simple, for each pair $(X,Y)$ we can add at most one element to $\B$. 
Therefore, there can only be $\ocal(n^2)$ additions to $\B$ before the algorithm either terminates or enters a partial failure. 

Hence: there are at most $\ocal(n^2)$ partial failures; during each partial failure, there are at most $\ocal(n^2)$ removals and one addition to $\B$; and between any two consecutive partial failures, there are at most $\ocal(n^2)$ additions to $\B$. We conclude that $\B$ changes at most $\ocal(n^4)$ times.
\end{proof}

\begin{restatable}[Termination]{thm}{termination}
\label{thm:termination}
The basis-updating algorithm terminates, for any simple grammar $\G$ without dead symbols 
and any pair $(\gamma,\delta)$ of words of nonterminals. Its time complexity is polynomial on the size of $\G$, the valuation of $\G$, and the seminorms of $\gamma,\delta$.
\end{restatable}

\begin{proof}
Let $n$ be the number of non-terminals in $\G$, $d$ be the maximum number of transitions of a non-terminal in $\G$ (the degree of $\G$), and $v$ be the maximum among the valuation of $\G$ and the seminorms of $\gamma$, $\delta$. 
Whenever the expansion of a node $(X\alpha',Y\beta')$ produces a child node, it fits into one of two cases. 
One case is that the child is obtained by applying a BPA rule (\Cref{subsec:bpa1,subsec:bpa2}). 
By the reasoning in \Cref{thm:coinductive-congruence} (using the fact that $\B$ is functional and simple), the child belongs to the canonical norm-reducing sequence of either $X\alpha',Y\beta'$ in the parent node, or some $X\alpha,Y\beta$ in $\B$. 
The other case is that a pair associated with $X,Y$ is added to $\B$ and the child is obtained by following a pair of matching transitions $\ltsred Xa{\gamma''}$, $\ltsred Ya{\delta''}$. 
Let us call these \emph{unruly} children (since they do not follow from BPA rules).
We can observe that the words in the unruly child node are taken from $\gamma'',\delta'',\gamma''\alpha',\delta''\beta'$.
Comparing with $X\alpha',Y\beta'$, the seminorm of the resulting children increases by at most $\max(\seminormof{\gamma''},\seminormof{\delta''})\leq v$.

At any point during the basis-updating algorithm, there may be at most $\ocal(n^2)$ nodes whose expansion produces unruly children (at most one for each pair in $\B$). 
Hence there are at most $\ocal(n^2d)$ unruly children. 
Each word belonging to an unruly child node can have a seminorm of at most $\ocal(n^2v)$, since it can occur after at most $\ocal(n^2)$ unruly expansions that each increase the seminorm by at most $v$. 

Now let us analyse the behaviour of the algorithm between two consecutive changes in $\B$. During this phase, any node produced by the algorithm is obtained from the application of BPA rules to either the root node or one of the unruly children. 
Therefore, by the reasoning in \Cref{thm:coinductive-congruence}, any word produced in this stage must belong to the canonical norm-reducing sequence of either the root node, some unruly child node, or some pair $X\alpha,Y\beta$ in $\B$. There are $\ocal(n^2d)$ such words, each having seminorm of at most $\ocal(n^2v)$. Hence, there are at most $\ocal((n^2d\times n^2 v)^2) = \ocal(n^8d^2v^2)$ different pairs that can be produced by the algorithm. 
In summary, between any two consecutive changes in $\B$ there can be at most $\ocal(n^8d^2v^2)$ expansion steps. 
Since, by \Cref{lem:basis-changes}, the basis changes at most $\ocal(n^4)$ times, the algorithm must terminate after at most $\ocal(n^{12}d^2v^2)$ iterations, which is polynomial in the size of $\G$, the valuation of $\G$ and the seminorms of $\gamma,\delta$. 

Finally, we need to bound the complexity of each iteration. 
It is obvious that most iterations can be done efficiently, as long as we use appropriate data structures (such as hash tables) to store the nodes in the tree (for the effect of loop detection, \Cref{subsec:loop}) and the pairs in $\B$.
There are two exceptions requiring further justification, one of which being partial failures (which may occur in \Cref{subsec:emptyvsnonempty,subsec:partialfailure,subsec:unnormednormed,subsec:normednormed}). 
Notice that if a node is visited in a partial failure, causing the partial failure to be propagated up the tree, then that node will be removed at the end of that chain of partial failures. 
Therefore, each node is visited at most once during all partial failures occurring during the algorithm. 
The total complexity of partial failures is thus bounded by the number of all nodes considered during the algorithm, which is bounded by the number of iterations $\ocal(n^{12}d^2v^2)$. 
The other exception is checking whether $\ltsred X u \beta$ during \Cref{subsec:normednormed}. 
Here we need to follow a chain of transitions of size $|u|=\normof Y$, which is bounded by the valuation of $\G$ (and thus by $v$). 
We only need to do this each time we are trying to add a BPA1 pair in $\B$, and this happens at most $\ocal(n^4)$ times by \Cref{lem:basis-changes}. 
Thus, the total complexity of these checks is bounded by $\ocal(n^{4}v)$.
\end{proof}

\noindent
In particular, the number of iterations of the algorithm is at most $\ocal(n^{12}d^2v^2)$, where $n$ is the number of nonterminals in $\G$, $d$ is the maximum number of transitions of a nonterminal in $\G$, and $v$ is the maximum among the valuation of $\G$ and the seminorms of $\gamma$, $\delta$.



\section{Context-free session types}
\label{sec:session-types}

Session types are an approach for modeling communication protocols~\cite{DBLP:conf/concur/Honda93,DBLP:conf/esop/HondaVK98,DBLP:conf/parle/TakeuchiHK94}. 
Among the several extensions 
of session types, we consider context-free session types as presented by Almeida~\etal~\cite{DBLP:conf/tacas/AlmeidaMV20}. 
This framework is simple but expressive enough to move beyond tail recursion (regular session
types~\cite{DBLP:conf/esop/HondaVK98}). 
Since its introduction~\cite{DBLP:conf/icfp/ThiemannV16}, context-free session types have
been augmented with constructors such as functions, records and variants~\cite{DBLP:journals/iandc/AlmeidaMTV22}, higher-order messages, functional polymorphism~\cite{DBLP:journals/corr/abs-2203-12877,DBLP:journals/tcs/CostaMPV24}, type-level abstraction, and type-level
application~\cite{DBLP:conf/esop/PocasCMV23}. 
Each of those extensions can still be converted into simple grammars. 
Thus, the results in this section can be adapted to more general settings. 

\subsection{Type syntax and type formation}
\label{sec:type-syntax-and-type-formation}
We assume a finite set of \emph{message types} including $\Int$ (for integers),
$\Bool$ (for booleans) and so on. We let $\MT$ denote a generic message type. We
also assume a countable set of \emph{type references}, denoted by $\aT$, $\bT$,
and so on. The syntax of \emph{session types} is as follows. 
\begin{align*}
  \MT & \grmeq \Int \grmor \Bool \grmor \cdots
  \\
  \TT & \grmeq \IN M \grmor \OUT M 
        \grmor \intchoice\typec{\recordf{\ell}{T_\ell}L}
        \grmor \extchoice\typec{\recordf{\ell}{T_\ell}L}
        \grmor \Skip \grmor \semit TT \grmor \aT \grmor \REC a T
\end{align*}

\noindent
We use $\IN M$ to represent the action of \emph{receiving} a value of type $\MT$, and
$\OUT M$ for \emph{sending} a value of type $\MT$. Type
$\intchoice\typec{\recordf{\ell}{T_\ell}L}$ represents an \emph{internal choice},
corresponding to the action of selecting a choice according to some label
$\typec\ell$ in some set $L$ and continuing as type $\typec{T_\ell}$. Dually, type
$\extchoice\typec{\recordf{\ell}{T_\ell}L}$ represents an \emph{external choice}, where
some choice $\typec\ell$ is received and communication continues with
$\typec{T_\ell}$. Context-free session types introduce \emph{sequential composition}
$\semit TU$ for the action of executing type $\TT$ followed by type $\UT$. Type
$\Skip$ is the neutral element of sequential composition; a lone $\Skip$
represents \emph{absence of communication}. Finally, we introduce \emph{recursion} $\REC a T$
where $\aT$ is a type reference. Recursion is used for representing potentially
infinite behaviour, as it is often the case with communication protocols.

Let us look at some examples.
\begin{itemize}
\item Type $\typename{query} = \semit{\IN\Int}{\OUT\Bool}$ specifies a channel that receives an integer and replies with a boolean.
\item Type $\typename{pingpong} = \REC a{\semit{\IN\Int}{\semit{\OUT\Bool}{a}}}$ specifies a channel that continuously receives an integer and replies with a boolean.
\item Type $\typename{mathserver} = \REC a {\extchoices{
\Label{add} \colon \semit{\IN\Int}{\semit{\IN\Int}{\semit{\OUT\Int}{a}}}, 
\Label{isprime} \colon \semit{\IN\Int}{\semit{\OUT\Bool}{a}},
\Label{quit} \colon \Skip}}$ specifies a mathematical server offering three choices: $\typec{\Label{add}}$ (receives two integers, sends an integer, and goes back to the start), $\typec{\Label{isprime}}$ (receives an integer, sends a boolean, and goes back to the start), or $\typec{\Label{quit}}$ (ends communication).
\item
Type $\typename{tree} = \REC a {\extchoices{
\Label{leaf} \colon \Skip, 
\Label{node} \colon \semit{a}{\semit{\IN\Int}{a}}}}$ describes a channel for
receiving a tree of integer values. 
When choice $\typec{\Label{node}}$ is selected, the channel receives the left subtree, then the (integer) value at that node, and then the right subtree.
\end{itemize}

\noindent
An object produced by grammar for $\TT$ is called a \emph{pretype}. Not every
pretype is a type. Intuitively, we need to exclude syntax that does not
correspond to actual communication. The standard approach is to define
predicates for termination and contractivity.

\begin{figure}[t]
  \begin{mathpar}
    \declrel{Terminated types}{$\isdone T$}
    \\
    \inferrule*[right=$\checkmark$-Skip]
    { }
    {\isdone{\Skip}}

    \inferrule*[right=$\checkmark$-Seq]
    { \isdone T \\ \isdone U }
    { \isdone{\semit TU} }

    \inferrule*[right=$\checkmark$-Rec]
    { \isdone T }
    { \isdone{\REC a T} }
    \\
    \declrel{Contractive types}{$\iscontrt T$}
    \\
    \inferrule*[right=C-In]
    { }
    { \iscontrt{\IN M} }

    \inferrule*[right=C-Out]
    { }
    { \iscontrt{\OUT M} }

    \inferrule*[right=C-IntC]
    { }
    { \iscontrt{\intchoice\typec{\recordf{\ell}{T_\ell}L}} }

    \inferrule*[right=C-ExtC]
    { }
    { \iscontrt{\extchoice\typec{\recordf{\ell}{T_\ell}L}} }

    \inferrule*[right=C-Skip]
    { }
    { \iscontrt{\Skip} }

    \inferrule*[right=C-Seq1]
    { \isdone T \\ \iscontrt U }
    { \iscontrt{\semit TU} }

    \inferrule*[right=C-Seq2]
    { \isnotdone T \\ \iscontrt T }
    { \iscontrt{\semit TU} }

    \inferrule*[right=C-Rec]
    { \iscontrt T }
    { \iscontrt{\REC A T} }
    \\
    \declrel{Well-formed types}{$\istypecta \Gamma T$}
    \\
    \inferrule*[right=T-In]
    { }
    { \istypecta \Delta{\IN M} }

    \inferrule*[right=T-Out]
    { }
    { \istypecta \Delta{\OUT M} }

    \inferrule*[right=T-IntC]
    { \istypecta \Delta {T_\ell} \\ \typec\ell\in L }
    { \istypecta \Delta{\intchoice\typec{\recordf{\ell}{T_\ell}L}} }

    \inferrule*[right=T-ExtC]
    { \istypecta \Delta {T_\ell} \\ \typec\ell\in L }
    {\istypecta \Delta{\extchoice\typec{\recordf{\ell}{T_\ell}L}}}

    \inferrule*[right=T-Skip]
    { }
    {\istypecta \Delta\Skip}

    \inferrule*[right=T-Seq]
    { \istypecta \Delta T \\ \istypecta \Delta U }
    { \istypecta \Delta{\semit TU} }

    \inferrule*[right=T-Var]
    { \aT \in \Delta }
    { \istypecta \Delta a }

    \inferrule*[right=T-Rec]
    { \iscontrt{\REC a T} \\ \istypecta {\Delta,\aT} T }
    { \istypecta \Delta{\REC a T} }
  \end{mathpar}
  \caption{Session types: rules for termination ($\isdone T$), contractivity ($\iscontrt T$), and type formation ($\istypecta \Delta T$).}
  \label{fig:session-types}
\end{figure}


The is-terminated predicate ($\isdone T$, \cref{fig:session-types}) materializes the intuition of a pretype that provides no operation; 
it comprises only sequential composition of terminated pretypes and recursions of terminated pretypes. 
For example, $\Skip$, $\semit \Skip \Skip$, and $\REC a {(\semit \Skip {\REC b \Skip})}$ are terminated, but $\semit \Skip {\IN \Int}$ and $\REC a a$ are not.

Contractivity ($\iscontrt T$, \cref{fig:session-types}), ensures that a pretype eventually rewrites to a (productive) type constructor. 
The is-terminated predicate is decidable~\cite{DBLP:conf/tacas/AlmeidaMV20,DBLP:journals/tcs/CostaMPV24}, which justifies the inclusion of its negation $\isnotdone T$ in rule {\sc C-Seq2}. 
For example, $\REC a {\semit{\IN \Int} a}$ is contractive, but $\REC a {\semit \Skip a}$ is not.

Type formation ($\istypecta \Delta T$, \cref{fig:session-types}) states that $\TT$ is a type under context $\Delta$. 
A \emph{context} is an (unordered) set of type variables, and the comma operator $\Delta, \aT$ extends a context by adding a new variable. 
We say that $\TT$ is a \emph{type} if $\istypecta {} T$, \ie if $\TT$ is a type under the empty context (this is only true if $\TT$ has no free type references). 
Rules for type formation are mostly straightforward. 
We remark only that in rule {\sc T-Rec}, we check the formation of $\REC a T$ by adding $\aT$ to the context and verifying that $\REC a T$ is contractive. 

\subsection{Type equivalence and conversion to simple grammar}
\label{subsec:type-equiv}
\emph{Type equivalence} is denoted by $\isequiv TU$ and can be defined in several
equivalent ways~\cite{DBLP:journals/tcs/CostaMPV24}. For example, it can be
defined by means of coinductive derivation rules, or by means of bisimulation on
a labelled transition system. Briefly, we desire type equivalence to extend
syntactic equality (if $\TT=\UT$ then $\isequiv TU$) with the monoidal axioms of
sequential composition and the equirecursive treatment of recursive types: 
\begin{gather*}
  \isequiv{\semit \Skip T}T
  \text{ and }
  \isequiv {\semit T \Skip}T
  \tag{neutral element}
  \\
  \isequiv{\semit {(\semit TU)}V}{\semit T{(\semit UV)}} 
  \tag{associativity}
  \\
  \isequiv{\semit{\intchoice\typec{\recordf{\ell}{T_\ell}L}}U}{\intchoice\typec{\recordf{\ell}{\semit{T_\ell}U}L}}
  \text{ and }
  \isequiv{\semit{\extchoice\typec{\recordf{\ell}{T_\ell}L}}U}{\extchoice\typec{\recordf{\ell}{\semit{T_\ell}U}L}}
  \tag{distributivity}
  \\
  \isequiv{\REC a T}{\subs T a {\REC a T}}
  \tag{unfolding}
\end{gather*}

\noindent
There are several procedures to convert session types into simple grammars~\cite{DBLP:journals/iandc/AlmeidaMTV22,DBLP:conf/tacas/AlmeidaMV20,DBLP:journals/tcs/CostaMPV24,DBLP:conf/esop/PocasCMV23,DBLP:conf/concur/SilvaMV23,DBLP:conf/icfp/ThiemannV16}.
We adapt the translation of Poças \etal~\cite{DBLP:journals/tcs/CostaMPV24,DBLP:conf/esop/PocasCMV23} which preserves and reflects type equivalence into simple grammar bisimilarity.
\begin{defi}[Conversion from types to words on a simple grammar]
Procedure $\word(\cdot)$ receives a type $\TT$ and returns a word of nonterminal symbols $\word(\TT)$, while constructing a simple grammar $\G$ as a side effect. The procedure creates fresh nonterminal symbols as needed; $\varepsilon$ denotes the empty word. It is coinductively defined on $\TT$ as follows:
\begin{itemize}
\item $\word(\IN M) = X$ for a fresh nonterminal symbol $X$ with production $\ltsred X{\IN M}\varepsilon$.
\item $\word(\OUT M) = X$ for a fresh nonterminal symbol $X$ with production $\ltsred X{\OUT M}\varepsilon$.
\item $\word(\intchoice\typec{\recordf{\ell}{T_\ell}L}) = X$ for a fresh nonterminal $X$ with productions $\ltsred X{\intchoice\typec{\ell}}{\word(\typec{T_\ell})}$.
\item $\word(\extchoice\typec{\recordf{\ell}{T_\ell}L}) = X$ for a fresh nonterminal $X$ with productions $\ltsred X{\extchoice\typec{\ell}}{\word(\typec{T_\ell})}$.
\item $\word(\Skip)=\varepsilon$.
\item $\word(\semit TU) = \word(\TT)\word(\UT)$.
\item If $\isdone {\REC aT}$, then $\word(\REC aT)=\varepsilon$.
\item If $\isnotdone {\REC aT}$, then $\word(\REC aT)=X$ for a fresh nonterminal symbol $X$. Let $Y\delta=\word(\TT)$. Then $X$ has a production $\ltsred X a \gamma\delta$ for each production $\ltsred Y a \gamma$. Moreover, $\word(\aT)=X$.
\end{itemize}
In order to make sure that the construction of $\word(\TT)$ eventually terminates, we keep track of all types visited during the construction, and we only add a fresh nonterminal $Y$ to our grammar if the type visited is syntactically different from all types visited so far. 
Therefore, we reuse the same symbol $Y$ with the same productions each time we revisit a type.
\end{defi}

As an example, consider type $\typename{tree} = \REC a {\extchoices{\Label{leaf} \colon \Skip, \Label{node} \colon \semit{a}{\semit{\IN\Int}{a}}}}$ from \cref{sec:type-syntax-and-type-formation}. 
The computation of $\word(\typename{tree})$ returns word $X$ and constructs a simple grammar with nonterminals $\{X,Y\}$, terminals $\{\extchoice\typec{\Label{leaf}}, \extchoice\typec{\Label{node}}, \IN \Int\}$ and the production rules below.
\begin{align*}
	\ltsred{X}{\extchoice\typec{\Label{leaf}}}{\varepsilon} && 
	\ltsred{X}{\extchoice\typec{\Label{node}}}{XYX} &&
	\ltsred{Y}{\IN \Int}{\varepsilon}
\end{align*}
For the purposes of this paper, we may assume by definition that types $\TT,\UT$ are equivalent ($\isequiv TU$) if their corresponding conversions to simple grammars yield bisimilar words, that is, $\isbisim{\word(\TT)}{\word(\UT)}$. 
For simplicity, we assume that the sets of nonterminals built during the construction of $\word(\TT)$ and $\word(\UT)$ are disjoint, so we can trivially fuse the two simple grammars into one.
We define the following measure to estimate the encoding size of a type.

\begin{defi}[Size]
We define the size of a type $\TT$, denoted $|\TT|$, inductively on $\TT$ as follows: 
$$|\IN M|=|\OUT M|=1,\qquad
|\intchoice\typec{\recordf{\ell}{T_\ell}L}|=|\extchoice\typec{\recordf{\ell}{T_\ell}L}|=1+\sum_{\ell\in
  L}|\typec{T_\ell}|,$$ 
$$|\Skip|=1, \qquad |\semit TU| = 1+|\TT|+|\UT|,\qquad |\aT|=1,\qquad
|\REC a T| = 1+|\TT|.$$
\end{defi}

\begin{restatable}[Linear conversion of types into grammars]{lem}{wordlinear}
  \label{lem:wordlinear}
  Let $\TT$ be a type and $\word(\TT)$ its conversion to a simple grammar $\G$.
  Then the number of nonterminals of $\G$, the degree of $\G$, the seminorm of
  $\word(\TT)$ and the valuation of $\G$ are all less than or equal to $|\TT|$.
\end{restatable}

\begin{proof}
We begin by noticing that $|\TT|$ corresponds to the number of distinct subexpressions of $\TT$ (or alternatively, to the number of nodes in an abstract syntax tree representation of $\TT$). 
By inspection on the rules defining the procedure $\word(\TT)$, we notice that at most one nonterminal is created for each subexpression of $\TT$. 
Hence at most $|\TT|$ nonterminals are generated. 
A similar argument can be made for the number of terminals. 
Regarding the number of productions, when computing $\word(\REC a T)$ we create one new production for each production for $\word(\TT)$. 
In all other cases we create at most one production for each subexpression of $\TT$. 
Therefore, the maximum number of transitions of a non-terminal in $\G$ (\ie the degree of $\G$) is again bounded by $|\TT|$. 
The next step is to prove that $\seminormof{\word(\TT)}\leq|\TT|$. This is done by induction on $\TT$. 

\begin{itemize}
\item Suppose $\TT$ is $\IN{\MT}$. Then $\seminormof{\word(\TT)}=1=|\TT|$. 
The case where $\TT$ is $\OUT{\MT}$ is similar.
\item Suppose $\TT$ is $\intchoice\typec{\recordf{\ell}{T_\ell}L}$. 
Then $\word(\TT) = X$ for a fresh nonterminal symbol $X$ with productions $\ltsred X{\intchoice\typec{\ell}}{\word(\typec{T_\ell})}$. 
By induction, $\seminormof{\word(\typec{T_\ell})}\leq |\typec{T_\ell}|$ for each $\typec{\ell}$. 
If $X$ is unnormed, then $\seminormof{\word(\TT)}=\seminormof{X}=0$, which is trivially less than $|\TT|$. 
Otherwise, let $\ltsred X{\intchoice\typec{\ell}}{\word(\typec{T_\ell})}$ be a norm-reducing transition. 
We have that $\seminormof{\word(\TT)}=\seminormof{X}=1+\seminormof{\word(\typec{T_\ell})}\leq 1 + |\typec{T_\ell}| \leq 1+\sum_{\ell\in L}|\typec{T_\ell}|=|\TT|$, as desired. 
The case where $\TT$ is $\extchoice\typec{\recordf{\ell}{T_\ell}L}$ is similar.
\item Suppose $\TT$ is $\Skip$. Then $\seminormof{\word(\TT)}=0 < 1=|\TT|$.
\item Suppose $\TT$ is $\semit{U}{V}$. 
Then $\word(\TT)=\word(\UT)\word(\VT)$ and, by induction, $\seminormof{\word(\UT)}\leq|\UT|$, $\seminormof{\word(\VT)}\leq|\VT|$. 
If $\word(\UT)$ is unnormed, then we get $\seminormof{\word(\UT)\word(\VT)} = \seminormof{\word(\UT)} \leq |\UT|$.
Otherwise, $\seminormof{\word(\UT)\word(\VT)}=\seminormof{\word(\UT)}+\seminormof{\word(\VT)}\leq |\UT|+|\VT|$. 
In either case, we have $\seminormof{\word(\TT)} < 1 + |\UT| + |\VT| = |\TT|$.
\item The case that $\TT$ is $\aT$ is not applicable since $\aT$ is not a type (under the empty context). 
\item Suppose $\TT$ is $\REC{a}{U}$, with $\isdone\TT$. Then $\seminormof{\word(\TT)}=\seminormof{\Empty}=0 < |\TT|$.
\item Suppose $\TT$ is $\REC{a}{U}$, with $\isnotdone\TT$. Then $\word(\REC{a}{U})$ is some nonterminal symbol $X$. 
If $X$ is unnormed, then again $\seminormof{\word(\TT)}=0 < |\TT|$. 
Finally, suppose that $X$ is normed. 
This last case is not trivial, since we cannot directly apply the induction hypothesis to $\UT$, as $\aT$ may occur free in $\UT$. 
Let us use $\botT$ to denote the type $\typec{\intchoices{}}$, \ie an empty choice. Notice that $|\botT|=1$ and $\word(\botT)=\bot$ is a nonterminal symbol without productions (we slightly abuse notation here, reusing the symbol $\bot$). 
Consider the type $\subs{U}{a}{\botT}$ obtained by replacing all free occurrences of $\aT$ in $\UT$ by $\botT$. 
Let $\G'$ be the grammar obtained when computing $\word(\subs{U}{a}{\botT})$. The only difference between $\G$ and $\G'$ is that $X$ is replaced by $\bot$ in the right-hand side of all productions. 
Since $X$ is normed in $\G$, any of its norm-reducing sequences in $\G$ never visits a word containing $X$ itself. Thus, any such sequence is also norm-reducing in $\G'$. We conclude that $X$ is normed in $G'$, and moreover $\seminormof{\word(\TT)}^{\G}$ = $\seminormof{X}^{\G} = \seminormof{X}^{\G'} = \seminormof{\word(\subs{U}{a}{\botT})}^{\G'}$. 
Notice that $\subs{U}{a}{\botT}$ is a type since $\aT$ no longer appears free. 
By induction hypothesis, $\seminormof{\word(\subs{U}{a}{\botT})}^{\G'} \leq |\subs{U}{a}{\botT}| = |\UT| < |\TT|$.
\end{itemize}

\noindent
We just need to conclude that the valuation of $\G$ is at most $|\TT|$. 
Each of the words appearing on the right-hand side of productions of $\G$ is of the form $\word(\typec{T_1})\ldots\word(\typec{T_n})$, where $\typec{T_1}$,\ldots,$\typec{T_n}$ are non-overlapping subexpressions of $\TT$. 
By the previous bound, we have $\seminormof{\word(\typec{T_1})\ldots\word(\typec{T_n})}\leq\seminormof{\word(\typec{T_1})}+\ldots+\seminormof{\word(\typec{T_n})}\leq |\typec{T_1}|+\ldots+|\typec{T_n}|\leq |\TT|$. 
Thus, the valuation of $\G$ is at most $|\TT|$.
\end{proof}

\begin{restatable}{thm}{typeequivalence}
\label{thm:polytime-type-equivalence}
There is a polynomial-time algorithm for determining whether two context-free
  session types, $\TT$ and $\UT$, are equivalent.
\end{restatable}

\begin{proof}
The algorithm for type equivalence converts both types into words $\word(\TT)$, $\word(\UT)$
over a common simple grammar $\G$, and then applies the basis-updating algorithm from \cref{sec:algorithm}. 
Clearly, the conversion to a simple grammar runs in polynomial time. 
By \cref{lem:wordlinear}, the size and valuation of $\G$, as well as the seminorms of $\word(\TT)$, $\word(\UT)$, are all bounded by $\max\{|\TT|,|\UT|\}$. 
By \cref{thm:termination}, the time complexity of the basis-updating algorithm is polynomial on the size of $\G$, the valuation of $\G$, and the seminorms of $\word(\TT)$, $\word(\UT)$. 
In particular, the number of iterations of the basis-updating algorithm is at most $\ocal(n^{16})$, where $n=\max\{|\TT|,|\UT|\}$. 
Therefore, the algorithm for type equivalence has running time polynomial in $|\TT|$ and $|\UT|$. 
\end{proof}

\subsection{Numerical experiments}
The best known alternative to the context-free session type equivalence algorithm we have described in \cref{thm:polytime-type-equivalence} is that of Almeida \etal~\cite{DBLP:conf/tacas/AlmeidaMV20}, which is based on the 2-EXPTIME algorithm for context-free grammars. 
Its reference implementation is written in Haskell and freely available as part of the \freest programming language~\cite{freest}. 
Having in mind its replacement, we also implemented our algorithm in Haskell. We conclude this section with an empirical comparison of the time performance of both implementations.

To obtain a diverse and robust sample for our experiments, we used the random data generation facilities provided by the QuickCheck library~\cite{DBLP:conf/icfp/ClaessenH00}.
The recursive algorithm used to randomly generate equivalent types is induced from the following properties of the type equivalence relation:

\begin{restatable}[Properties of type equivalence~\cite{DBLP:conf/tacas/AlmeidaMV20}]{prop}{propertiesoftypeequiv}
\label{prop:properties-of-type-equiv}
\hfill
\begin{enumerate}[align=left]
\item \label{item:equiv-skip} $\isequiv \Skip \Skip$;
\item \label{item:equiv-in} $\isequiv {\IN M} {\IN M}$;
\item \label{item:equiv-out} $\isequiv {\OUT M} {\OUT M}$;
\item \label{item:equiv-intchoice} $\isequiv {\intchoice\typec{\recordf{\ell}{T_\ell}L}} {\intchoice\typec{\recordf{\ell}{U_\ell}L}}$ if $\isequiv {\typec{T_\ell}} {\typec{U_\ell}}$ for each $\typec{\ell}$;
\item \label{item:equiv-extchoice} $\isequiv {\extchoice\typec{\recordf{\ell}{T_\ell}L}} {\extchoice\typec{\recordf{\ell}{U_\ell}L}}$ if $\isequiv {\typec{T_\ell}} {\typec{U_\ell}}$ for each $\typec{\ell}$;
\item \label{item:equiv-semi} $\isequiv {\semit {T_1} {T_2}} {\semit {U_1} {U_2}}$ if $\isequiv {T_1} {U_1}$ and $\isequiv {T_2} {U_2}$;
\item \label{item:equiv-rec} $\isequiv {\REC a T} {\REC a U}$ if $\isequiv T U$;
\item \label{item:equiv-id} $\isequiv T {\semit U \Skip}$ and $\isequiv T {\semit \Skip U}$ if $\isequiv T U$;
\item \label{item:equiv-distrib-int} $\isequiv {\semit {\intchoice\typec{\recordf{\ell}{T_\ell}L}} T} {\intchoice\typec{\recordf{\ell}{\semit {U_\ell} U}L}}$ if $\isequiv {T_\ell} {U_\ell}$ for each $\typec{\ell}$ and $\isequiv T U$;
\item \label{item:equiv-distrib-ext} $\isequiv {\semit {\extchoice\typec{\recordf{\ell}{T_\ell}L}} T} {\extchoice\typec{\recordf{\ell}{\semit {U_\ell} U}L}}$ if $\isequiv {T_\ell} {U_\ell}$ for each $\typec{\ell}$ and $\isequiv T U$;
\item \label{item:equiv-sym} $\isequiv T U$ if $\isequiv U T$;
\item \label{item:equiv-assoc} $\isequiv {\semit {T_1} {(\semit {T_2} {T_3})}} {\semit {(\semit {U_1} {U_2})} {U_3}}$ if $\isequiv {T_1} {U_1}$, $\isequiv {T_2} {U_2}$ and $\isequiv {T_3} {U_3}$;
\item \label{item:equiv-recrecl} $\isequiv {\REC a {\REC b T}} {\REC a {(\subs U b a)}}$ if $\isequiv T U$;
\item \label{item:equiv-recrecr} $\isequiv {\REC a {\REC b T}} {\REC b {(\subs U a b)}}$ if $\isequiv T U$;
\item \label{item:equiv-recfree} $\isequiv {\REC a T} U$ if $\isequiv T U$ and $\aT \notin \free(\TT)$;
\item \label{item:equiv-recsubs} $\isequiv {\subs T a {T'}} {\subs U a {U'}}$ if $\isequiv T U$ and $\isequiv {T'} {U'}$;
\item \label{item:equiv-unfold} $\isequiv {\REC a T} {\subs U a {\REC a U}}$ if $\isequiv T U$.
\end{enumerate}
\end{restatable}

\noindent
To randomly generate a pair of equivalent types, we arbitrarily select one of the pairs in \Crefrange{item:equiv-skip}{item:equiv-out} for the base case or one of the pairs in \Crefrange{item:equiv-intchoice}{item:equiv-unfold} for the recursive case. 
To include the possibility of generating pairs of non-equivalent types, we proceed similarly but include also the following properties:

\begin{restatable}[Properties of type non-equivalence]{prop}{propertiesoftypenonequiv}
\label{prop:properties-of-type-non-equiv}
\hfill
\begin{enumerate}
	\item \label{item:nonequiv-skip-in} $\isnotequiv {\Skip} {\IN M}$;
	\item \label{item:nonequiv-skip-out} $\isnotequiv {\Skip} {\OUT M}$;
	\item \label{item:nonequiv-in-out} $\isnotequiv {\IN M} {\OUT M}$
	\item \label{item:nonequiv-skip-intchoice} $\isnotequiv {\Skip} {\intchoice\typec{\recordf{\ell}{T_\ell}L}}$;
	\item \label{item:nonequiv-skip-extchoice} $\isnotequiv {\Skip} {\extchoice\typec{\recordf{\ell}{T_\ell}L}}$;
	\item \label{item:nonequiv-intchoice-extchoice} $\isnotequiv {\intchoice\typec{\recordf{\ell}{T_\ell}L}} {\extchoice\typec{\recordf{\ell}{T_\ell}L}}$;
	\item \label{item:nonequiv-intchoice-subset} $\isnotequiv {\intchoice\typec{\recordf{\ell}{T_\ell}L}} {\intchoice\typec{\recordf{\ell}{T_k}K}}$ with $L \subset K$;
	\item \label{item:nonequiv-extchoice-subset} $\isnotequiv {\extchoice\typec{\recordf{\ell}{T_\ell}L}} {\extchoice\typec{\recordf{\ell}{T_k}K}}$ with $L \subset K$.
\end{enumerate}
\end{restatable}

\begin{proof}
By definition.
\end{proof}

\noindent
Using these generators, we obtained a sample comprised of $\num{1000}$ pairs of types, of which $\num{500}$ are `YES'-pairs and $\num{500}$ are `NO'-pairs. 
The sum of the sizes of the types in each pair ($|\TT|+|\UT|$) ranges from $\num{2}$ to $\num{500}$, with a roughly even distribution.

We measured the time that the implementations of our algorithm and that of Almeida \etal~took for each pair in the sample, with a timeout of $\num{30}$ seconds. 
Since the first step of both algorithms is to convert the input pair to a grammar, we start counting only after this step. 
To obtain robust measurements we used the Criterion microbenchmarking library~\cite{criterion}, which runs each experiment several times and performs statistical analysis to detect outliers, reduce measurement overhead and minimize interference from external processes. 
The microbenchmarks were compiled using GHC version 9.6.7 with the \texttt{-O2} flag enabled and were executed on an AMD Ryzen 7 PRO 6850U processor at $\num{2.7}$ \unit{\GHz} with $\num{32}$ \unit{\giga\byte} of RAM running Ubuntu. 
\begin{figure*}
	\centering
	\begin{subfigure}[b]{0.48\textwidth}
		\centering
		\includegraphics[width=\textwidth]{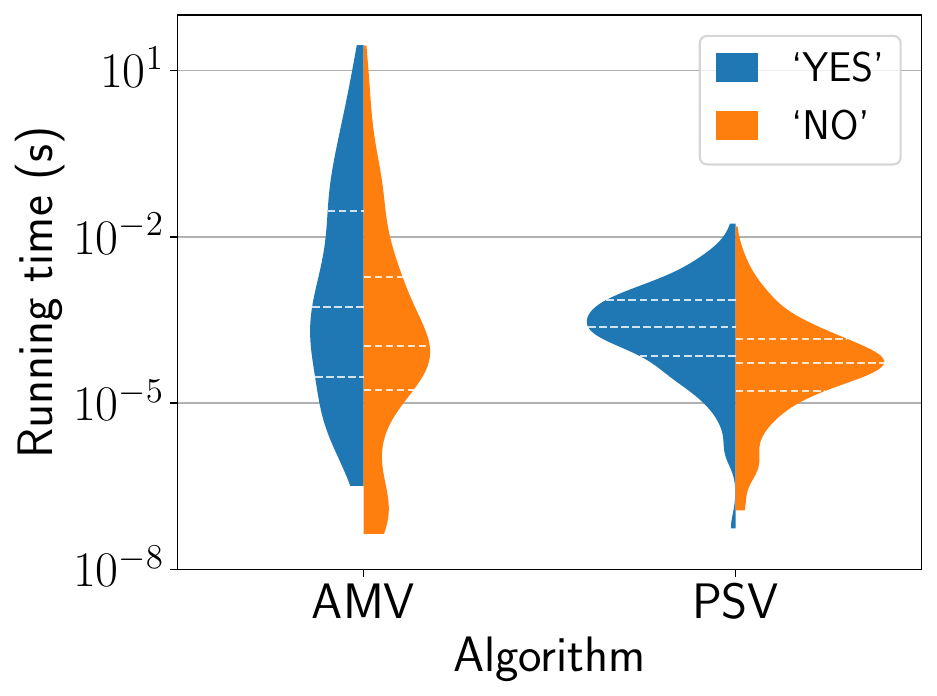}
		\caption[Network2]%
		{{\small Running time distribution per algorithm, excluding timeouts.}}    
		\label{fig:empirical-distribution}
	\end{subfigure}
	\hfill
	\begin{subfigure}[b]{0.48\textwidth}  
		\centering 
		\includegraphics[width=\textwidth]{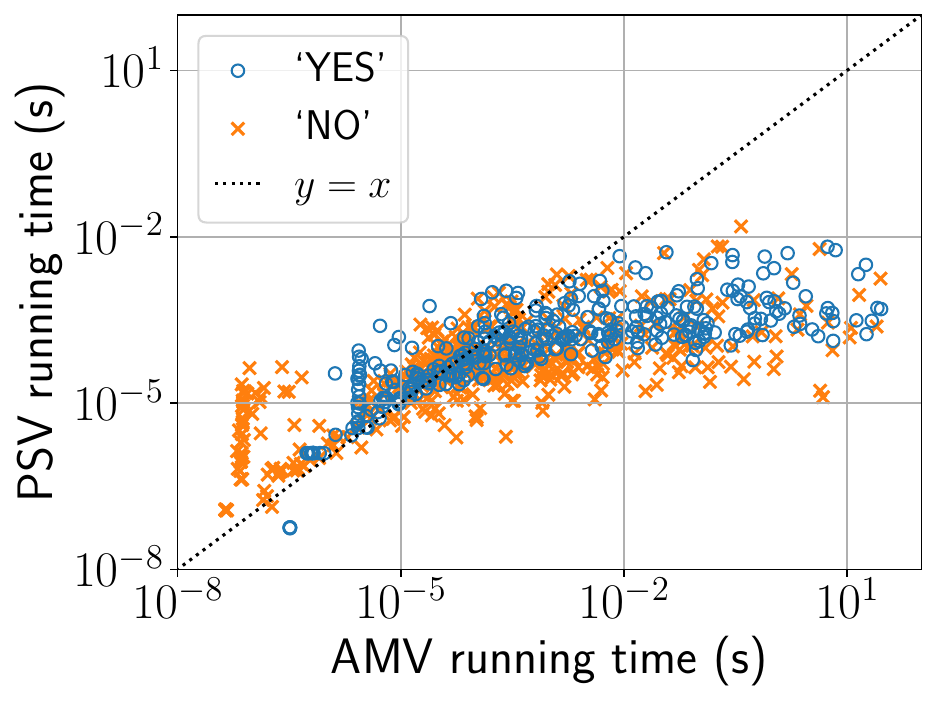}
		\caption[]%
		{{\small PSV running time vs.~AMV running time, excluding timeouts.}}    
		\label{fig:empirical-comparison}
	\end{subfigure}
	\vskip\baselineskip
	\begin{subfigure}[b]{0.48\textwidth}   
		\centering 
		\includegraphics[width=\textwidth]{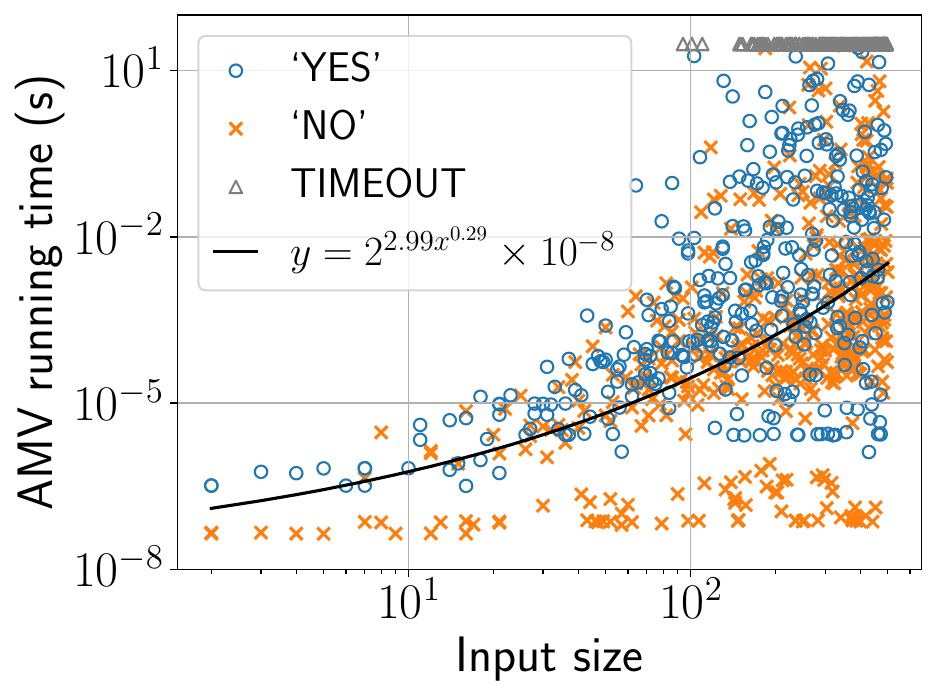}
		\caption[]%
		{{\small AMV running time per input size.}}    
		\label{fig:empirical-amv}
	\end{subfigure}
	\hfill
	\begin{subfigure}[b]{0.48\textwidth}   
		\centering 
		\includegraphics[width=\textwidth]{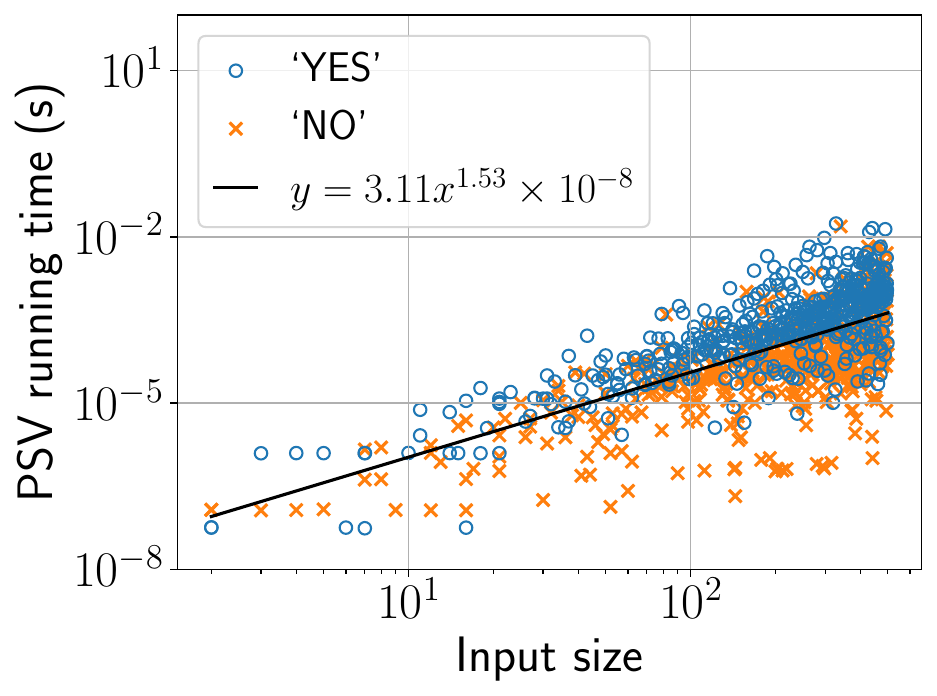}
		\caption[]%
		{{\small PSV running time per input size.}}    
		\label{fig:empirical-psv}
	\end{subfigure}
	\caption{Results of the empirical comparison. PSV denotes our algorithm, AMV denotes that of Almeida \etal~\cite{DBLP:conf/tacas/AlmeidaMV20}. All times are in seconds and all scales are logarithmic.}
	\label{fig:empirical}
\end{figure*}

We begin our analysis by comparing the number of timeouts: whereas none were found with our algorithm, the algorithm of Almeida \etal~exhibited $\num{181}$ timeouts. 
This comprises $\num{18} \unit{\percent}$ of the entire sample, and $\num{36} \unit{\percent}$ of the `YES'-pairs (for `NO'-pairs, neither algorithm reached a timeout).

The distributions of running times in seconds per algorithm and output can be compared on the violin plots in \Cref{fig:empirical-distribution}. 
`NO'-pairs generally exhibit lower running times than `YES'-pairs of the same size in both algorithms, since we can determine non-equivalence earlier by finding unmatched transitions, whereas to determine equivalence we must exhaust all branches of the derivation tree. 
Our algorithm exhibits a more concentrated distribution, with slightly lower medians and significantly lower maxima for both `YES' and `NO'-pairs, and a lower minimum for `YES'-pairs. 
The algorithm of Almeida et al., however, achieves a lower minimum for `NO'-pairs. 

We compared the results for each pair directly, plotting the running times obtained from the algorithm of Almeida \etal~against those obtained from our algorithm, as shown in \Cref{fig:empirical-comparison}. 
The dotted line in the figure separates the inputs where our algorithm performed better (below) from those where it performed worse (above). 
We found that our algorithm performs better in $\num{64} \unit{\percent}$ of the cases, with significant speedups ranging from $\num{5.17e-08}$ \unit{\second} to $\num{28.64}$ \unit{\second}, the average being $\num{0.68}$ \unit{\second}. 
In the remaining $\num{36} \unit{\percent}$ of the cases where our algorithm performed worse, we found less significant slowdowns ranging from $\num{2.53e-08}$ \unit{\second} to $\num{8.28e-04}$ \unit{\second}, the average being $\num{5.98e-05}$ \unit{\second}. As shown on the lower left corner of the plot, the algorithm of Almeida \etal~achieves a better, almost constant time for a small group of `NO'-pairs (roughly $\num{e-7}$ \unit{\second} \vs~$\num{e-6}$ to $\num{e-5}$ \unit{\second}). 
Upon inspection, we found that these pairs translate to words in which the transitions do not match early on in the derivation tree. 
We suspect that, in such cases, our algorithm is slower due to the initial overhead of pruning the root node to maintain the pruning convention, which takes longer as the size of the grammar grows. 
It might be possible to optimize our algorithm to better handle these cases.

To better understand the relationship between the performance of the algorithms and the size of the input, we plotted the sum of the sizes of the types in each pair against the measured running time for each implementation. The results are depicted in \Cref{fig:empirical-amv,fig:empirical-psv}. Predictably, we found a positive correlation between the two variables in both algorithms, with variance also increasing along with the size. This is especially noticeable in the case of the algorithm of Almeida \etal~(\Cref{fig:empirical-amv}). 
Comparing the two figures, we see that for the group of `NO'-pairs mentioned above, the algorithm of Almeida \etal~takes roughly constant time, whereas for the same inputs the running time of our algorithm slowly grows in proportion to the size. 

The plots in \Cref{fig:empirical-amv,fig:empirical-psv} also suggest a super-polynomial running time for the algorithm of Almeida \etal and a polynomial running time for our algorithm, as was expected in theory. 
To approximate the empirical time complexity of each algorithm, we performed regression analysis on the data, using the least-squares method to obtain models that predict the running time in seconds $y$ as a function of the sum of the sizes of the input types $x$. 
For the algorithm of Almeida et al., we fitted an exponential model 
$y = 2^{bx^a}\times10^{-8}$ \unit{\second}, 
obtaining $a = 0.29$ and 
$b=2.99$, which suggests an empirical average-case time complexity of $2^{\ocal(x^{0.29})}$. 
For our algorithm, we fitted a polynomial model $y = bx^a\times 10^{-8}$ \unit{\second}, obtaining $a = 1.53$ and $b = 3.11$, which suggests an empirical average-case time complexity of $\ocal(x^{1.53})$. 
We notice that the empirical average-case time complexity has a significantly lower degree than the theoretical worst-case complexity of $\ocal(n^{12}d^2v^2)$ iterations yielded by \cref{thm:termination} (where $n$, $d$ and $v$ are all upper bounded by $x$, the size of the types, as in \cref{thm:polytime-type-equivalence}). 
This suggests that our theoretical analysis may in principle be improved, and that worst-case inputs are not that common for randomly generated problem instances.



\section{Conclusion}
\label{sec:conclusion}

This paper presents a single-exponential running time algorithm for the simple
grammar bisimilarity problem.
We have implemented and thoroughly tested the basis-updating algorithm,
scheduled to appear in a
forthcoming release of the \freest programming language~\cite{freest}. 
Numerical experiments show that our implementation outperforms the current
implementation in \freest, which is based on the 2-EXPTIME algorithm for
context-free grammars.

The question whether a polynomial-time algorithm exists for simple grammar bisimilarity remains open. 
In our basis-updating updating algorithm, between any two consecutive changes in $\B$, new nodes are generated by applying BPA rules to existing nodes, \ie~the algorithm is essentially checking for coinductive congruence during these stages. 
This is the key point at which our analysis yields an exponential running time, since the number of new nodes generated is polynomial in the maximum seminorm among all words, which itself may grow exponentially. 
In order to reach a polynomial-time algorithm (or prove that none exist, conditioned on some complexity-theoretic assumption), it is thus worthwhile to gain a deeper understanding of the \emph{coinductive congruence decision problem}: 
\begin{prob}[Coinductive congruence]
Given a simple grammar $\G=(\V,\T,\pcal)$, a basis $\B$ over $\G$, and words of nonterminals $\gamma,\delta\in\V^\ast$, determine if $\iscongcoi{\gamma}{\delta}$.
\end{prob}

\noindent
We can restrict $\B$ to satisfy ``suitable'' assumptions such as those kept invariant by our algorithm (\Cref{def:basis-prop,lem:basis-invariant}). 
It might be possible to explore cyclic properties on the structure of congruent pairs of words to design an improved algorithm, in an approach similar to Hirshfeld~\etal~\cite{DBLP:journals/tcs/HirshfeldJM96} or Jančar~\cite{DBLP:journals/corr/abs-1207-2479}. 
Besides its theoretical importance, a polynomial-time algorithm would have a practical application in other formulations of context-free session types, such as those where infinite types are represented by means of (mutually recursive) equations rather than the $\REC a T$ recursion operator~\cite{DBLP:journals/tcs/CostaMPV24}.

Another significant open question in the area is closing the complexity gap for context-free grammar bisimilarity (between EXPTIME-hardness and 2-EXPTIME). 
It is not clear whether the techniques presented in this paper can be applied to the general setting. 
The main issue is that the uniqueness results in \Cref{lem:charnormedunnormed,lem:charnormnorm} no longer hold.
In fact, given $X$, $Y$, there might be finite but exponentially many non-bisimilar solutions to $\isbisim{X\alpha}{Y\beta}$. 
Natural avenues for future research are understanding the structure of such solutions and, for example, extending our basis-updating algorithm to allow for exponentially many BPA pairs $(X\alpha,Y\beta)$ in the basis for given nonterminals $X,Y$.


\bibliographystyle{alphaurl}
\bibliography{references}

\end{document}